\pdfoutput=1 

\documentclass[preprint]{aastex}
\usepackage{url}
\newcommand{\e}{et al.\ }

\slugcomment{Accepted by the Astronomical Journal}

\shorttitle{Near-Infrared Variability in the ONC}
\shortauthors{Rice, Reipurth, Wolk, Vaz, Cross}

\begin{document}

\title{Near-Infrared Variability in the Orion Nebula Cluster}

\author{Thomas S. Rice\altaffilmark{1},
Bo Reipurth\altaffilmark{2},
Scott J. Wolk\altaffilmark{3},\\
Luiz Paulo Vaz\altaffilmark{4},
and
N. J. G. Cross\altaffilmark{5},
}

\vspace{0.5cm}

\affil{1: Department of Astronomy, University of Michigan, 
          311 West Hall, 1085 South University Avenue, Ann Arbor, MI 48109, USA; tsrice@umich.edu}
\affil{2: Institute for Astronomy at Manoa and NASA Astrobiology Institute, 
          University of Hawaii, 640 N. Aohoku Place, Hilo, HI 96720, USA;
          reipurth@ifa.hawaii.edu}
\affil{3: Harvard-Smithsonian Center for Astrophysics, 
          60 Garden Street, Cambridge, MA 02138, USA; swolk@cfa.harvard.edu}
\affil{4: Departamento de F\'\i sica, ICEx, Universidade Federal de
  Minas Gerais, C.P. 702, 30.123-970, Belo Horizonte, MG, Brazil; vaz@fisica.ufmg.br }
\affil{5: Scottish Universities' Physics Alliance, Institute for
  Astronomy, School of Physics, University of Edinburgh, Royal
  Observatory, Blackford Hill, Edinburgh, EH9 3HJ, UK; njc@roe.ac.uk}
  \email{reipurth@ifa.hawaii.edu}

\begin{abstract}

% Near-infrared variability in young stellar objects (YSOs) with protoplanetary disks reveals information about the inner disks and star-disk interface. 
Using the United Kingdom Infrared Telescope on Mauna Kea, 
we have carried out a new near-infrared $J$, $H$, $K$ monitoring
survey of almost a square
degree of the star-forming Orion~Nebula~Cluster with observations on
120 nights over three observing
seasons, spanning a total of 894 days. 
We monitored $\sim$ 15,000 stars down to $J \approx 20$ using the 
WFCAM instrument, and have extracted 1203 significantly variable stars from our data. 
By studying variability in young stellar objects (YSOs) in the $H-K, K$ color-magnitude diagram, we are
able to distinguish between physical mechanisms of variability.
Many variables show color behavior indicating either dust-extinction or
disk/accretion activity, but we find that when monitored for longer
periods of time, a number of stars shift between these two
variability mechanisms.
Further, we show that the intrinsic timescale of disk/accretion
variability in young stars is longer than that of dust-extinction variability.
We confirm that variability amplitude is statistically correlated with evolutionary
class in all bands and colors.

Our investigations of these 1203 variables have revealed 73 periodic AA Tau type
variables, many large-amplitude and long-period ($P > 15$ day) YSOs, including three stars showing
widely-spaced periodic brightening events consistent with circumbinary disk activity, and four new eclipsing binaries.
These phenomena and others indicate the activity of long-term
disk/accretion variability processes taking place in young stars.
We have made the light curves and
associated data for these 1203 variables available online.

\end{abstract}

%% Keywords should appear after the \end{abstract} command. The uncommented

%% example has been keyed in ApJ style. See the instructions to authors
%% for the journal to which you are submitting your paper to determine
%% what keyword punctuation is appropriate.


%\clearpage

\section{Introduction} 

%!TEX root = varm42.tex

The Orion Nebula Cluster (ONC) is the richest young cluster in the
solar vicinity (for detailed reviews, see \citealt{muench08} and
\citealt{odell08}). At a distance of only $\sim$414$\pm$7~pc \citep{menten07}, 
it is possible to study the cluster in great detail.
The classic study of stars towards the ONC is due to \citet{parenago54},
followed by the work of \citet{herbig86}, \citet{jones88}, 
and \citet{hillenbrand97}, and most recently by catalogues of
stars based on optical and near-infrared HST data 
\citep{dario09, dario10, dario12, robberto13}.  
Major progress has been made in
identifying the population of young stars at various evolutionary
stages in the ONC. Large numbers of H$\alpha$ emission line stars have
been found centered on the Orion Trapezium (e.g., \citealt{haro53}, 
\citealt{parsamian82}, \citealt{wiram91}, 
\citealt{szegedi13}, \citealt{pettersson14}).  
Membership from near- and mid-infrared
studies has been determined by, e.g., \citet{hillenbrand00},
\citet{lada00}, \citet{muench02}, \citet{robberto05},
\citet{andersen11}, \citet{megeath12}, and \citet{billot12}.
Similarly, X-ray studies have identified a large population of young
stars (e.g., \citealt{getman05}, \citealt{prisinzano08}, \citealt{ignazio13}).

The first variable star recognized in the ONC, now known as AF~Ori, was found
around 1848 by W.C. Bond \citep{herbig82}.  During the 1920s and 1930s,
attention was drawn to numerous irregular variables in the region of
the Orion Nebula, and surveys to discover and characterize these stars
were carried out (e.g.,  \citealt{pickering04}, \citealt{ludendorff28},
\citealt{brun35}, \citealt{hoffmeister49}, \citealt{parenago54}, \citealt{rosino56}).  A variety
of classification schemes based on light curve behavior were devised
(for further references, see the monographs by \citealt{glasby74} and
\citealt*{hoffmeister84}), but none provided an important
key to understand the nature of these stars.  The spectroscopic
identification of T Tauri stars as a class (\citealt{Joy45}, \citealt{herbig62}) and
the recognition of T Tauri associations \citep{ambartsumian49} eventually
provided the basis for a physical understanding of the large
populations of variable stars associated with dark clouds.

The photometric behavior of young stars in associations is best
described as irregular, with an enormous range in amplitude, time
scale and frequency of fluctuations. This is discussed in more detail
in subsequent sections. It gradually became evident
that the variability of young stars is difficult to characterize, and
in particular that the type of variability, e.g., the RW Aur stars,
does not define homogeneous groups of stars (e.g., \citealt{herbig60}).
Spectroscopy, on the other hand, emerged as a powerful tool to probe
the nature of young stars, reducing interest in photometric studies.
In recent years, wide-field optical and infrared array detectors have
become much more common, and as a result large-scale photometric
surveys have again attracted interest.

Optical surveys have mainly focused on periodic variables, due to cool
or hot spots on the stars \citep[e.g.,][]{stassun99, herbst00, herbst02, rodriguez09}. 
A few studies have also focused on the irregularly variable stars 
\citep[e.g.,][]{parihar09}, and lately infrared studies, including those performed with
Spitzer and Herschel, have produced a wealth of results
(\citealt{carpenter01}, \citealt{morales11}, \citealt{billot12}).
\citet{carpenter01} studied a $0\fdg8 \times 6 \degr$ strip in the Orion A molecular
cloud centered on the ONC and complete to $K_s = 14.8$.
This $\sim 50$-day study showed variability arising from starspots, dust
extinction, accretion variability, and changes of the inner disk. 
However, no previous infrared monitoring campaign in the ONC has had more than
two months of continuous coverage, which limits our understanding
of long-term variability, especially if young stars change their
short-term behavior over long timescales.
Strong suggestions of such long-term behavior were seen in the $JHK$ monitoring
study of a small number of YSOs in the Cygnus OB7 star-forming region
presented by \citet{rice12} and \citet{wolk13b}.

We have carried out
monitoring observations of the Orion Nebula Cluster in the $J$, $H$,
$K$ near-infrared bands on 120 nights spanning 894 days from late
2006 to mid-2009.
These observations, obtained using the WFCAM instrument on the United
Kingdom Infrared Telescope (UKIRT), have allowed us to identify 1203
variable stars within a 53~$\times$~53 arcmin region centered on the
Trapezium stars.
The evolutionary class of nearly all of these 1203 variables has been
determined by \citet{megeath12}, allowing us to derive NIR
variability properties as a function of class.
As established in previous $JHK$ surveys of young stars \cite[cf.][]{carpenter01, rice12, wolk13b},
different variability mechanisms impart different color signatures in
$JHK$ data: 
starspot-induced variability is nearly colorless,
variable dust extinction makes stars redder as they get \textit{fainter},
and accretion variability or other changes in inner disk geometry cause
stars to get redder as they get \textit{brighter}.
We identify these different motions in $H-K, K$
color-magnitude space and measure the fraction of stars showing either
dust-reddening or disk/accretion variability, noting that the fraction
of disk/accretion variables increases with longer monitoring
campaigns.

The UKIRT observing campaign
is described in Section~\ref{sec_data_obs} along with the data
reduction.
Our strategy for selection of variables and other methodology is described in
Section~\ref{variable_selection}:
to select variable stars we use the Stetson index \citep{stetson96}; we
identify periods through the Lomb-Scargle periodogram \citep{press89} and the 
fast chi-squared algorithm (f$\chi^2$, \citealt{palmer09}); and we
cross-match many of our targets to catalogs in the literature in order
to obtain further information, such as the \textit{Spitzer}-derived
evolutionary classes from \citet{megeath12}.
In Section~\ref{sec:analysis}, we show that variability mechanisms can
be constrained via the color-color and color-magnitude diagrams; we
derive the fraction of stars that show blueing behavior (associated
with accretion/disk activity) versus reddening behavior (associated
with dust occultation), and find that this ratio evolves significantly with the monitoring duration; many
individual systems were seen to switch between color variability modes
during this campaign.
In Section~\ref{sec:timescales}, we show that different color variability modes have different timescales, with blueing (accretion/disk activity) variability showing longer typical timescales.
We report a number of exceptional variables in
Section~\ref{sec:exceptional}, including a newly-identified class of stars with widely-separated 
periodic brightening events, ``dipper'' objects \citep[cf.][]{morales11, stauffer14} 
and periodic AA Tau variables, and other stars with long IR periods; Section~\ref{sec:eb} identifies four previously unreported eclipsing binary systems.
We connect our results with other recent infrared monitoring surveys in
the concluding commentary in Section~\ref{sec:discussion}.

%!TEX root = varm42.tex

\section{OBSERVATIONS AND DATA}
\label{sec_data_obs}

\subsection{Observations: The UKIRT Campaign}
\label{subsec_obs}

\begin{figure}

\plotone{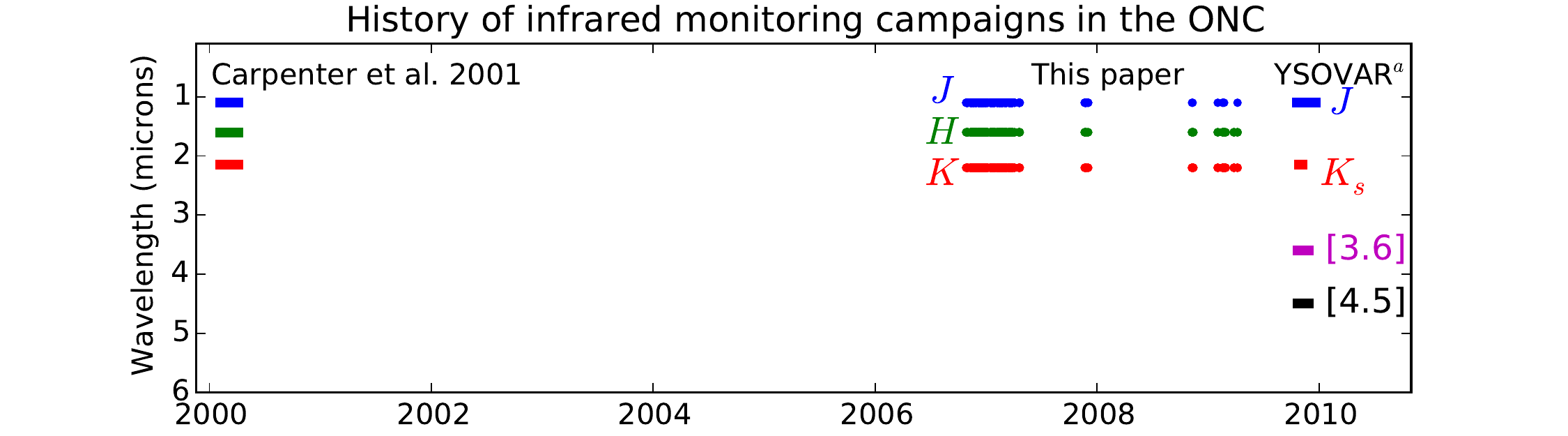}
\caption{ A history of infrared monitoring campaigns in the Orion
  Nebula Cluster published to date, with each band ($J$, $H$, $K$,
  $[3.6]$, $[4.5]$) shown in a different color.
  The survey presented in this paper lasted from October 2006 to
  April 2009.  \newline
  Notes\textendash $~^a$ YSOVAR refers to \citet{morales11}. \label{fig:comparison_observing_log} }
\end{figure}

% Global paragraph
Our data comprise $J$, $H$, $K$ observations of a 0.9 $\times$ 0.9 degree
field towards the Orion Nebula Cluster that were taken with the Wide Field
Camera (WFCAM) instrument on the 3.8 m United Kingdom Infrared
Telescope (UKIRT), which sits atop Mauna Kea, Hawaii at 4194 m
elevation. 
Observations have been taken on 120 nights between October 2006 and April 2009 over three
observing seasons: 
October 26, 2006 -- April 21, 2007;
November 21, 2007 -- December 2, 2007;
and November 8, 2008 -- April 7, 2009.
This time coverage is longer than any previous infrared monitoring
campaign in the ONC; see Figure \ref{fig:comparison_observing_log} for
a time coverage comparison.

\begin{figure}
\plotone{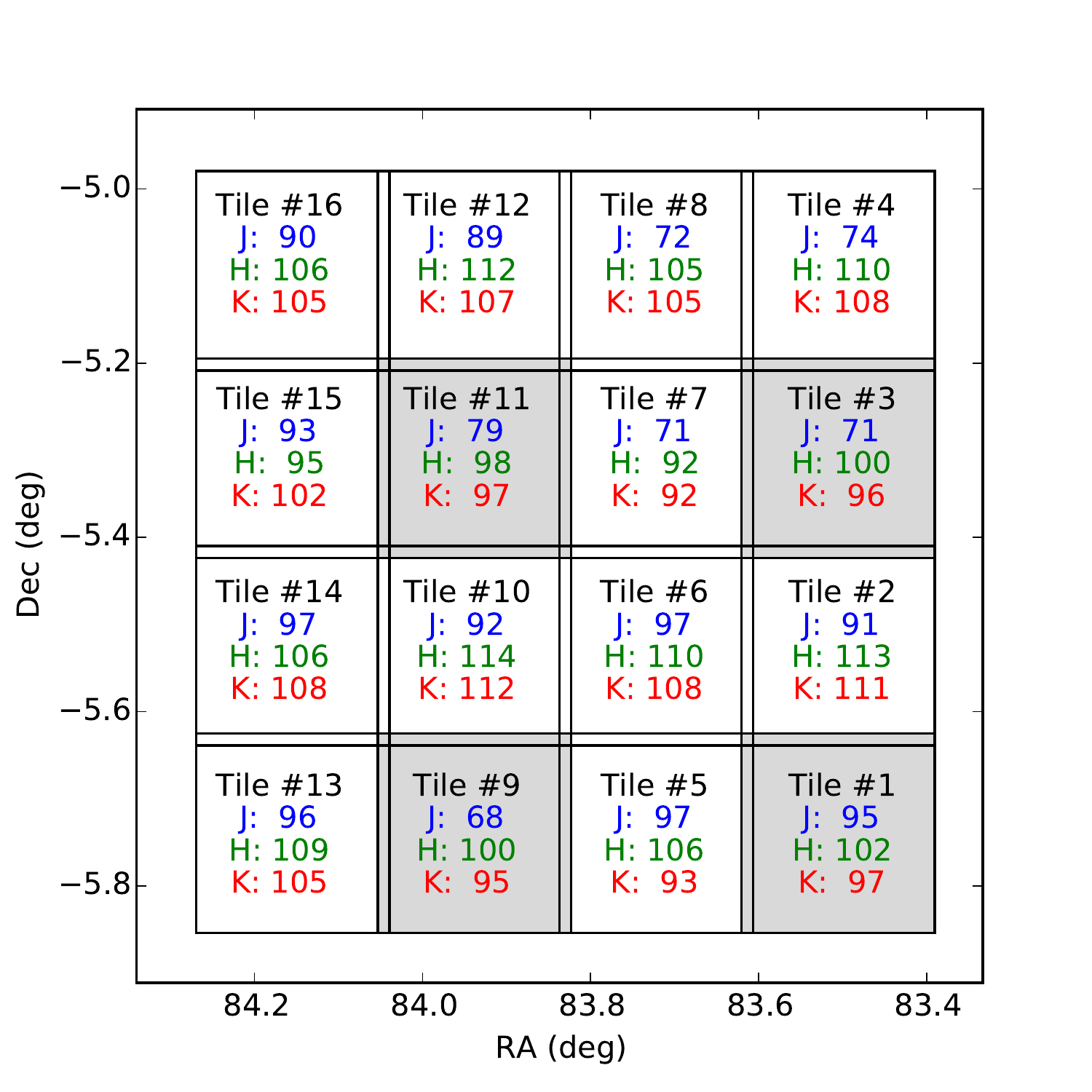}
\caption{An illustration of the observed tiles in our field, with the
  effective number of observations in each band written in each tile, i.e.,
 the number of observations that
remain after the data cleaning process described in \S
\ref{subsec:data}.
The four gray tiles illustrate a single detector ``pawprint'' of WFCAM; to image the
entire square degree, four pointings of this pawprint are required.
The spacings between WFCAM chips cause some overlap in the pawprints, so a small area of the field is imaged multiple times per night.
\label{fig:observing_log_map} }
\end{figure}

% Instrument description
The WFCAM instrument, described in \citet{casali07},  consists of
four Rockwell Hawaii-II PACE arrays of $2048 \times 2048$ pixels. 
The detectors are widely spaced such that four pointings are required
to image a contiguous square field on the sky (Figure \ref{fig:observing_log_map});
this $2\times2$ observing strategy ultimately defines the $53' \times 53'$ field of view of our survey.
These ONC observations use the same instrument and configuration
outlined in \citeauthor{rice12} (\citeyear{rice12}; see that paper for a more detailed
description of this observing setup).
WFCAM was first used to study YSO variability by
\citet{alves08}.

\begin{figure}

  \plotone{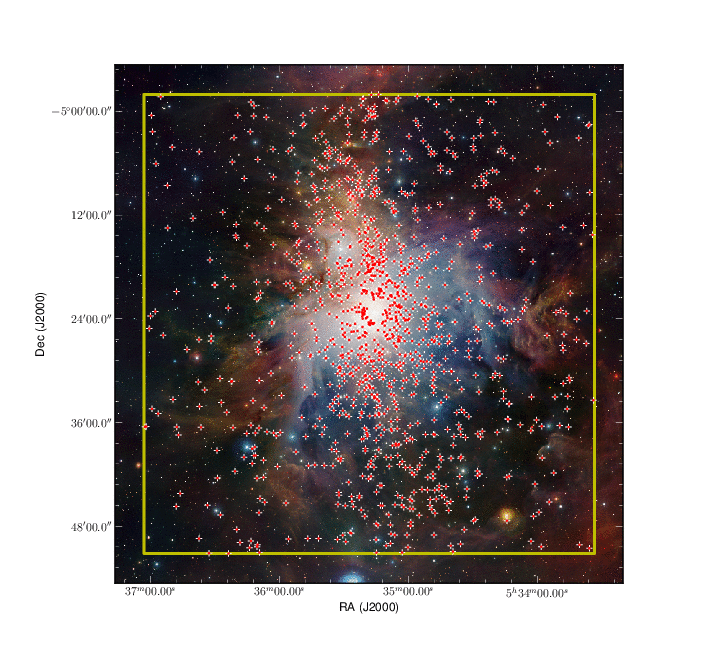}

  \caption{
    \label{fig1}
    Spatial distribution of the 1203 variable stars selected in \S \ref{variable_selection}.
    Each red marker on this map shows the position of one variable. 
    Our observations are centered on
    the Orion Nebula Cluster and cover a $\sim 0.9 \times 0.9
    \textrm{~deg}^2$ field.
    The brightest stars in the ONC, including the well-known Trapezium stars
    at the center, are saturated in our images, preventing us from
    studying their variability.
    The highest density of variables runs along the north-south ridge corresponding to a filament of active star formation.
    For comparison, a detailed map of dust optical depth in this region can be found in \citet{lombardi14}.
    Credit for underlying M42 near-infrared ($Z$, $J$, $K_s$) image:
    ESO/J. Emerson/VISTA. 
    Acknowledgment: Cambridge Astronomical Survey Unit
    }
\end{figure}

% Observing description
The WFCAM observations are centered on the coordinates
$\alpha_{2000}=$ 05:35:21.0, $\delta_{2000}=$ $-05$:24:35, spanning $\alpha$ 05:33:33.7 to 05:37:08.4 and
$\delta$ $-04$:57:57 to $-05$:51:14.
This region is displayed in Figure \ref{fig1}.
During each of the three observing seasons, data were taken about once per night; 
each night's observations were identical in pointing and filter choice.
Because four pointings from the four-detector array are required to
cover the target field, and we observed at three wavelengths, one
night's data comprises $4 \times 4 \times 3 = 48$ image ``frames''.
As discussed in section \ref{subsec:data}, we have removed a number of these frames
during processing due to poor data quality.
The weather during these winters was sometimes poor, so much
of our data cleaning strategy revolves around mitigating the effect of
this bad weather on our light curves by removing suspicious data that
are likely to have large errors.

The observations came from UKIRT WFCAM programmes U/06B/H54, U/07B/H60B, 
and U/09A/H33 and were combined into a single WSA combined programme WSERV5.
The data in this paper come from the data release WSERV5v20120208,
and can be accessed at \url{http://surveys.roe.ac.uk/wsa}.

\begin{figure}
\plotone{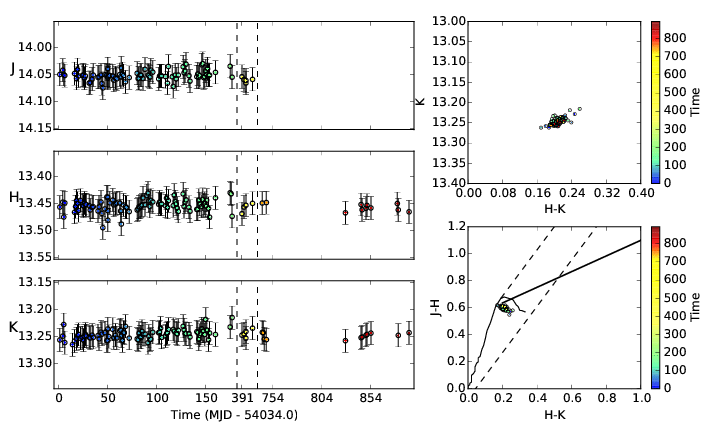}
\caption{ An example light curve of a constant (reference) star, to demonstrate the level of variability detectable in this study. Errors in each band are about 0.02 mag, depicted by whiskers around the photometry points. The vertical dashed lines mark where ``empty'' stretches of time were omitted from the plot. In the bottom-right panel, a color-color diagram is shown, with the solid lines denoting a main sequence locus \citep[thin]{koornneef83} and a CTTS locus \citep[thick]{meyer97}, and dashed lines showing reddening vectors using the ISM reddening law of \citet{rieke85}, extending from the extrema of the main sequence and CTTS curves. 
These solid, dotted, and dashed lines appear in all light curves and color-color diagrams throughout this paper.}
\label{fig:reference_lightcurve}
\end{figure}

\begin{figure}
\plotone{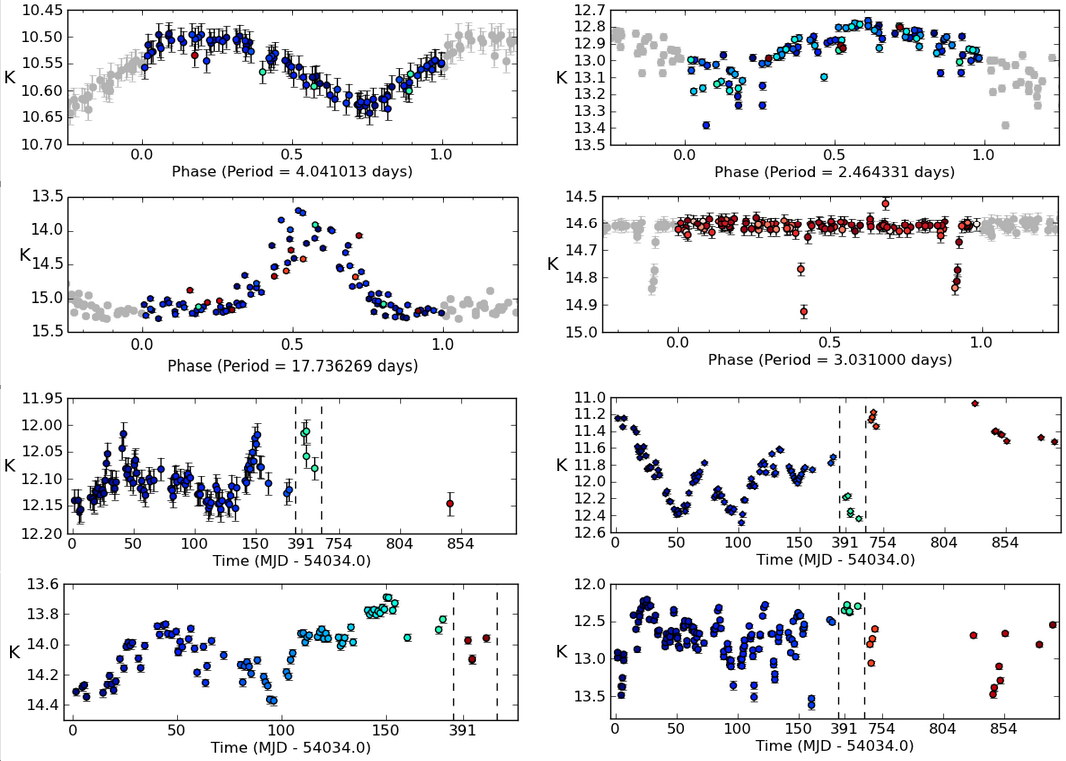}
\caption{ Eight example light curves highlighting the
  variety of periodic and nonperiodic behaviors seen in our data.
The top four panels show periodic variables folded on their own
periods; the bottom four panels show nonperiodic variables displayed
without folding.
\label{fig:example_lcs} }
\end{figure}

\subsection{Data: Processing and Cleaning}
\label{subsec:data}

Before using the images obtained from UKIRT to study
variability in the ONC, we needed to extract the photometry 
and then ``clean'' it to ensure its integrity.
The data have been reduced and processed for analysis in several stages: 
first, the raw images in minimum schedulable blocks (MSB) were reduced to catalogs and calibrated at the 
Cambridge Astronomy Survey Unit 
\citep[CASU;][]{irwin04, hodgkin09}
Second, the reduced, calibrated images and extracted catalogs were transferred to the Wide Field Astronomy Unit in Edinburgh where they were ingested into the WFCAM Science Archive (WSA) and multiple MSBs' data were processed together to produce deep images, multi-band catalogs, links from unique sources to epochs, variability tables and external surveys 
\citep[see][]{hambly08, cross09}
Third, the catalogs were analyzed and cleaned after retrieval from WSA.
We briefly describe these reduction and processing steps in this subsection.

The data reduction, calibration and archiving that form the first and second step were done as part of the VISTA Data Flow System designed to process near-infrared data from WFCAM and VISTA 
\cite[VDFS;][]{emerson04}, in a similar fashion to the data used in \citet{rice12}.
Stellar magnitudes are calibrated against local Two Micron All-Sky Survey (2MASS) standard sources with high signal-to-noise 
\citep[see][]{hodgkin09}.
The WSA produces a catalog $J$, $H$, $K$ magnitudes and colors for each epoch 
(see \citealt{cross09}, \S 2.2 for details),
with a maximum epoch duration of 720 minutes.
Additionally a single master catalog is created from merging $J$, $H$, $K$ catalogs extracted from deep stacks of all epochs.
As a result of the size of this survey, a large number of stars have data quality issues (such as saturation, blending, bad pixels, and other effects).
These stars are given error flags by the WSA pipeline whenever they
arise, and we have used these error flags to help determine which
light curves have a quality high enough to allow automatic light curve
classification (see Appendix \ref{sub:autosubj}).

Some regions in individual nightly catalogs have been removed by WFAU due
to obvious quality issues. 
We rejected 137 detector frames (there are 4 in a normal multiframe)
because they had a dead detector or all the channels were bad, and one
channel because it had a value of 0 for the seeing, indicating that a non-sensical catalogue was produced. 
We also rejected 738 detector frames when we recalibrated the data
(small adjustments to improve the light curves) because the
adjustments were $>0.05$ mag. 
This is usually an indicator that there is a problem with the frame.

\begin{figure}
\plotone{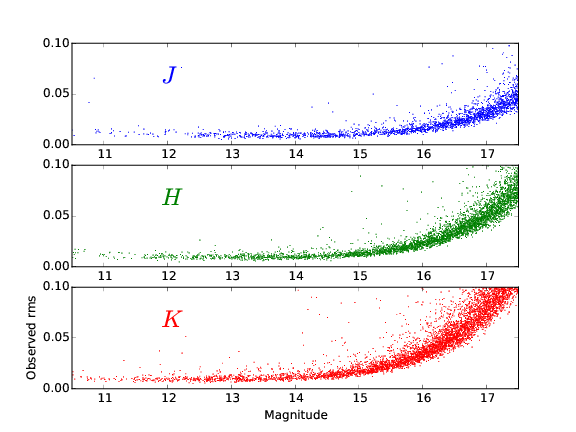}
\caption{
Observed rms error as a function of magnitude, measured for nonvariable stars.
The photometry has roughly 2\% error down to a limit of $J=16$, $H=15.3$, and
$K=14.8$, and better than 5\% error to $J=17.3$, $H=16.7$, and
$K=16.2$. 
We restricted our highest-quality subset of variable stars (``$Q=2$'') to data
within $11 \le J \le 17$, $11 \le H \le 16$ and $11 \le K \le 16$,
among other criteria.
\label{fig:sensitivity_per_band} }
\end{figure}

\begin{figure}
\plotone{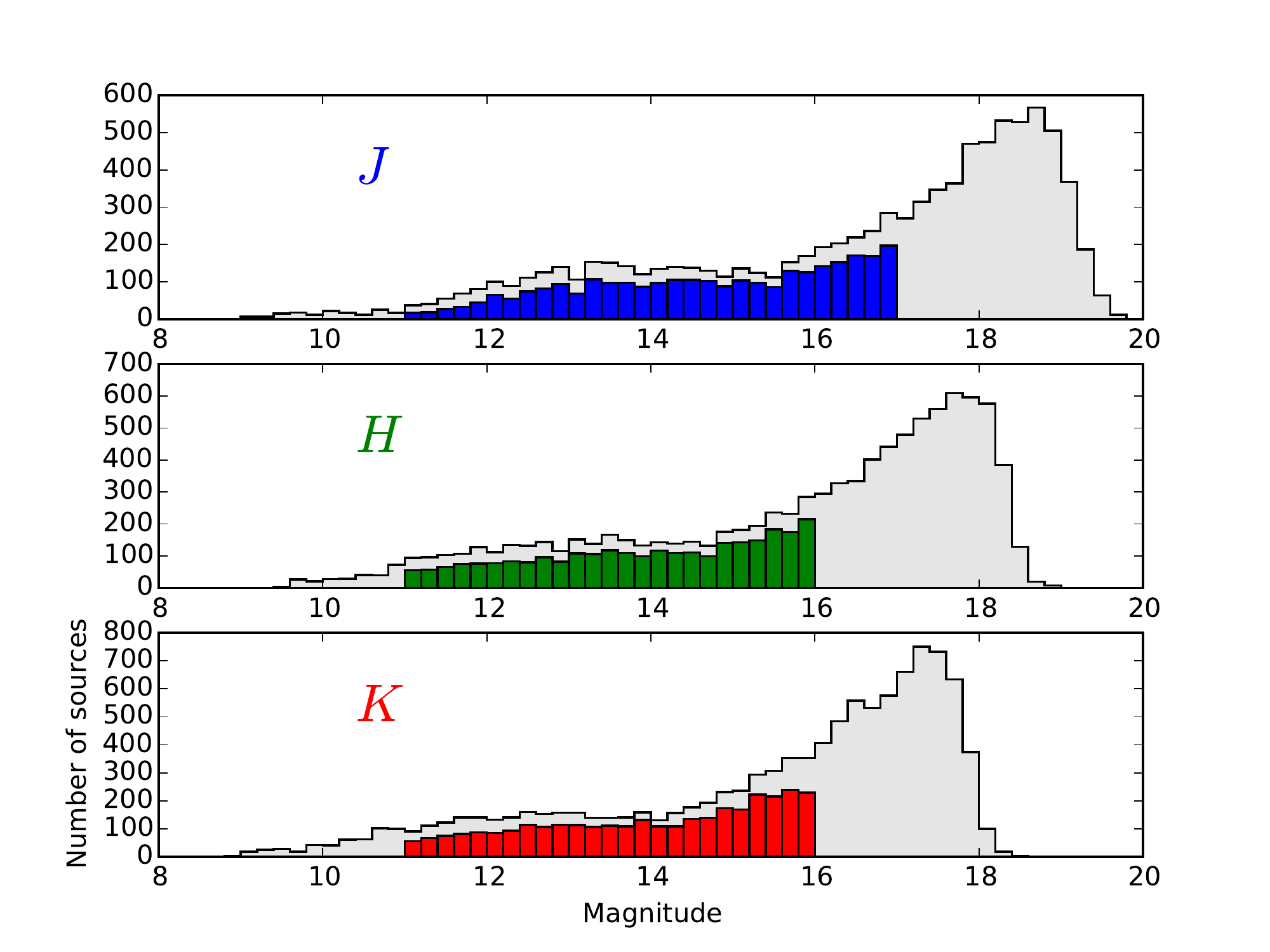}
\caption{
Distribution of mean magnitudes among sources, per band.
The empty (light gray) region shows the distribution for all sources detected at least 50 times; the filled regions show only stars that meet the stricter $Q=1$ or $Q=2$ quality requirements.
\label{fig:flux_per_band} }
\end{figure}

After obtaining the night-by-night catalog of photometry for all
sources, we have applied a systematic correction to the pipeline-estimated photometric
errors. 
The correction, derived by \citet{hodgkin09} and following 
\citet{rice12} appears as follows:
\begin{equation}
M^2 = c  E^2 + s^2
\end{equation}
\noindent where $M$ is the measured (i.e., realistic) total error, $E$
is the
pipeline-estimated photometric error, and $c = 1.082$ and $s = 0.021$; $c$ and $s$ are empirically determined.
The observational error as a function of magnitude is plotted in Figure \ref{fig:sensitivity_per_band} for each band.

We removed individual exposures showing unusually high errors in photometry
relative to typical observations.\footnote{ In this context, an
  ``exposure'' refers to photometry at one band
that was collected using one pointing on one night; a typical
exposure covers four tiles/``frames'', spaced as in Figure
\ref{fig:observing_log_map},  though some exposures had one or more
frames removed by WFAU as described above.
Thus, a typical night had $4 \times 3 = 12$ exposures, for 4 pointings
and 3 bands.}
To quantify the fidelity of a given exposure, we have selected a
subset of likely constant ``reference'' stars using a version of the
Stetson index (described in \S \ref{sec:stetson} as part of our variability
criteria); reference stars were also required to be between magnitudes 13 and 15.5, 
in order to guarantee high signal-to-noise while widely avoiding saturation effects.
An example reference star light curve is shown in Figure \ref{fig:reference_lightcurve}.
At least 20 reference stars were selected in each tile to ensure that no
single star could disqualify an exposure.
These stars' median observed magnitude was tabulated, and then each
night's photometry was compared to this median magnitude for each night.
We have assessed photometric fidelity on each night by measuring the fraction of reference
stars that fall within 0.05 mag of their median magnitude.
In the vast majority of observations, fewer than 10\% of reference
stars (i.e., $>90\%$ fidelity) deviate by 0.05 mag; we have chosen to discard all exposures in
which more than 20\% of reference stars deviate from their typical magnitude by
at least 0.05 mag (i.e., fidelity less than 80\%) as unreliable for our analysis.

Next, after noting that most ``clean'' light curves still had one or a
few flagged datapoints in each band, we removed those minority errant datapoints
(in this context, data ``flags'' note data
quality issues such as deblending, edge effects, saturation,
diffraction spikes, etc.).
This procedure analyzed each band (of $J$, $H$, and $K$)
independently: 
for each band on each star, the number of flagged versus flag-free
datapoints is counted.
If a star has $> 90\%$ flag-free data in a given band, then any
flagged datapoints are removed.
Finally, all datapoints with error $> 0.5$
mag are removed, because such data occasionally clutter
light curves that otherwise contain good data.

From the above procedure, we extract 14,728 lightcurves meeting the minimum quality requirement ($Q=0$) or better; a distribution of their median magnitudes is shown in Figure \ref{fig:flux_per_band} for each band.
Example light curves of various types can be seen in Figure \ref{fig:example_lcs}.

%\clearpage

%!TEX root = varm42.tex

\section{VARIABLE SELECTION}
\label{variable_selection}

% Global paragraph
In this study we focus on stars with significant variability.
We have chosen to use the Stetson index
\citep[and citations therein]{stetson96}
to quantify variability seen in our $JHK$ light curves; this index
is ideal for measuring variability that is correlated between bands,
and so it makes the best use of our three-band data\footnote{
While the Stetson index de-emphasizes anti-correlated variability in principle,
we find that stars showing obvious anticorrelated events
 have generally correlated variability in practice, leading to large
Stetson indices; thus, the use of Stetson is useful even for stars with some anticorrelated variability. 
See Appendix \ref{sec:lowvars} for further
discussion.}.
In order to characterize the periodicity of these stars' variability,
we have used the Lomb-Scargle Periodogram
(see \citealt{press89} for our implementation),
which searches for sinusoidal periods in irregularly time-sampled data, 
as well as the Fast Chi-Squared (f$\chi^2$) method developed by
\citet{palmer09}
in specific cases where it was clear that the variability was not
sinusoidal (e.g., eclipsing binaries).

\subsection{Stetson Index}
\label{sec:stetson}

The Stetson variability index\footnote{We rename the $J$ Stetson index to ``$S$'' to
  avoid confusion with the $J$ wavelength band, following previous
  authors' usage.} \citet{stetson96} is defined as follows:

\begin{equation}
  S = \frac{ \sum_{i=1}^{p} g_i ~\textrm{sgn} \left(P_i\right)\sqrt{\left|P_i\right|}}
      {\sum_{i=1}^{n} g_i},
\end{equation}

\noindent where $p$ is the number of pairs of simultaneous observations of a star, and
$P_i = \delta_{j(i)}\delta_{k(i)}$ is the product of normalized residuals of two observations (where the normalized residual is defined

\begin{equation}
  \delta_i = \sqrt{\frac{n}{n-1}}\frac{m_i - \overline{m}}{\sigma_i}
\end{equation}

\noindent for a given band; $n$ is the number of observations in that band, $m_i$ represents the magnitude at a given epoch, $\sigma_i$ is the observational uncertainty on $m_i$, and $\overline{m}$ is the average magnitude of the star). 
Most stars in our analysis have many simultaneous $J$, $H$ and $K$ measurements.
In the Stetson index, three simultaneous observations are treated with $g_i = \frac23$ for each pair of measurements, giving a total weight 2 for the set of three pairs.
Many stars lack complete data in three bands (for example, they are
too faint at $J$ or too bright at $K$); for these stars, a two-band or
one-band Stetson index has been calculated where appropriate, with
weights set at $g_i = 1$.

The Stetson index is designed to identify significant variables at $S \ge 1.0$ when noise properties of data are well characterized; truly nonvariable stars will cluster around $S = 0$.
In practice, previous authors have found selection criteria of $S \gtrsim 0.5$ 
\citep{carpenter01, morales11, flaherty13}
to be useful for determining minimal variability criteria\footnote{Low-amplitude variable candidates are discussed in Appendix \ref{sec:lowvars}}.

Because of the large range of photometric quality issues that arise in
a major dataset such as ours, we have categorized all our sources into one
of three quality ``bins''.
Briefly, the stars are divided into $Q=2$, $Q=1$, or $Q=0$.
$Q=2$ stars have completely clean data in all three bands; 
$Q=1$ stars have one or two bands with completely clean data;
$Q=0$ stars have some quality issue in all three bands.

Each quality bin has different variability selection criteria, due to the varying confidence in light curve fidelity.
In all cases, the minimum variable criteria are $S \ge 1.0$ and at least 50 detections in at least one band; stars in lower ``quality'' bins were required to meet further criteria.
$Q=2$ and $Q=1$ stars have been extracted algorithmically, while $Q=0$ stars required visual light curve inspection to be included in our list of variables. 
Full details are provided in Appendix \ref{sub:autosubj}.
We were very conservative in selecting $Q=0$ variables, taking
care to only include light curves that are clearly not false signatures from instrumental or weather-related effects.
No $Q=0$ stars (variable or non-variable) are included in any
statistical analyses presented later in this paper.

\begin{figure}
  \plotone{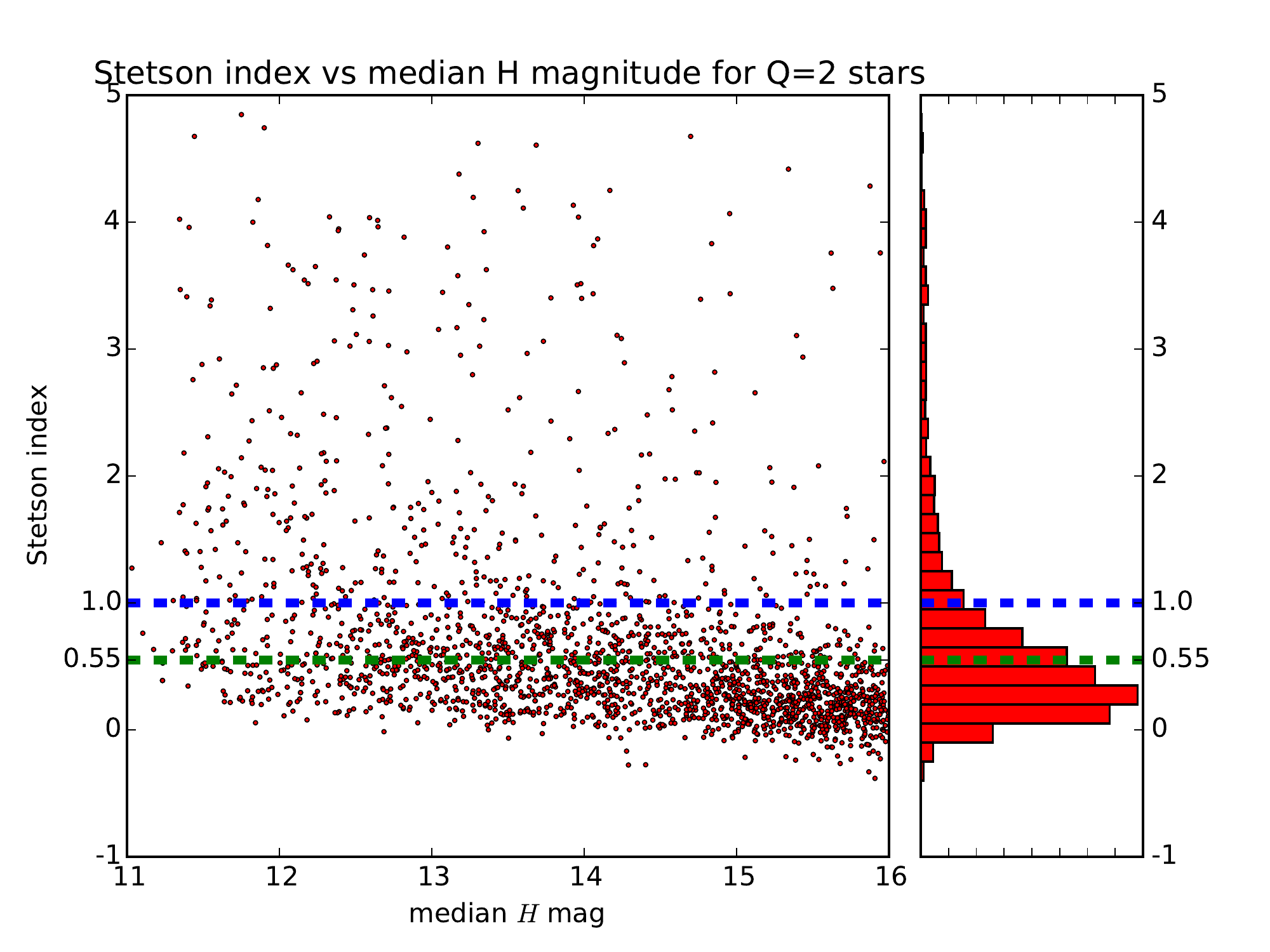}
  \caption{
    \label{fig:stetson_vs_hmag}
Distribution of the three-band Stetson index versus median $H$ magnitude, for all 2,348 $Q=2$ stars. 
The distribution peaks at a median value of $S=0.487$; we select $S \ge 1.0$ as our strong variable star definition.
Stars between $1.0 < S < 0.55$ are excluded from our analysis but are discussed briefly in Appendix \ref{sec:lowvars}.
150 stars in this sample have $S > 5$ and are not shown in this plot.
}

\end{figure}

Using the Stetson Index and the selection criteria described in
Appendix \ref{sub:autosubj}, we identify 1203 significant variable stars, out of 14,728 stars tested.
The breakdown of these 1203 variables into quality classes is
described in Table \ref{tab:variablebreakdown} as $Q=2$, $Q=1$ and $Q=0$. 
The distribution of Stetson indices among the $Q=2$ sources is
clustered around a median index $S = 0.487$ as shown in Figure \ref{fig:stetson_vs_hmag}.

\subsection{Periodicity Analysis: LSP, F$\chi^2$}
\label{subsection:period}

Much observed variability among stars is periodic. 
To detect periodic variability in our data, we have applied the Lomb-Scargle periodogram to the $J$, $H$, and $K$ band light curve of every single variable candidate.
The Lomb-Scargle periodogram was developed to identify and characterize periodic, sinusoidal signals in irregularly time-sampled data, and has been used widely in other studies of variability in young stars 
(see, e.g., \citealt{attridge92}, \citealt{stassun99}, \citealt{rebull01}).

Among light curves that suggest a periodic (often high-frequency)
non-sinusoidal signal, we have used the Fast Chi-Squared (f$\chi^2$)
method developed by \citet{palmer09} to get a more accurate period fit
than the Lomb-Scargle periodogram can provide.
This is most effective for eclipsing binaries, W UMa ``contact
binary'' variables, as well as stars that show any sort of abrupt
``dip'' or ``burst'' that is periodic but surrounded by a flat,
``quiescent'' light curve.

\subsubsection{Automatic period detection algorithm}
\label{subsubsection:autoperiod}

\begin{figure}
 \plottwo{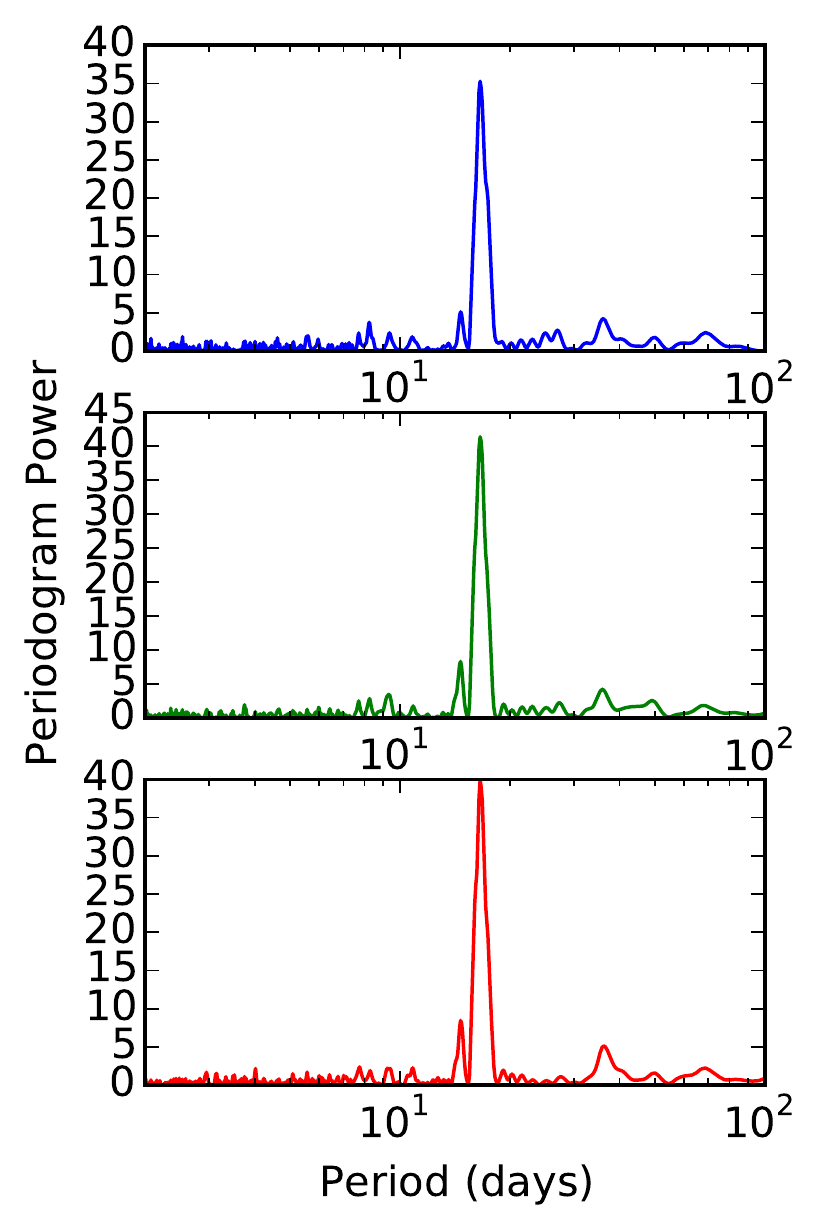}{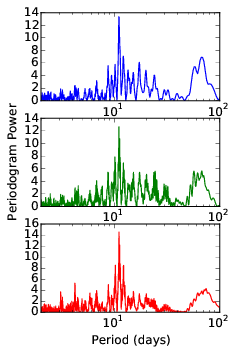}
 \caption{
 \label{fig:lsp}
 Three-band Lomb-Scargle periodograms for two stars, plotted in period space from 2 to 100 days. 
 From top to bottom: $J$ (blue), $H$ (green), $K$ (red).
 Left column: periodogram of star ONCvar 40, a rotating star (light curve shown in Fig. \ref{lc:rotators}). 
 It shows a strong, sharp peak at 16.6 days, indicating a clean periodic signal.
 Right column: periodogram of star ONCvar 553 (light curve shown in Fig. \ref{fig:aatau_tenpanel}).
 Its 11.0 day period is significant, but has notable structure, indicating that the lightcurve has substantial irregular variability superposed on a periodically varying signal.
 }
 \end{figure}

In order to efficiently detect strong periodic signals in our large
dataset of light curves, we have created a filter to automatically
classify stars as periodic or non-periodic based on their Lomb-Scargle
periodogram strength. 
The Lomb-Scargle periodogram returns a plot of strength versus
frequency; periodogram strength can be transformed into a
corresponding ``false-alarm probability,'' assuming Gaussian errors
around perfect sinusoidal signals.
Example periodograms for two stars are shown in Figure \ref{fig:lsp}.

We have calculated Lomb-Scargle periodograms in the range $2-50$ days
for all candidate variable stars; 
if a star showed periodogram strength exceeding 15 in a single band
(i.e., false alarm probability, or FAP, $< 0.0002$) or exceeding 12 at the same period, within a
tolerance of 5\%, in two or more bands (i.e., FAP $< 0.0041$ in each
band independently), then we classified it as either a ``periodic
variable'' (if it was in $Q=2$ or $Q=1$ and was selected as a variable) or a
``subjective periodic candidate'' (if it was in $Q=0$ and was selected as a candidate variable, pre-inspection).
These filters have been tuned by hand after much trial-and-error and
inspection of individual light curves, and are meant to detect the strongest,
repeating periodic variables rather than being sensitive to every
statistically-significant periodic signal.
Because of the large sample size of our dataset, this 
optimization is preferable to imposing a target false-alarm
probability from the top down.

The lower limit of 2 day periods comes from a Nyquist limit (i.e., a fundamental sampling limit) on our ability to identify short-period sinusoidal signals in our data; excluding period fits below 2 days guarantees that true periods above 2 days are not aliased to below 2 days, at the potential cost of aliasing some (poorly-sampled) sub-2-day periods upwards.
This aliasing effect arises from the fact that a given set of discretely sampled points can typically be fit by two different sine waves: one with a frequency above the Nyquist frequency of half the sampling frequency, and one with a frequency below the Nyquist frequency.
The upper period limit of 50 days was chosen to ensure that periodic variability was reliably repeatable, and would be observed to repeat multiple times in the data. 

\subsection{Results of Variability Selection}
\label{sec:results_of_variability}

\begin{deluxetable}{ l | c c | c | c |  c}
  \tablewidth{0pt}
  \tablecaption{ Breakdown of variables into period \& quality bins
\label{tab:variablebreakdown} }
  \startdata
      & Periodic & Non-Periodic & P+NP & Non-variable & Total \\ \hline
    Q=2 & 231 & 318 & 549 & 1799 & 2348\\
    Q=1 & 116 & 192 & 308 & 936  & 1244\\
    Q=0 & 193 & 153 & 346 & 10790$^a$ & 11136\\ \hline
    total & 540 &  663 & 1203 & 13525 & 14728\\
    \enddata
  \tablecomments{$^a$ For $Q=0$, the ``Non-variable'' column enumerates stars not selected as variable, but which are not necessarily non-variable; see Appendix \ref{sub:autosubj}.}

\end{deluxetable}

% here's the deluxetable version.
% \begin{deluxetable}{ l | c c | c}
%   \tablecaption{ Breakdown of variables into period \& quality bins
%     \label{tab:variablebreakdown} }
% \tablewidth{0pt}
% \tablehead{
%     & Periodic & Non-Periodic & P+NP \\
% }
% \startdata
%     Q=2 & 231 & 318 & 549 \\
%     Q=1 & 116 & 192 & 308 \\
%     Q=0 & 192 & 153 & 345 \\ 
%     total & 539 &  663 & 1203 \\
% \enddata

% \end{deluxetable}

The breakdown of the 1203 variables into quality bins and periodic/nonperiodic bins is shown in Table \ref{tab:variablebreakdown}. 
In this section, we briefly characterize the overall variability and periodicity properties of our variables, as well as compare our set of variable stars to a number of catalogs from the literature, allowing us to comment on the accuracy of our period search, identify cross-matched identifications in SIMBAD and other datasets, and assign \textit{Spitzer} photometry and evolutionary classes to 90\% of our stars.
The coordinates, SIMBAD cross-matched names, median WFCAM photometry, \textit{Spitzer} photometry, and evolutionary classes of all 1203 variables are listed in Table \ref{table1}.
Stellar classes are taken from \citet{megeath12} and are
used extensively in our analysis in \S \ref{sec:analysis}.
SED classification suffers from a degeneracy between the extinction
and source inclination (both poorly constrained), which in particular
affects the determination of source luminosity \citep{fischer12}
and possibly also the YSO spectral class. This should be kept in mind
in the following discussion.

%!TEX root = varm42.tex

\begin{deluxetable}{ccccccrrrrrrrc}
\tabletypesize{\scriptsize}
\rotate
\tablecaption{ Basic Properties of 1203 Significantly Variable Stars 
\label{table1} }
\tablewidth{0pt}

\tablehead{
\colhead{ ID } & \colhead{ $\textrm{RA}_{J2000}$ } & \colhead{ $\textrm{Dec}_{J2000}$ } & \colhead{ SIMBAD alt. ID } & \colhead{ Q } & \colhead{ P } & \colhead{ $J$ } & \colhead{ $H$ } & \colhead{ $K$ } & \colhead{ [3.6] } & \colhead{ [4.5] } & \colhead{ [5.8] } & \colhead{ [8.0] } & \colhead{ Class${^1}$ } \\
}
\startdata
1 & 05:33:33.90 & -05:33:26.5 & V* V385 Ori & 1 & 1 & 11.91$\pm$.02 & 11.20$\pm$.02 & 10.90$\pm$.02 & 10.40 & 10.15 & 9.79 & 9.24 & D \\
2 & 05:33:34.43 & -05:14:17.7 & 2M05333443-0514177 & 2 & 0 & 12.54$\pm$.02 & 11.76$\pm$.02 & 11.28$\pm$.02 & 10.79 & 10.50 & 10.04 & 9.05 & D \\
3 & 05:33:35.71 & -05:09:23.5 & 2M05333571-0509235 & 2 & 0 & 12.43$\pm$.02 & 11.68$\pm$.02 & 11.39$\pm$.02 & 11.16 & 11.11 & 10.99 & 10.95 & ND \\
4 & 05:33:35.74 & -05:50:33.6 &  & 2 & 0 & 13.59$\pm$.02 & 13.00$\pm$.02 & 12.77$\pm$.02 & 12.60 & 12.60 & 12.45 & 12.36 & ND \\
5 & 05:33:35.88 & -05:01:32.4 & V* VY Ori & 1 & 0 & 11.63$\pm$.02 & 10.84$\pm$.02 & 10.29$\pm$.02 & 9.61 & 9.20 & 8.84 & 8.00 & D \\
6 & 05:33:36.37 & -05:01:40.4 & 2M05333636-0501405 & 2 & 1 & 13.26$\pm$.02 & 12.70$\pm$.02 & 12.50$\pm$.02 & 12.40 & 12.33 & 12.30 & 12.25 & ND \\
7 & 05:33:37.05 & -05:23:06.9 & V* V725 Ori & 1 & 1 & 12.64$\pm$.02 & 11.95$\pm$.02 & 11.71$\pm$.02 & 11.59 & 11.55 & 11.45 & 11.36 & ND \\
8 & 05:33:37.24 & -05:26:52.8 & V* HU Ori & 0 & 1 & 9.66$\pm$.02 & 9.26$\pm$.02 & 9.04$\pm$.02 & 8.94 & 8.96 & 8.94 & 8.90 & ND \\
9 & 05:33:38.56 & -05:13:12.4 & Haro 4-98 & 2 & 0 & 12.81$\pm$.02 & 12.12$\pm$.02 & 11.79$\pm$.02 & 11.16 & 10.84 & 10.42 & 9.68 & D \\
10 & 05:33:40.37 & -05:50:09.4 &  & 2 & 0 & 14.45$\pm$.02 & 13.55$\pm$.02 & 13.23$\pm$.02 & 13.05 & 13.04 & 12.92 & \ldots & ND \\
11 & 05:33:40.82 & -05:03:05.2 &  & 2 & 0 & 15.64$\pm$.02 & 14.84$\pm$.02 & 14.55$\pm$.02 & 14.30 & 14.24 & \ldots & \ldots & ND \\
12 & 05:33:41.67 & -05:24:04.2 & Haro 4-121 & 2 & 0 & 12.57$\pm$.02 & 11.85$\pm$.02 & 11.55$\pm$.02 & 11.13 & 10.83 & 10.39 & 9.51 & D \\
13 & 05:33:41.82 & -05:26:25.7 &  & 2 & 1 & 16.03$\pm$.03 & 13.97$\pm$.02 & 12.76$\pm$.02 & 11.97 & 11.74 & 11.51 & 11.43 & ND \\
14 & 05:33:44.72 & -05:21:04.0 & V* V730 Ori & 1 & 0 & 11.44$\pm$.02 & 10.87$\pm$.02 & 10.72$\pm$.02 & 10.67 & 10.66 & 10.55 & 10.28 & ND \\
15 & 05:33:44.93 & -05:31:08.5 & V* V386 Ori & 1 & 1 & 11.19$\pm$.02 & 10.47$\pm$.02 & 10.16$\pm$.02 & 9.92 & 9.88 & 9.79 & 9.78 & ND \\
16 & 05:33:45.26 & -05:30:49.8 & V* VZ Ori & 0 & 0 & 10.43$\pm$.02 & 9.64$\pm$.02 & 9.05$\pm$.02 & 7.98 & 7.42 & 6.83 & 5.83 & D \\
17 & 05:33:45.45 & -05:36:32.4 & 2M05334545-0536323 & 1 & 0 & 11.97$\pm$.02 & 11.30$\pm$.02 & 10.91$\pm$.02 & 10.18 & 9.99 & 9.44 & 8.15 & D \\
18 & 05:33:45.87 & -05:32:58.1 & V* V1896 Ori & 2 & 1 & 15.69$\pm$.02 & 14.31$\pm$.02 & 12.95$\pm$.02 & 11.67 & 10.33 & 9.72 & 8.45 & P \\
19 & 05:33:46.09 & -05:34:26.5 & V* V1667 Ori & 1 & 1 & 11.78$\pm$.02 & 11.15$\pm$.02 & 10.99$\pm$.02 & 10.87 & 10.86 & 10.80 & 10.83 & ND \\
20 & 05:33:46.68 & -05:23:25.5 & V* V1897 Ori & 1 & 0 & 13.77$\pm$.02 & 12.25$\pm$.02 & 10.79$\pm$.02 & 9.52 & 8.85 & 8.41 & 7.71 & D \\
21 & 05:33:47.64 & -05:25:48.5 & V* V1898 Ori & 2 & 0 & 12.45$\pm$.02 & 11.82$\pm$.02 & 11.51$\pm$.02 & 11.16 & 11.02 & 10.65 & 9.61 & ND \\
22 & 05:33:47.92 & -05:31:30.1 &  & 2 & 1 & 15.05$\pm$.02 & 14.53$\pm$.02 & 14.19$\pm$.02 & 13.91 & 13.78 & \ldots & \ldots & ND \\
23 & 05:33:47.97 & -05:27:38.4 & 2M05334797-0527385 & 2 & 1 & 14.40$\pm$.02 & 13.60$\pm$.02 & 13.01$\pm$.02 & 12.41 & 12.11 & 11.81 & 11.08 & D \\
24 & 05:33:48.22 & -05:13:26.1 & V* V1900 Ori & 2 & 1 & 13.02$\pm$.02 & 12.24$\pm$.02 & 11.70$\pm$.02 & 10.99 & 10.70 & 10.23 & 9.49 & D \\
25 & 05:33:48.27 & -05:27:05.3 &  & 1 & 0 & 17.94$\pm$.06 & 16.38$\pm$.03 & 14.59$\pm$.02 & 12.19 & 11.10 & 10.30 & 9.59 & D \\
26 & 05:33:49.18 & -05:23:02.3 & 2M05334918-0523023 & 2 & 1 & 12.94$\pm$.02 & 12.36$\pm$.02 & 12.06$\pm$.02 & 11.79 & 11.72 & 11.48 & \ldots & ND \\
27 & 05:33:49.54 & -05:36:20.8 & V* V354 Ori & 2 & 0 & 13.07$\pm$.02 & 11.97$\pm$.02 & 11.14$\pm$.02 & 10.06 & 9.53 & 9.02 & 8.34 & D \\
28 & 05:33:50.74 & -05:00:39.4 & V* V1902 Ori & 2 & 0 & 13.28$\pm$.02 & 12.57$\pm$.02 & 12.19$\pm$.02 & 11.97 & 11.84 & 11.69 & 11.26 & D \\
% 29 & 05:33:51.31 & -05:23:16.4 & V* V1903 Ori & 1 & 0 & 14.31$\pm$.02 & 12.11$\pm$.02 & 10.32$\pm$.02 & 8.36 & 7.66 & 7.15 & 6.28 & D \\
% 30 & 05:33:51.79 & -05:33:03.5 & V* V1905 Ori & 2 & 0 & 15.15$\pm$.02 & 13.28$\pm$.02 & 12.08$\pm$.02 & 10.59 & 9.92 & 9.30 & 8.38 & D \\
\enddata
\tablecomments{Sample; the remainder of the table is available electronically. Column Q refers to quality class, and column P denotes whether a star is periodic (1) or not (0). $J$, $H$, $K$ values listed are median values, with representative errors on the median measurement. $^1$ P=``Protostar'', D=``Disked'', ND=``Non-disked'' (classes derived from \citet{megeath12}).}
\end{deluxetable}

\subsubsection{Global Variability Properties}

By considering the total number of stars (variable and nonvariable) that met the $Q=2$ or $Q=1$ criteria, we can study the fraction of stars in our input catalog that are variable, periodic or not. 
This approach minimizes selection biases, because all $Q=2$ and $Q=1$
stars have been subject to nearly the same data quality, magnitude, PSF, and spatial cuts, the only difference being the number of bands which passed those tests.
Results are statistically indistinguishable when the $Q=1$ and $Q=2$
groups were considered separately, so for global considerations we
combine them while excluding $Q=0$ data.
There are 3592 $Q=1\&2$ stars of which 857 are variables with $S \ge 1$.
Overall, the fraction of our 3592 stars that are variable is thus 24.2\%.
Of these, the fraction of variables that we found periodic is 0.41.
Therefore, the percentage of stars in the overall dataset that are periodic variables is 9.9\%.

Our results can be compared to other NIR monitoring campaigns of this
region: 
\citet{carpenter01}, who covered a much larger area ($0\fdg8 \times
6 \degr$, oriented north-south) at $JHK_s$ to a shallower magnitude of
$K_s = 14.8$;
and \citet{morales11}, who studied a similarly-sized field
at $[3.6]$ and $[4.5]$ \textit{Spitzer} bands that overlaps with most
of our observed field.
\citet{carpenter01} found 233 periodic variables among 1235
variables in their 17,808 star sample (i.e., $\sim 7\%$ of their
overall sample was variable, a fraction $\sim 0.19$ of their
variables were periodic, giving $\sim 1.3 \%$ of the overall star sample
as periodic);
\citet{morales11} identified a higher proportion, 501 periodic variables among
1259 variables in a 2069-star sample (i.e., $\sim 60\%$ of the
pre-selected sample were variable, a fraction $\sim 0.40$ of their
variables were periodic, giving $\sim 24.2\%$ of the pre-selected star
sample as periodic).
The clear differences between the three surveys are due to several factors: a shallower survey will probe only the upper, less-populated part of the initial mass function and will not penetrate as deep into higher extinction regions, while a survey at longer wavelengths is likely to target younger sources that are more likely to be variable.
Beyond this, the Morales-Calder\'on survey focused on a pre-selected star sample.

Because our variability and periodicity cuts are conservative and
specifically exclude very low amplitude variability and short
periods, these fractions can be taken as lower limits for the
intrinsic periodicity and variability fractions of stars in our field.

\subsubsection{Literature Cross-Matching}
\label{sec:crossmatching}

Because the ONC is an extremely well-studied region, we have been able
to access numerous catalogs containing previously-published data
relating to the variable stars we present in this paper.
We have obtained previously measured periods, \textit{Spitzer}-derived
evolutionary classes, and XMM-Newton-derived evolutionary classes, from the literature.

We compare the periods found via our technique described in \S \ref{subsubsection:autoperiod} to six catalogs of periods from the literature:
\citet{carpenter01},
\citet{morales11},
\citet{herbst02},
\citet{parihar09},
\citet{rodriguez09}, 
and the General Catalog of Variable Stars \citep{samus09}.
The Herbst, Parihar, and Rodr\'iguez-Ledesma campaigns were all done in the optical, finding periods spanning $\frac23$ to 22.2 days, 0.9 to 19.1 days, and 0.09 to 16.9 days, respectively.
The Carpenter study was a NIR ($J$, $H$, $K_s$) project, while Morales-Calder\'on used the $[3.6]$ and $[4.5]$ \textit{Spitzer} IRAC bands.

\begin{figure}
  \plotone{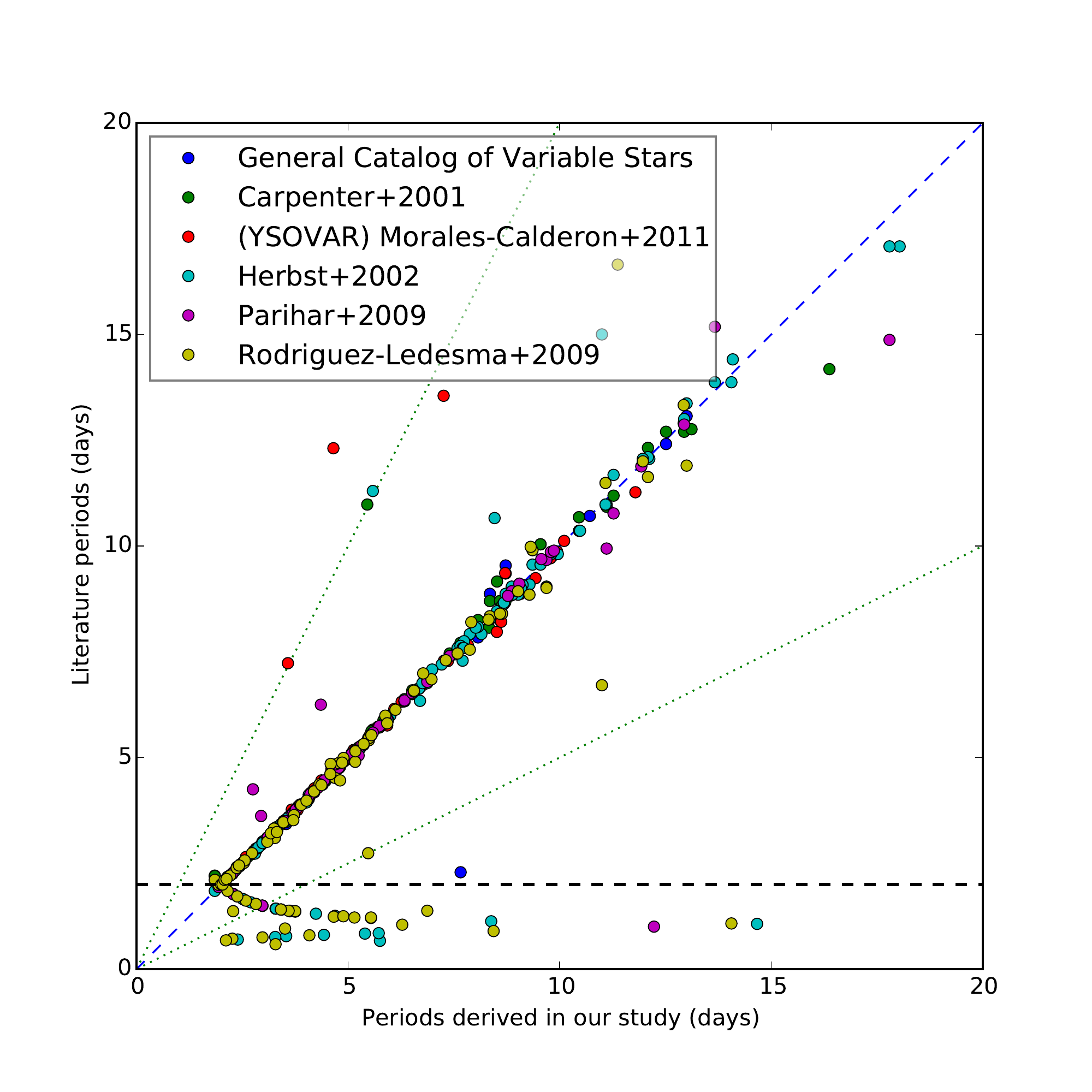}

  \caption{
    \label{fig:period_lit_comparisons}
    Comparison of previously-published periods with periods derived in this study, for the 297 stars in which the data overlap. A strong general agreement is seen between literature periods and our periods, except where the literature periods are shorter than two days; such periods get aliased to longer periods in our data. 
    The two nearly horizontal curves that appear below two days on the y-axis are due to such aliasing, which is expected given the time resolution of our study; a black dashed line drawn at $y=2$ is drawn to indicate that periods below this are affected by aliasing.
In addition to the overlapping periods, we find 242 periodic variables unknown to have periods in the literature. 
We expect that the accuracy of these newly reported periods is comparable to that seen among our periods for the previously-known periodic stars.  
}

\end{figure}

Of the 539 periodic variables we discuss in this paper, a comparison
with the literature shows that 297 of these stars are already known to
be periodic, while 242 are previously unknown periodic stars. 
Where our period data overlap with the literature, they largely agree, as shown in Figure \ref{fig:period_lit_comparisons}. 
There is disagreement where the literature periods are shorter than two
days because our sampling is not designed to be sensitive to these
shorter periods.
For overlapping stars, the fraction of our periods that agree with published periods within $\pm 10\%$ is tabulated in Table \ref{table:period_accuracy}.
The agreement is greatest for the two infrared studies (\citealt{carpenter01} and \citealt{morales11}); 
this is indicative that infrared and optical variability do not systematically arise from the same mechanism(s).

\begin{deluxetable}{ccl}
  \tablecaption{Period agreement for each catalog
    \label{table:period_accuracy} }

  \tablewidth{0pt}
  \tablehead{ 
    \colhead{ } &
    \colhead{ Fraction that agree } & 
    \colhead{ Number of matches } \\
    \colhead{Period catalog} &
    \colhead{ within $\pm 10\%$} &
   \colhead{ to this catalog }
  }
  \startdata
  GCVS & 89.58\% & of 48 \\
  Carpenter+01 & 94.59\% & of 74 \\
  Morales-Calder\'on+11& 90.00\% & of 40 \\
  Herbst+02 & 79.74\% & of 153 \\
  Parihar+09 & 77.05\% & of 61 \\
  Rodriguez-Ledesma+09 & 66.67\% & of 96 \\
  \enddata
\end{deluxetable}

The overlap is not perfect:
194 of our non-periodic variables were previously reported in the literature as periodic.
Because the bulk of previous period studies in the ONC were done at optical wavelengths, and some of them had far finer time resolution than our $\sim$nightly survey, the non-recovery of these 194 periods is likely attributable to 
\textit{a)} the observation that infrared variability is often more irregular (or simply different) than optical variability (see simultaneous optical and infrared monitoring of YSOs in \citealt{cody14}), due to the emission of the disk at infrared wavelengths, and
\textit{b)} our survey is not designed to recover periods below two days.
Indeed, among these 194 stars, 44 have literature periods below two days (22\% of the sample), so case \textit{b)} accounts for almost one-fourth of the missing periodic stars.

In order to determine evolutionary classes for our variable stars, we made use of the 
\citet{megeath12} study.
This study identifies YSOs based on their infrared data from 2MASS ($J$, $H$, $K_s$ near-infrared bands)
and \textit{Spitzer} (IRAC $[3.6]$, $[4.5]$, $[5.8]$, $[8.0]$ and MIPS $[24]$ mid-infrared bands).
Stars with significant infrared excesses are identified as YSOs, and those meeting a spectral index criterion of 
$\alpha \ge -0.3$, where $\alpha = d\lambda F_\lambda / d\lambda$, are further identified as protostellar YSOs.
The subsequent catalog classifies all sources as ``Protostar'', ``Disked'', ``non-Disked'', or ``data too imprecise to determine''.
As noted in the first paragraph of Section \ref{sec:results_of_variability}, this class determination suffers from inclination degeneracy.
We have cross-matched our variable stars to this \citet{megeath12} catalog and
found that 90\% have a corresponding
evolutionary class.
Of our 1203 variables, 47 match to protostars, 641 to disked stars,
and 390 to diskless stars, with the remaining 125 either matching to
stars with \textit{Spitzer} colors too imprecise to determine a class,
or failing to match to a \textit{Spitzer} source altogether\footnote{These non-matches did not have a counterpart in the Megeath catalog within 1 arcsecond. Most of the non-matches are in crowded, easily confused regions, and the UKIRT seeing is typically better than \textit{Spitzer}'s images, so we are not concerned that we have included spurious sources in our variability catalog.}.
A comparison of periodicity (among $Q=2$ variables) to evolutionary class
shows 27\% of protostars, 34\% of disked stars, and 55\% of diskless
stars as periodic in our data.

We have also cross-matched our sources to the X-ray identified YSOs in
\citet{ignazio13}, which was an XMM-Newton study of the ONC. 
This catalog includes X-ray identified YSOs, and assigned them classes
(``Protostar'', ``disked'', and ``Class III''') based on the \citet{megeath12}
classification; through this, young stars can be identified independent of the
presence of a disk. 
Class III status was assigned to X-ray sources with no IR-detected disk.
Of the 850 sources in this X-ray catalog, 270 matched to stars in
our study, of which 166 had estimated evolutionary classes.

%\clearpage

\section{Analysis and Results}

%!TEX root = varm42.tex

\label{sec:analysis}

In addition to the overall variability statistics presented in the previous section, we have analyzed these stars' variability in light of prior information about the stars themselves as well as near-infrared variability mechanisms.
As discussed in \S \ref{sec:crossmatching}, the \citet{megeath12} catalog provides \textit{Spitzer}-derived evolutionary classes for most of our variables; we analyze the dependence of variability on evolutionary class in \S \ref{sec:variability_by_class}, 
and the dependence of period distribution on class in \S \ref{sec:periods_by_class}.
\citet{carpenter01} found that a star's slope in the $H-K, K$ color-magnitude diagram can be used to constrain or identify the physical mechanism causing its variability; in \S \ref{sec:colorslope} we analyze the color slope behavior of the bulk population as well as individual stars.

To give some context about the physical properties of such young stars and their disks, a reference such as \citet{eisner09} can be consulted. The PMS stars RY Tau, DG Tau, and RW Aur A were measured to have inner disk walls in the range $0.14-0.18$ AU for these three systems, with disk wall temperatures of $1260-1850$ K.
Such systems have orbital periods of $5-20$ days for their inner disk walls; other typical, lower-mass stars have inner walls at 0.1 AU, with a roughly 11 day orbital period.

\begin{figure}
  \plotone{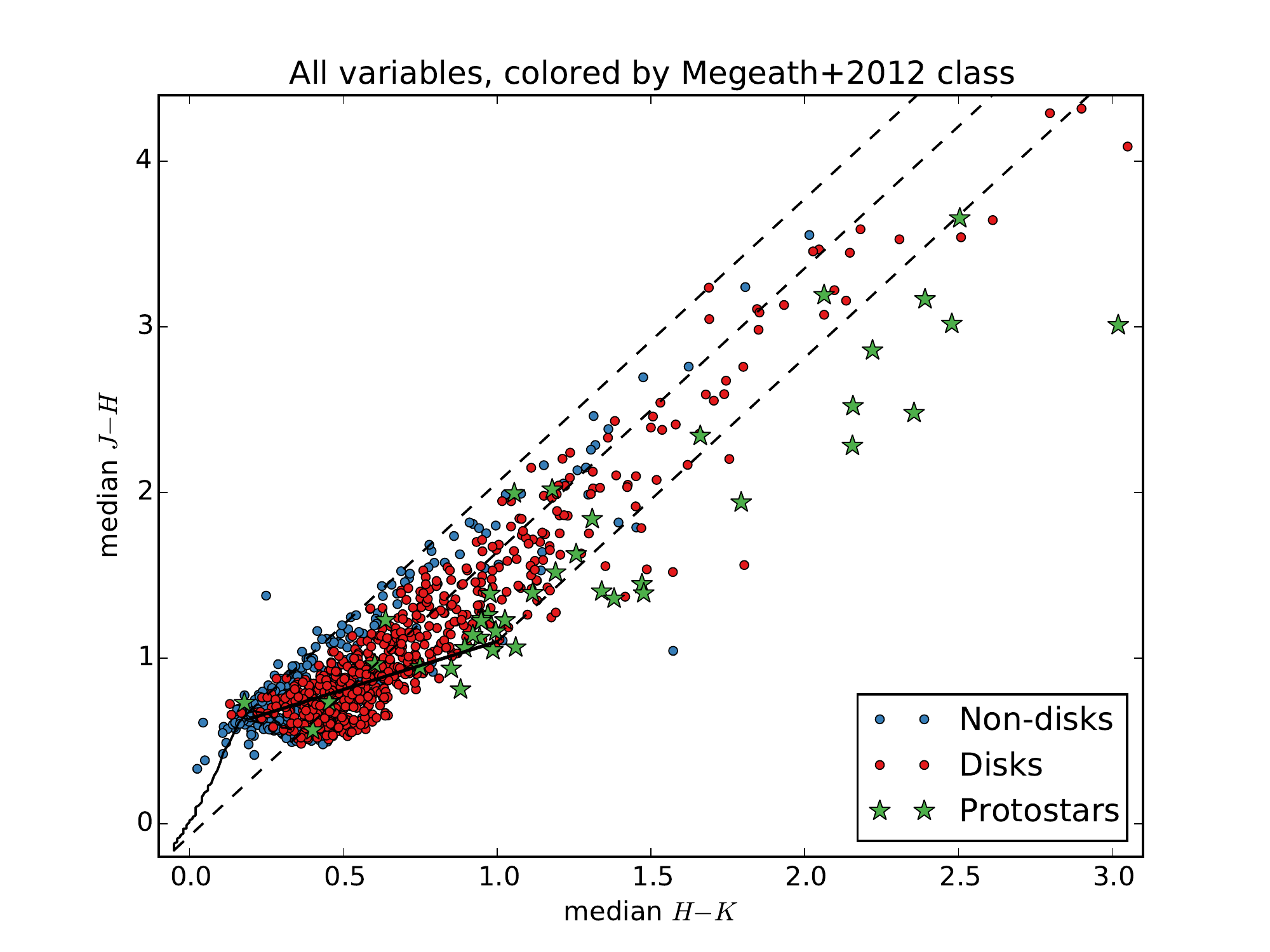}
  \caption{ Median $J-H, H-K$ color-color diagram of the 1,077 variable stars discussed in this paper that have an evolutionary class from \citet{megeath12}.
As in Figure \ref{fig:reference_lightcurve}'s color-color diagram,
the solid lines in the bottom-left show the main sequence \citep{koornneef83} and the CTTS locus of \citet{meyer97}, while dashed lines extending to the upper right show reddening vectors of typical ISM dust \citep{rieke85}.
Stars with disks tend to have larger near-infrared excesses than diskless stars, with protostars showing the largest excesses.
Note that only the median values of $J-H$ and $H-K$ are plotted for each star; in reality many stars change their position in color-color space, with less-evolved stars showing the greatest color changes (see Figure \ref{fig:delta_hist}).
\label{fig:cc_by_class}
}
\end{figure}

A color-color diagram showing the median $J-H, H-K$ values of our
variables, by evolutionary class, shows that disked stars tend
to have larger near-infrared excesses than diskless stars (Figure
\ref{fig:cc_by_class}).
Note that such a color-color diagram only captures the median color of
a star in our observations; in reality, individual stars move
substantially throughout color-color space, making near-infrared
excesses variable \citep{rice12}.

\subsection{Variability as a Function of Class}
\label{sec:variability_by_class}

\begin{figure}
  \plotone{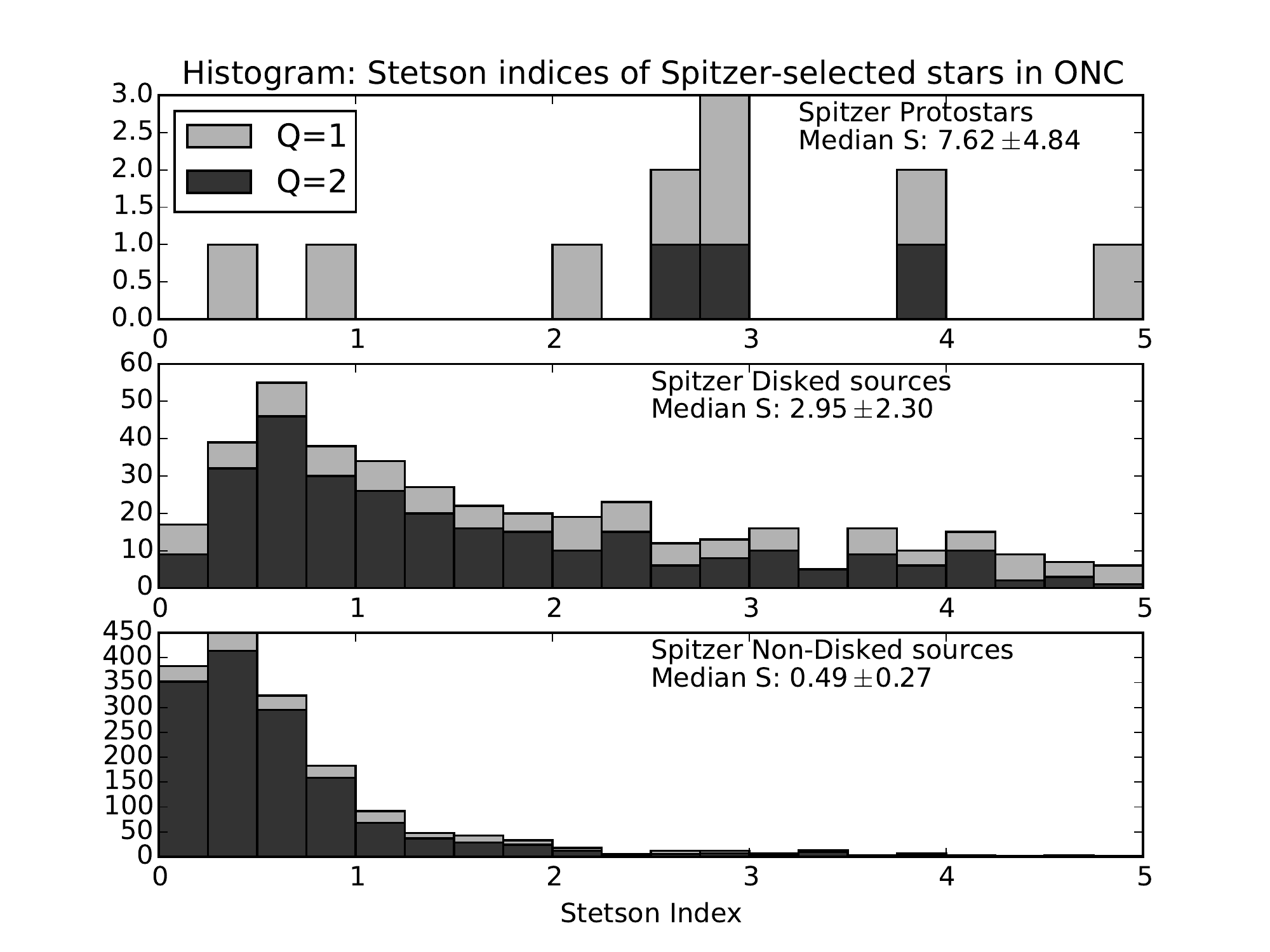}
  \caption{ Stetson index for each class. 
Median Stetson indices are computed for the aggregate Q=1 + Q=2.
The x-axis is limited to the range $0-5$ to highlight the differences between the distributions;
in the top panel, 16 sources lie beyond $S=5$; in the middle panel, 232 sources lie beyond $S=5$; in the bottom panel, 22 sources lie beyond $S=5$.
\label{fig:stetson_by_class} }
\end{figure}

Young stars are known to be more variable than main-sequence stars (see e.g., \citealt{morales11}; \citealt{herbst02}); using the evolutionary classes associated with the stars in our sample, we can refine this observed trend and investigate the variability associated with each class.
As discussed in \S \ref{sec:crossmatching}, we obtain evolutionary classes for 90\% of our variables from the \citet{megeath12} catalog; 
many non-variables in our sample also have classes.
A large fraction of these are identified as protostars or disked stars: among 2373 $Q=1+2$ stars (both variable and nonvariable) with Megeath identification, 27 are protostars and 639 are disked, while 1707 are nondisked. 
The Stetson index statistics presented below are from this sample.

The disked sources have a median Stetson index of $3.0 \pm 2.3$
\footnote{The presented uncertainties on all median values are calculated as the median absolute deviation.}
(Figure \ref{fig:stetson_by_class}), while the protostars are fewer in number and have a median value of $S= 7.6 \pm 4.8$.
Since the Stetson index is an aggregate of behavior across all bands, we analyze variability strength at different wavelengths \textit{among variable stars} by inspecting amplitudes in each band\footnote{
When calculating $\Delta mag$ for a given star, extreme ($3 \sigma$) outliers in a star's lightcurve are iteratively clipped so that errant data cannot bias the magnitude ranges upwards for low-amplitude variables.}.
In the following analysis we restrict ourselves to the 522 $Q=2$ variables with Megeath classes, of which 11 are protostars, 288 are disked, and 223 are nondisked.
Again we find the protostars are far more variable:
median $\Delta K$ ($K_{max}-K_{min}$)
for the variable protostars
is 0.61 $\pm$ 0.34 
while the corresponding value for the other disked variables is 0.21$\pm$ 0.10. 
Non-disked stars are typically much less variable: median $\Delta K = 0.09 \pm 0.02$.
If the $Q=1$ protostars are included alongside the $Q=2$ protostars, giving 27 protostars in this analysis, the median
$\Delta K$ ($K_{max}-K_{min}$
is 0.492 $\pm$ 0.29, which is consistent with the $Q=2$-only case.

\begin{figure}
\plottwo{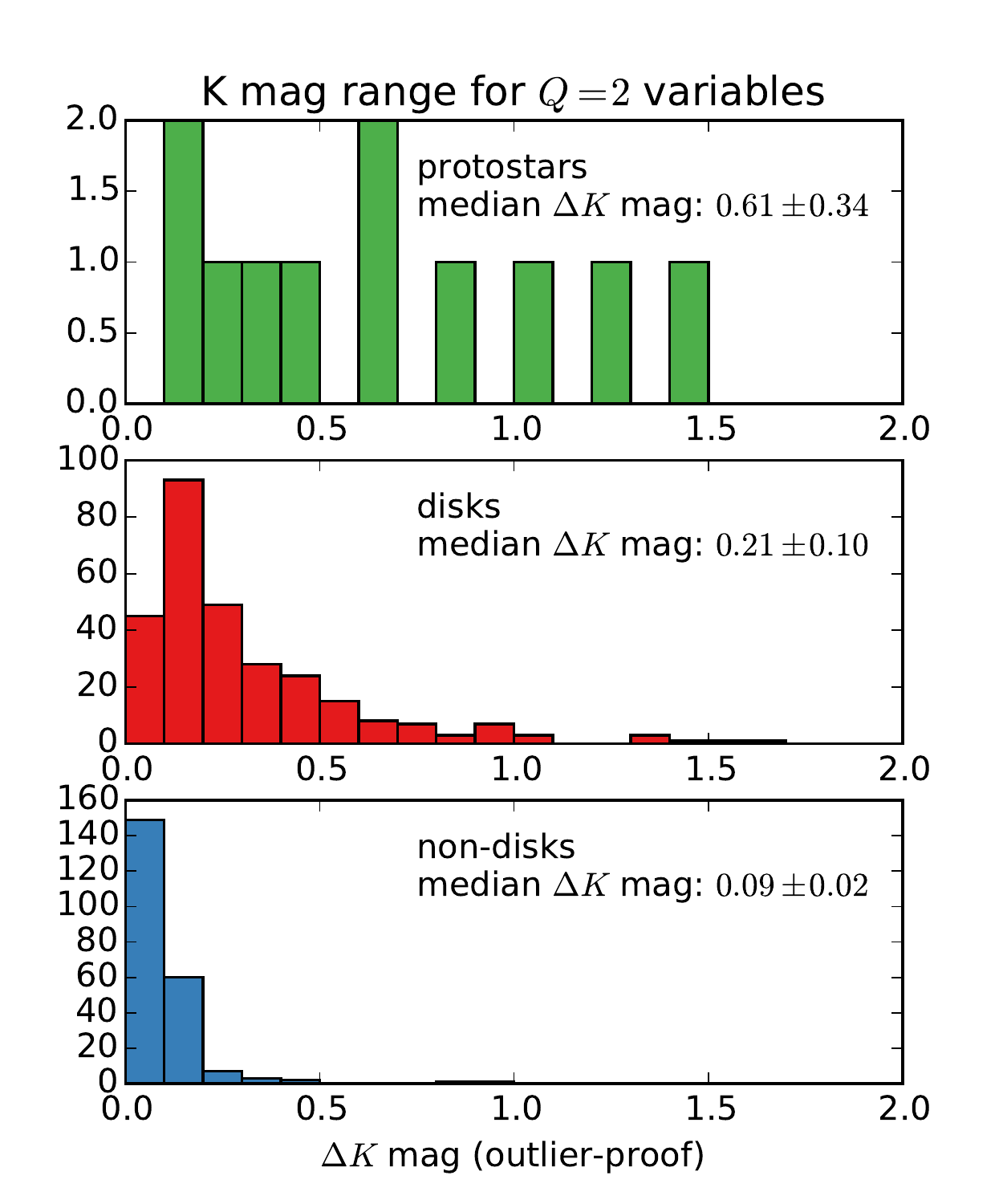}{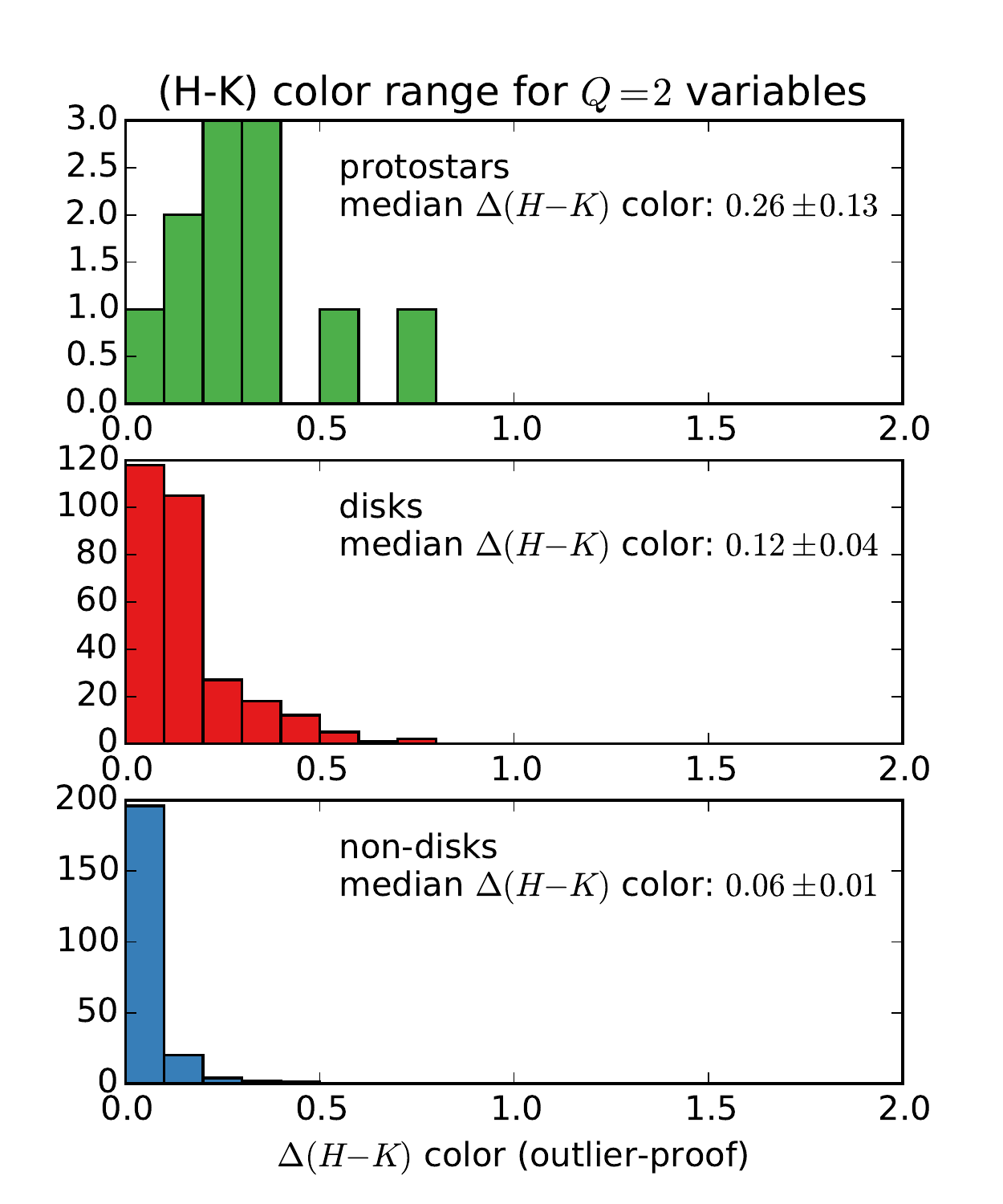}
\caption{
Distributions of variability amplitude at $K$ band and $H-K$ color for protostars, disked
stars, and diskless stars. 
Typical amplitudes are much higher for less-evolved stars versus
more-evolved stars. 
The median $K$ amplitudes for each class are 0.61,
0.21, and 0.09, for protostars, disked stars, and diskless stars,
respectively;
median $H-K$ amplitudes are 0.26, 0.12, and 0.06 respectively.
\label{fig:delta_hist} }
\end{figure}

The color terms $\Delta (J-H)$ and $\Delta (H-K)$, measured as the difference between maximum and minimum color values after iteratively clipping 3$\sigma$ outliers, show similar trends:
color shifts are smaller for more evolved stars.
As an example we show the results for changes of the $H-K$ color in Figure \ref{fig:delta_hist}.   
We primarily use the $\Delta (H-K)$ color term to probe changes in the disk excess.\footnote{$J-K$ is equally as useful as $H-K$, but as the redder stars in our sample are very faint and noisy at $J$, the $H-K$ index is more reliable.}
Indeed, we find the median $\Delta (H-K)= 0.26  \pm 0.13$ for the protostars,  $\Delta (H-K)= 0.12  \pm 0.04$ for the disked variables and  $\Delta (H-K)= 0.06  \pm 0.01$ for the diskless variables.

The \textit{Spitzer} sample is able to identify diskless objects, but it is limited by its inability to distinguish diskless YSOs from stars that have evolved onto the main sequence.
To better understand the variability properties of young diskless stars without contamination by main-sequence stars, we use an \textit{XMM-Newton} X-ray selection, as 
X-ray data provide a sample of YSOs unbiased by the presence of a disk.
The central portion of the monitored field was subjected to three XMM pointings averaging about 50 ks each.
X-ray detections of star-like sources with IRAC [3.6]-[4.5] color $<$ 0.3 are likely to be Class III YSOs \citep{ignazio13}\footnote {We do not use the deeper Chandra data available because the Chandra-observed region is much smaller than the XMM-observed region and hence more subject to spatial bias.}, so they are identified as such in their \textit{XMM-Newton}-selected catalog.
270 \textit{XMM} X-ray sources were identified with $Q=1+2$ stars in our monitored field; 166 of these sources are identified as young stars in the \textit{XMM} catalog. 
This includes three Class I objects, and a roughly even split between the number of Class II ($N=90$) and Class III ($N=73$) objects detected in X-rays.
Statistics below are limited to the $Q=2$ stars in these samples: N=28 (Class II) and N=49 (Class III).
The median Stetson indices and $\Delta K$ values observed for the Class II stars in this sub-sample are consistent with the \textit{Spitzer}-selected disk sample ($S = 2.54 \pm 1.74$ and $\Delta K = 0.20 \pm 0.10$)
confirming that the X-ray sample is relatively unbiased with respect to the NIR variability. 
Meanwhile, the disk-free young stellar objects (``Class III'') are far less variable by both criteria  (median $S=1.14 \pm 0.62$ and median $\Delta K=0.09 \pm 0.03$).  
The median $\Delta K =0.09$ value is fully consistent with variability arising only from starspots 
(e.g., \citealt{carpenter01}, \citealt{wolk13b}). 
In both the X-ray sample and the \textit{Spitzer}-selected sample, we find that less-evolved objects are by far the most variable.

It is further interesting that protostars are significantly more variable than evolved disked systems.
The distinguishing feature that separates protostars from evolved disked stars is the presence of a dense outer circumstellar envelope, which near-infrared observations at $K$ band (2.2 $\mu$m) are not sensitive to.
The $JHK$ bands should only be measuring the stellar photosphere, the star-disk interface, and the innermost edge of the dust disk.
Thus, the inner disk edge and/or the star-disk interface must have a fundamentally different structure between protostars and disked stars, with the protostar inner environment more unstable, leading to either accretion/disk geometry events, dust-occulting events, or (most likely) a combination of both.
The instability is strongest for the least evolved systems.
Note that, by definition, protostars are in the evolutionary phase where they accrete most of their mass; it is therefore expected that the least evolved young stellar objects show stronger and likely more frequent instabilities and variability due to the accretion process.
An alternative interpretation of this result is that the distribution and stability of the hottest circumstellar dust grains\footnote{The solids at the inner edge are refractory oxide grains (e.g., CaO, Al$_2$O$_3$, Ti$_2$O$_3$, MgO) \citep{pollack94}}, thought to be responsible for $K$-band excesses, is different for protostars when compared to typical disked objects.

In Tables \ref{tab:var_periodics} and \ref{tab:var_nonperiodics}, we
show the variability statistics, including derived color slopes, for
all periodic and nonperiodic stars in our dataset, respectively.
We sort each table by \citet{megeath12} class.

%!TEX root = varm42.tex

\begin{deluxetable}{cccccccccccc}
\tabletypesize{\scriptsize}
\rotate
\tablecaption{ Variability properties of periodic stars
		\label{tab:var_periodics} }
\tablewidth{0pt}

\tablehead{
\colhead{ ID } & \colhead{ $N_J$ } & \colhead{ $N_H$ } & \colhead{ $N_K$ } & \colhead{ $\Delta K$ } & \colhead{ $\Delta J-H$ } & \colhead{ $\Delta H-K$ } & \colhead{ $m(J-H,H-K)$ } & \colhead{ $m(J,J-H)$ } & \colhead{ $m(K,H-K)$ } & \colhead{ $S$ } & \colhead{ Period (days)} }
\startdata
\multicolumn{12}{c}{Protostars } \\
\hline 
18 & 91 & 113 & 111 & 1.01 & 0.24 & 0.50 & 0.75$\pm$.06 & 4.57$\pm$.67 & -5.85$\pm$1.12 & 11.05 & 8.4097 \\
149 & 1 & 94 & 91 & 0.83 & 0.00 & 0.68 & \ldots & \ldots & \ldots & 19.85 & 35.7089 \\
337 & 70 & 91 & 88 & 1.54 & 0.37 & 0.33 & \ldots & \ldots & \ldots & 39.01 & 7.2664 \\
341 & 84 & 101 & 93 & 0.38 & 0.24 & 0.16 & \ldots & \ldots & \ldots & 8.12 & 5.6621 \\
489 & 70 & 92 & 92 & 0.17 & 0.11 & 0.10 & \ldots & \ldots & \ldots & 2.82 & 4.3810 \\
640 & 97 & 109 & 108 & 0.61 & 0.27 & 0.23 & 0.81$\pm$.07 & 5.83$\pm$.88 & 5.63$\pm$1.45 & 12.38 & 30.0919 \\
780 & 94 & 115 & 112 & 0.65 & 0.11 & 0.20 & \ldots & \ldots & \ldots & 16.44 & 6.6787 \\
828 & 89 & 112 & 107 & 0.19 & 0.12 & 0.15 & \ldots & \ldots & \ldots &
2.68 & 4.1716 \\
\cutinhead{Disked stars}
1 & 91 & 113 & 111 & 0.13 & 0.06 & 0.16 & \ldots & \ldots & \ldots & 2.53 & 7.0741 \\
23 & 91 & 113 & 111 & 0.12 & 0.05 & 0.08 & \ldots & \ldots & \ldots & 4.11 & 3.9155 \\
24 & 70 & 100 & 96 & 0.29 & 0.14 & 0.15 & \ldots & \ldots & \ldots & 7.61 & 41.4687 \\
31 & 95 & 100 & 94 & 0.32 & 0.16 & 0.15 & \ldots & \ldots & \ldots & 7.97 & 5.1322 \\
43 & 73 & 110 & 108 & 0.19 & 0.12 & 0.14 & \ldots & \ldots & \ldots & 4.42 & 4.3725 \\
44 & 90 & 112 & 111 & 0.07 & 0.06 & 0.06 & \ldots & \ldots & \ldots & 1.71 & 18.2819 \\
65 & 91 & 113 & 110 & 0.76 & 0.27 & 0.36 & \ldots & \ldots & \ldots & 24.46 & 34.5571 \\
66 & 91 & 113 & 111 & 0.08 & 0.05 & 0.07 & \ldots & \ldots & \ldots & 1.32 & 2.7869 \\
115 & 90 & 111 & 112 & 0.10 & 0.07 & 0.08 & \ldots & \ldots & \ldots & 1.84 & 18.0349 \\
129 & 69 & 33 & 36 & 0.15 & 0.10 & 0.14 & \ldots & \ldots & \ldots & 5.94 & 8.5198 \\
130 & 71 & 91 & 89 & 0.93 & 0.29 & 0.35 & \ldots & 5.65$\pm$.28 & \ldots & 29.29 & 8.8413 \\
143 & 97 & 106 & 93 & 0.22 & 0.07 & 0.07 & \ldots & \ldots & \ldots & 5.92 & 1.9025 \\
145 & 97 & 109 & 106 & 0.26 & 0.11 & 0.14 & \ldots & \ldots & \ldots & 3.53 & 5.0552 \\
152 & 97 & 104 & 90 & 0.31 & 0.14 & 0.27 & \ldots & \ldots & -2.21$\pm$.48 & 6.22 & 4.7192 \\
157 & 96 & 108 & 105 & 0.15 & 0.07 & 0.07 & \ldots & \ldots & \ldots & 2.98 & 5.9983 \\
159 & 97 & 108 & 106 & 0.39 & 0.26 & 0.23 & \ldots & \ldots & \ldots & 15.57 & 8.3320 \\
161 & 70 & 100 & 102 & 0.11 & 0.06 & 0.07 & \ldots & \ldots & \ldots & 2.43 & 3.2721 \\
%162 & 165 & 199 & 200 & 0.22 & 0.27 & 0.13 & \ldots & \ldots & \ldots & 4.52 & 9.9560 \\
%171 & 81 & 32 & 43 & 0.24 & 0.31 & 0.57 & \ldots & \ldots & \ldots & 1.10 & 3.4512 \\
\enddata
\tablecomments{Sample; the remainder of the table is available electronically. $N_J$, $N_H$, and $N_K$ are the number of detections in each band; some stars fall on overlap regions and are observed twice per night. $m(x,y)$ denotes the slope in the specified color-color or color-magnitude diagram; a slope is only included if the orthogonal distance regression fit yielded a confidence of better than 10\% on the fit. $S$ denotes the Stetson Variability Index.}
\end{deluxetable}

%!TEX root = varm42.tex

\begin{deluxetable}{ccccccccccc}
\tabletypesize{\scriptsize}
\rotate
\tablecaption{ Variability properties of nonperiodic stars
		\label{tab:var_nonperiodics} }
\tablewidth{0pt}

\tablehead{
\colhead{ ID } & \colhead{ $N_J$ } & \colhead{ $N_H$ } & \colhead{ $N_K$ } & \colhead{ $\Delta K$ } & \colhead{ $\Delta J-H$ } & \colhead{ $\Delta H-K$ } & \colhead{ $m(J-H,H-K)$ } & \colhead{ $m(J,J-H)$ } & \colhead{ $m(K,H-K)$ } & \colhead{ $S$ } \\
}
\startdata
\multicolumn{11}{c}{Protostars } \\
\hline 
110 & 185 & 215 & 206 & 1.53 & 0.44 & 0.96 & \ldots & \ldots & \ldots & 6.12 \\
204 & 97 & 109 & 107 & 0.49 & 0.16 & 0.26 & \ldots & \ldots & -3.16$\pm$.28 & 7.62 \\
212 & 96 & 106 & 93 & 1.22 & 0.46 & 0.80 & 0.45$\pm$.02 & -2.69$\pm$.58 & -1.81$\pm$.07 & 13.73 \\
233 & 97 & 110 & 108 & 0.27 & 0.09 & 0.15 & \ldots & \ldots & \ldots & 5.67 \\
279 & 71 & 92 & 92 & 0.16 & 0.10 & 0.09 & \ldots & \ldots & \ldots & 2.66 \\
363 & 70 & 92 & 92 & 0.57 & 0.44 & 0.41 & \ldots & \ldots & \ldots & 16.46 \\
395 & 0 & 92 & 92 & 0.25 & \ldots & 0.14 & \ldots & \ldots & \ldots & 2.06 \\
415 & 97 & 110 & 108 & 0.39 & 0.16 & 0.32 & \ldots & \ldots & \ldots & 7.99 \\
443 & 93 & 99 & 95 & 1.11 & 0.66 & 1.22 & \ldots & \ldots & \ldots & 30.00 \\
447 & 96 & 108 & 105 & 0.22 & 0.07 & 0.21 & \ldots & \ldots & \ldots & 4.03 \\
494 & 68 & 92 & 92 & 1.46 & 0.48 & 0.32 & 1.97$\pm$.28 & -1.73$\pm$.10 & -8.92$\pm$1.36 & 28.91 \\
503 & 69 & 89 & 91 & 0.45 & 0.43 & 0.17 & \ldots & \ldots & \ldots & 6.40 \\
510 & 2 & 91 & 87 & 0.93 & 0.14 & 0.63 & \ldots & \ldots & \ldots & 105.12 \\
535 & 2 & 92 & 92 & 0.68 & 0.21 & 0.29 & \ldots & \ldots & \ldots & 27.19 \\
537 & 166 & 201 & 152 & 1.41 & 0.72 & 0.42 & \ldots & \ldots & \ldots & 20.71 \\
605 & 97 & 107 & 106 & 0.88 & 0.26 & 0.39 & \ldots & \ldots & -5.86$\pm$1.46 & 8.48 \\
633 & 25 & 105 & 105 & 0.43 & 0.20 & 0.29 & \ldots & \ldots & \ldots & 23.16 \\
660 & 0 & 0 & 92 & 0.47 & \ldots & \ldots & \ldots & \ldots & \ldots & 2.78 \\
663 & 97 & 106 & 93 & 0.35 & 0.16 & 0.24 & \ldots & \ldots & 3.04$\pm$.55 & 10.50 \\
683 & 91 & 110 & 106 & 0.60 & 0.28 & 0.52 & \ldots & \ldots & \ldots & 20.03 \\
688 & 0 & 104 & 105 & 0.60 & \ldots & 0.43 & \ldots & \ldots & \ldots & 44.56 \\
706 & 67 & 89 & 45 & 0.17 & 0.13 & 0.20 & \ldots & \ldots & \ldots & 4.80 \\
714 & 0 & 110 & 106 & 0.55 & \ldots & 0.38 & \ldots & \ldots & \ldots & 42.93 \\
723 & 97 & 110 & 108 & 0.79 & 0.45 & 0.48 & \ldots & \ldots & 2.20$\pm$.23 & 17.13 \\
727 & 71 & 92 & 91 & 0.42 & 0.40 & 0.45 & \ldots & \ldots & \ldots & 19.60 \\
732 & 0 & 0 & 100 & 0.40 & \ldots & \ldots & \ldots & \ldots & \ldots & 2.88 \\
784 & 159 & 214 & 208 & 0.37 & 0.16 & 0.16 & \ldots & \ldots & \ldots & 6.45 \\
801 & 68 & 100 & 95 & 0.63 & 0.31 & 0.39 & 0.45$\pm$.05 & 1.20$\pm$.23 & -1.44$\pm$.07 & 3.76 \\
\enddata
\tablecomments{Sample; the remainder of the table is available electronically. Columns are explained in Table \ref{tab:var_periodics}. }
\end{deluxetable}

\subsection{Period distributions by Class}
\label{sec:periods_by_class}

Periodic behavior is frequently found in variability studies of YSOs, and searches for periodicity have been used, e.g., to study the evolution of angular momentum loss throughout early stellar evolution.
Periods are regularly found for both disked and diskless objects;
well-sampled surveys of some star-forming regions have found a bimodal distribution of periods in which disked stars have periods of $\sim 8$ days, while mostly disk-less stars have periods under 4 days and diskless stars tend to have shorter periods overall\footnote{This bimodality has been hailed as evidence of magnetic coupling between the star and its disk (e.g., \citealt{attridge92}, Edwards \e 1992, \citealt{rebull01}).  
In this model, the star-disk coupling prevents the star from spinning up while the radius shrinks during the contraction phase of the Hayashi track 
\citep{hayashi65}. 
Once the disk dissipates, the stellar magnetic field is decoupled and the star is free to spin up as mass settles to the center of the star.}.
In the near-IR, the first significant study of periodic behavior was that of \citet{carpenter01}. 
They found a more or less continuous distribution of stellar periods spread between 2 and 10 days. 
The distribution peaks with about 45 stars having periods between 2-3 days, and falls smoothly to 15 stars with periods between 10-11 days. The peak value is probably augmented by stars with true periods shorter than the sampling frequency, which would imply a somewhat flatter distribution.
The bulk of the periodic stars in that sample did not have $K$ band excesses indicative of disks.

\begin{figure}
	\plotone{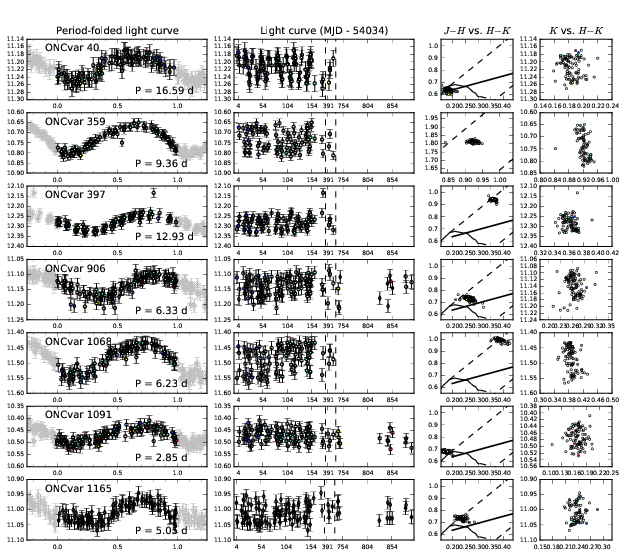}
	\caption{Light curves of seven stars that show clean, rotationally modulated light curves. \label{lc:rotators}}
\end{figure}

In Figure \ref{lc:rotators}, seven examples of cleanly rotating spotted stars are shown.

\begin{figure}
\plotone{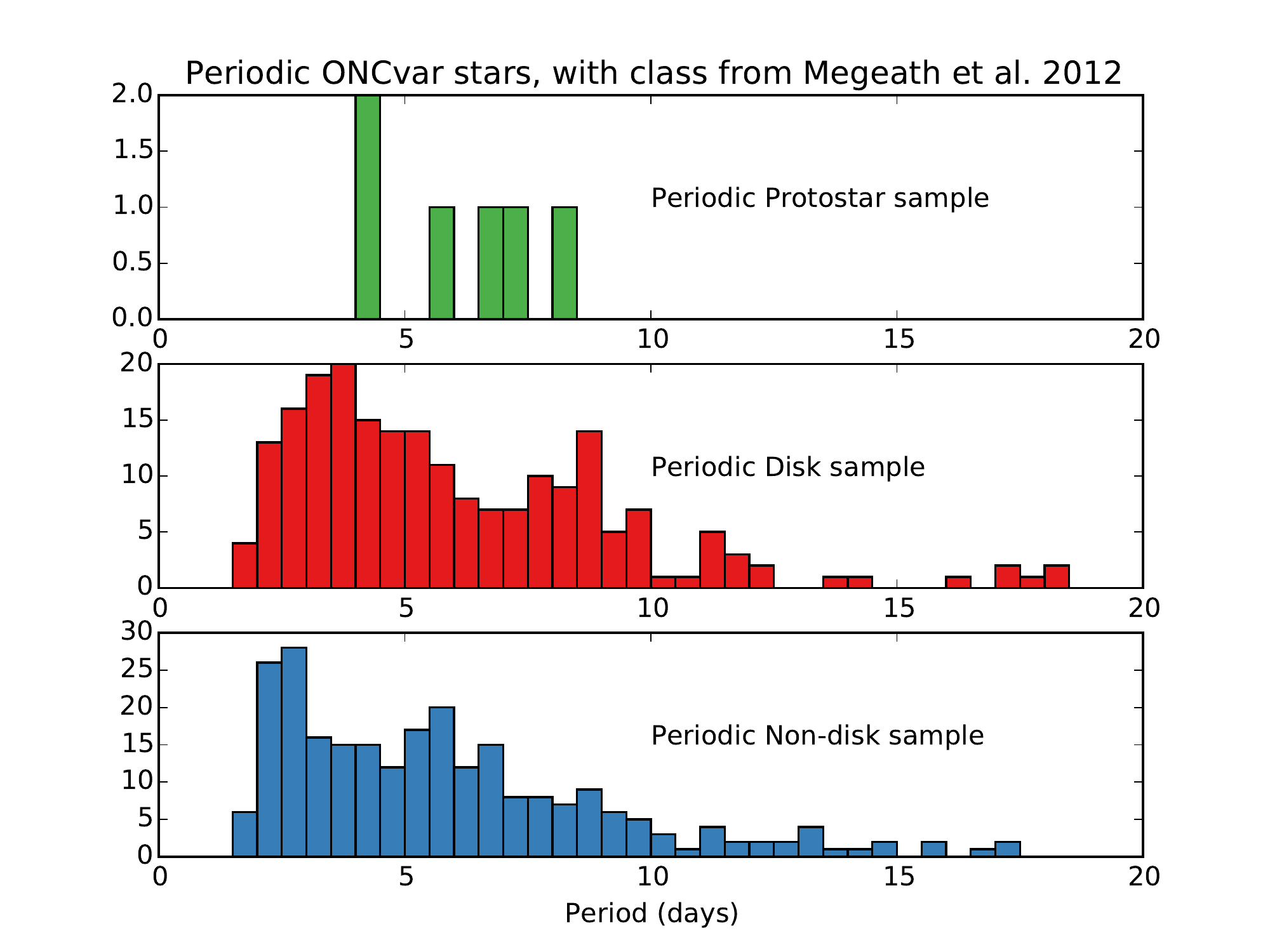}
\caption{
The distribution of periods below 20 days for protostars (top panel, green), disked stars (middle panel, red) and
non-disked stars (bottom panel, blue).
A modest secondary peak in the disked periods is seen at $\sim 8$ days; such a
bimodal distribution has been observed before for optical periods, and
is typically interpreted as evidence for magnetic disk locking.
The eight-day bump is not seen for non-disked stars.
A two-sided Kolmogorov-Smirnov test indicates a near-zero probability 
($1.277 \times 10^{-22}$)
that the disked periods vs. non-disked periods were drawn from the same distribution.
\label{fig:period_distributions} }
\end{figure}

% para 3
In Figure~\ref{fig:period_distributions}, we show the period distribution for 471 periods found using the Lomb-Scargle method as described in \S \ref{subsubsection:autoperiod}.
All periodic sources (regardless of quality status) with a \textit{Spitzer}-derived evolutionary class are included in this histogram, provided they have periods below 20 days.
While the distribution for the protostars is poorly sampled, the disked and diskless distributions are well populated and intriguing in a couple of respects.  
First of all, the disked and disk-free populations are distinct; the disk-free population has more short periods (below 5 days) with respect to longer periods than the disked population.   
The distribution seen among the non-disked sample is very similar to that found by \citet{carpenter01}.  
Here, we find about 5 times more stars with periods in the 2-3 day range as in the 10-15 day range.
One caveat about the period distribution we find, which is common to \citet{carpenter01}, is that some periods are likely to have been moved from below 2 days to above 2 days via aliasing.

% para 4
The distribution of periods for disked stars
is significantly different from that of the diskless stars: a two-sided Kolmogorov-Smirnov test gives a near-zero probability that these two groups of periods were drawn from the same distribution.
A small enhancement of disked objects at periods between $5 -10$ days marks a key difference between the disked and diskless distributions; there is no peak in the latter half of the distribution as seen in the $I$ band studies
(e.g., \citealt{herbst02}, \citealt{stassun00}).  
However, one difference seen between our disked sample and previous studies of disked star periods is that fully 1/4 of our disked sample have rotation periods between $2-5$ days.
In contrast, earlier studies found that disked stars (usually identified via $H-K$ excess) tend to be less likely to have periods below 5 days 
(e.g., \citealt{herbst00}, \citealt{rebull01})
 
% para 5
% It has been noted several times (e.g., \citealt{herbst00}, \citealt{rebull01}, \citealt{herbst05}) that low mass stars (M$<$0.25 M$_\odot$) do not show a bimodal distribution, but instead have a distribution that peaks near 1 day and ramps down, with 10 times as many stars with 1-day periods as 10-day periods. 
% A possible way to explain the difference between our result and this previous result may lie in our definition of ``disked'': the \citet{megeath12} disked YSO sample includes stars whose infrared excesses appear only at $[5.8]$ or redder. 
% The inner edges of such disks may lie beyond a radius at which magnetic locking between the disk and star can occur, hence diluting the $\sim 8$-day peak expected in the disked period distribution.
% Additionally, it should be kept in mind, as pointed out at the
% beginning of Section~\ref{variable_selection}, that there are uncertainties inherent in the
% SED classification of \citet{megeath12}, which may further
% complicate the issue.
  
% para 6
On the other end of the period distribution, stars with periods longer than 15 days are seen in both the disked and non-disked populations.
The diskless long-period stars in this sample could just be slow rotators.
But among disked stars, the longest periods are up to 90 days (see \S \ref{sec:longperiods}, e.g., Fig. \ref{lc:longperiods}) and are accompanied by color changes consistent with changes in the disk or dust along the line of sight. 
These long-period, color-changing light curves cannot be modulated by the star's rotation period unless the object rotates very slowly -- instead, they trace disk phenomena with relatively long orbital periods, such as orbital variation in disk extinction or dynamically modulated accretion.

\subsection{Color slopes constrain physical mechanisms}
\label{sec:colorslope}

\begin{figure}
\plotone{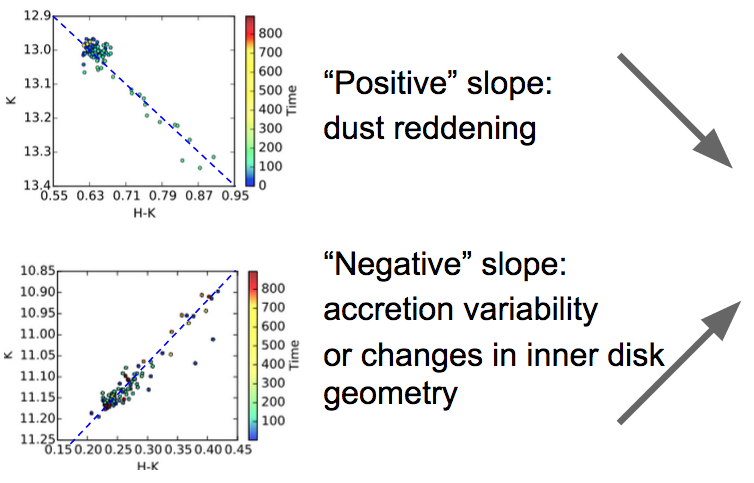}
\caption{
Illustration of two example tracks in color-magnitude space.
A ``positive'' slope (top; ONCvar 310 track shown as example) is associated with variable dust extinction, while a ``negative''
slope (bottom; ONCvar 1226 track shown) is associated with accretion variability or
changes in the inner disk geometry.
More variability mechanisms, such as cool and hot spots, exist, and are outlined in \citet{wolk13b}.
In this figure, the expected direction of dust reddening is derived from the extinction law given by \citet{rieke85};
similar results are obtained from more modern versions of the extinction law.
\label{fig:colorslope_illustration} }
\end{figure}

By studying the trajectories traced out by variable stars through color-color and
color-magnitude space, we are able to add additional constraints to the physical
mechanisms that cause their variability (Figure \ref{fig:colorslope_illustration}).
Accretion disk models generally indicate that motion up and to the right in a $H-K, K$ or $J, J-H$ color-magnitude diagram is a result of the additional flux re-radiated at long wavelengths as the total flux increases due to accretion, whereas motion \textit{down} and to the right is a result of dust reddening.
In the $JHK$ color-color diagram, the dust reddening tracks were well established by \citet{lada92} while a locus occupied by classical T Tauri stars was identified by \citet{meyer97}.
In the color-color diagram, unlike in the color-magnitude diagrams, the CTTS locus and reddening vectors are not orthogonal, and many young stars have color-color diagram trajectories with slopes between the CTTS locus and the reddening vector, as seen in Cyg OB7 \citep{rice12}.
Illustrations of the different color-dependent effects on the spectral energy distributions (SEDs) of young stars can be found in \citep{robitaille07}.

\citet{carpenter01} were the first to analyze NIR color space trajectories in young variable stars.
They used this approach as the basis
for their findings that most stars' variability is consistent with starspots and that stars with high color variability tend to show changes in dust-reddening (slopes between 30 and 90 degrees), 
with a small minority showing changes in disk geometry or accretion rate (slopes between -90 and -45 degrees). 
In the current study, we have examined the color trajectories over our longer-baseline data.

To determine the color ``slope'', we use the orthogonal distance
regression package included in SciPy (\texttt{scipy.odr}), which fits a line to values that
have comparable uncertainties in both the color and magnitude (or second color).
The ODR code returns both a slope and an estimated error on that slope.
We calculate such slopes for $\frac{\Delta K}{\Delta(H-K)}$,
$\frac{\Delta J}{\Delta (J-H)}$, and $\frac{\Delta(J-H)}{\Delta  (H-K)}$.
Following \citet{carpenter01}, we identify disk-activity variables via a negative slope between -45 and -90 degrees in the $H-K, K$ color-magnitude diagram, and we identify reddening variables with positive slopes between 30 and 90 degrees in the same diagram.\footnote{
In the color-magnitude diagrams, a ``positive slope'' is down and to the right because the magnitude scale is inverted following general astronomical convention.}
We noted in \citet{rice12} that the color space trajectories are often linear, and in \citet{wolk13b} 
we find that many are well-fit using this outlier-resistant algorithm. 

\subsubsection{Examples of typical color behavior}
\label{sec:typicalcolor}
\begin{figure}
\plotone{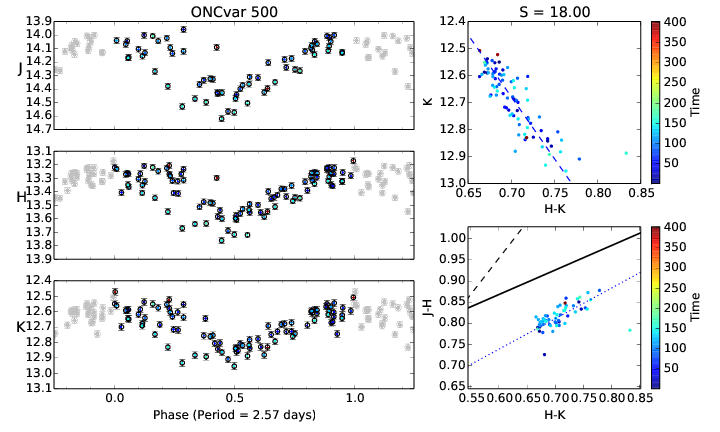}
\caption{ Three-band light curve folded on a period of 2.57 days, plus color-magnitude and color-color diagrams, for the variable star ONCvar 500.
This star exhibits periodic ``AA~Tau''-type variability (as defined in \S \ref{sec:dippers}) that shows a dust reddening signature in $H-K, K$ space.
\label{lc:500} }
\end{figure}

\begin{figure}
\plotone{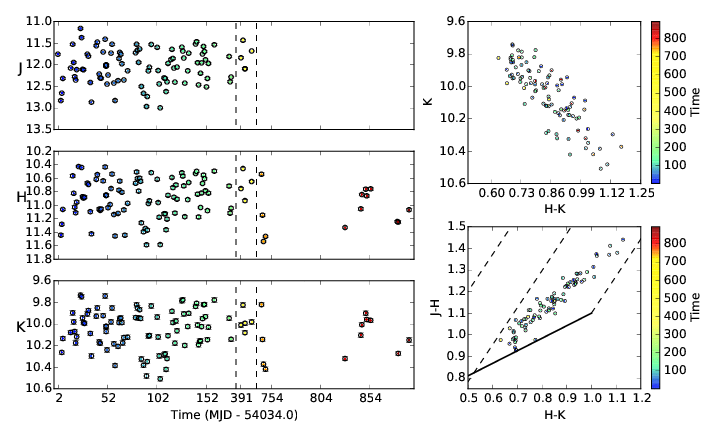}
\caption{ Three-band light curve, plus color-magnitude and color-color
  diagrams, for the variable star ONCvar 539.
  While the color-magnitude trend is consistent with (and dominated by) reddening, the observed deviations from the reddening vector are far in excess of observational errors, indicating that more than one process contributes to the measured color changes.
\label{lc:539} }
\end{figure}

\begin{figure}
\plotone{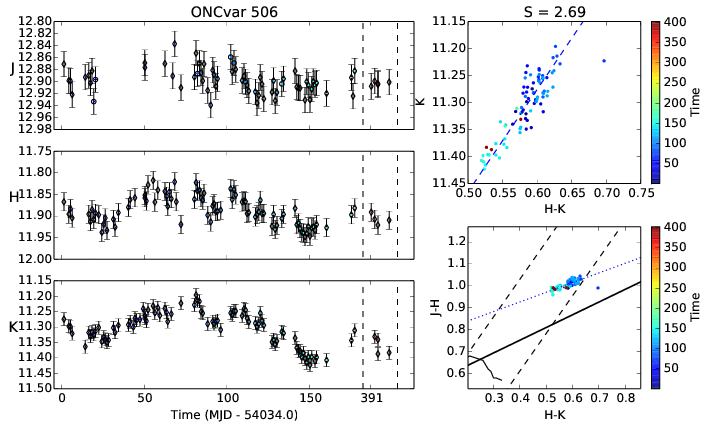}
\caption{ Three-band light curve, plus color-magnitude and color-color
  diagrams, for the variable star ONCvar 506.
This star shows mild color-magnitude variability and color-color variability that is consistent with changes in accretion rate or inner disk geometry (\citealt{carpenter01}).
\label{lc:506} }
\end{figure}

A few illustrative examples of stars that show well-defined color trajectories are shown in Figures \ref{lc:500}, \ref{lc:539}, and \ref{lc:506}.
Figure~\ref{lc:500} shows a fairly typical periodic disked source.  
The power in each band's periodogram is very high, but the amplitude varies from phase to phase by about 10\%.  
On the other hand, the phases themselves seem stable over more than 100 periods as evidenced by the points from season two which fit appropriately into the phase-folded light curve.
There is limited movement in color-color space, but dramatic motion in the direction of reddening ($\sim$ 0.5 A$_K$, $\sim$ 0.7 A$_J$ ) is visible in the color-magnitude diagrams. 
The direction of this motion is typically steeper than predicted by ISM reddening models (e.g., \citealt{rieke85}).  

Figure~\ref{lc:539} shows a typical aperiodic reddening object, albeit one with a large amplitude. 
While there is no clear periodic behavior, all the motion in the color-magnitude diagrams still lies generally along the reddening vectors. 
In this case the motion is very close to what one would predict from  ISM reddening models (e.g., \citealt{rieke85}). 
The color-color diagram, on the other hand, shows what \citet{rice12} called ``hybrid motion'', halfway between the expected slope of reddening and the CTTS locus.
Finally,  Figure~\ref{lc:506} shows an example of a star which has a changing interaction with its disk.   
This star happens to be a relatively low-amplitude example, with $\Delta K \sim 0.2$. 
Its motion in the color-color diagram is almost parallel to the CTTS locus;
in the $K$, $H-K$ color-magnitude diagram, it becomes bluer by $\Delta H-K \sim 0.15$ as it dims.
The $J$, $J-H$ trajectory is best fit with a vector which also gets bluer as it gets fainter, but the scatter is very high in this fit and it is hard to determine if there really is a preferred color change in $J$, $J-H$. 
  
\subsubsection{Aggregate color behavior}
\label{sec:aggregate}
\begin{figure}
\plotone{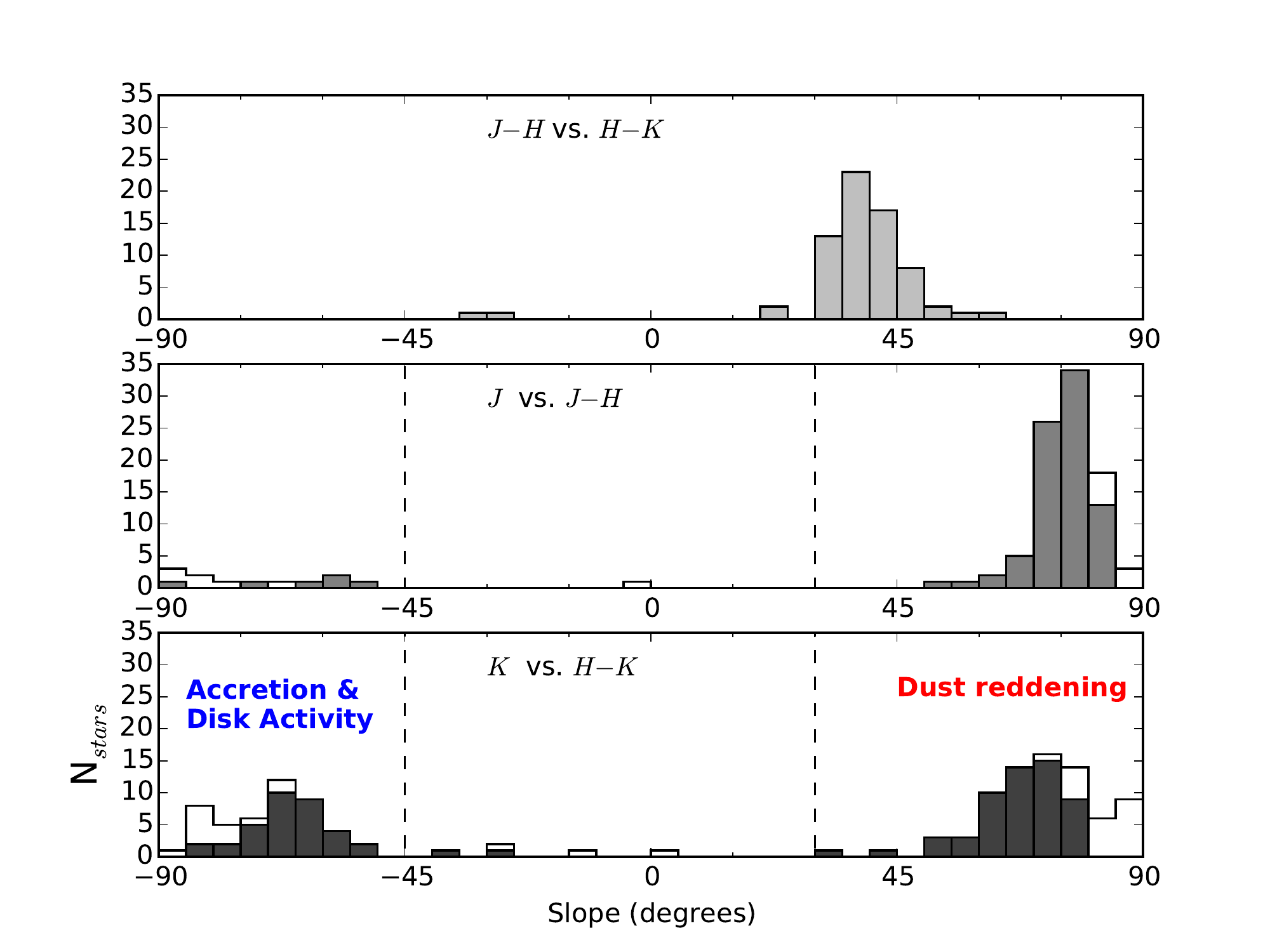}
\caption{Distributions of the fitted slopes in color-color and color-magnitude
  space.
As in \citet{carpenter01}, only stars showing rms variability in the
appropriate bands or color channels that exceeded noise by a factor of 1.5 are
shown, with the colored-in bars showing color slopes with better than
20\% accuracy.
Notably, these data show a much larger relative fraction of
accretion/disk-activity slopes versus dust-reddening slopes 
than did \citet{carpenter01}; see Figure 20 in that paper.
We show in \S \ref{sec:timescales} that this disparity is
primarily due to our longer observing baseline: the longer one
observes in the near-infrared, the more accretion/disk-activity
variables are seen.
\label{fig:colorslope_threepanel} }
\end{figure}

In Figure \ref{fig:colorslope_threepanel} we show histograms of the fitted slopes in color-color and color-magnitude space.
We include stars of all classes in this aggregate color analysis, and require a star to be at least $Q=1$ to be included, as well as an additional band-quality criterion\footnote{To be included in the $J-H$ vs. $H-K$ analysis, a star must be $Q=2$; to be included in the other color analyses, it must be $Q=1$ and have high-quality, flag-free data in the bands in question: $J$ and $H$, or $H$ and $K$.}.
The distributions indicate a strong bimodal behavior in $K$, $H-K$ and weaker bimodality in $J$, $J-H$.
We interpret these color slopes as indicative of disk-activity variability (bluer when fainter) and dust reddening variability (redder when fainter).
As in \citet{carpenter01}, only stars showing significant variability in the appropriate bands/color channels that exceeded noise by a factor of 1.5 are shown. 
The filled-in portion of the histogram represents those objects to which we fitted color slopes with better than 20\% accuracy.
These plots should be directly comparable to Carpenter \e (2001; their Figure 20). 
Our data show a much larger relative fraction of accretion/disk-activity slopes versus dust-reddening slopes.  
We find up to 44\% of the fitted sources show signs of disk activity, compared to between 15-20\% by \citealt{carpenter01}.

In addition to differences in slope, each variability mechanism has a different maximum variability amplitude.
For example, as summarized in \citet{wolk13b}, while extinction can add essentially infinite reddening, cold spots are limited to changes 
of $< 15\%$  and even hot spots are limited to changes of about 0.6 mag at $K$.  
Disk models are less well-calibrated but appear capable of changes of at least 0.75 mag at $K$. 
By comparing the observed distributions of color slopes in $\Delta(J-H)/\Delta
(H-K)$, $\Delta K / \Delta(H-K)$, and $\Delta J / \Delta(J-H)$, we can
infer information about which variability mechanisms are most common
in our data, as well as identify which color slopes are best for
distinguishing certain mechanisms.

\subsubsection{Compound variables}
\label{sec:compound}
We also note a significant number of stars that show
changes in the observed variability mechanism. 
Their trajectories in color-color space and color-magnitude space are not confined to a
single linear slope, but instead show both ``positive-sloped'' and
``negative-sloped'' tracks cleanly separated in time, indicating that
short-term monitoring may not be sufficient in identifying a given
star's primary mode of variability.  
Indeed, there may be no ``primary mode'' of variability in these objects.
We discuss a few examples below:\\

\begin{figure}

\plotone{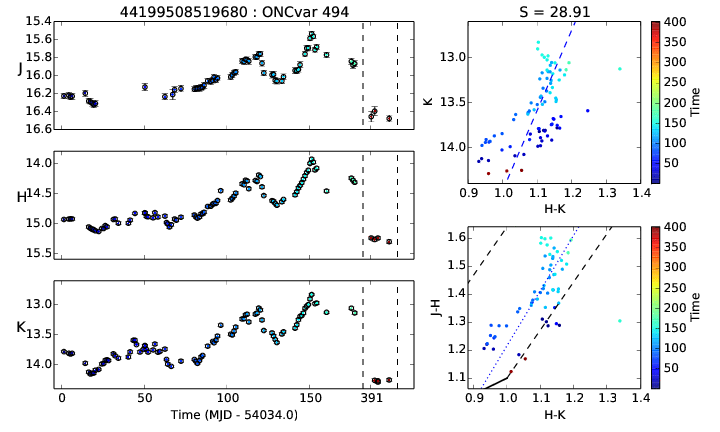}

\caption{ Three-band light curve, plus color-magnitude and color-color
  diagrams, for the variable protostar ONCvar 494, which shows ``compound
  variability'' in color-magnitude space (discussed in \S
  \ref{sec:colorslope}).
This star exhibits a $H-K, K$ track showing accretion/disk activity
from $t = 0$ to $t = 80$; around $t \sim 80$, the star gets bluer as
it brightens (an indicator that the dust reddening has decreased)
before resuming the accretion/disk activity until $t=130$. 
At $t=130$, the slope in $H-K, K$ shifts again, showing variability
indicating changes in dust reddening.
\label{lc:494} }
\end{figure}

% In [11]: megeath_column[ ukvar_spread.UKvar_ID == 494][0]
% Out[11]: 'P'
ONCvar 494 (Figure \ref{lc:494}), Protostar class, shows a long term rise by over a magnitude in $K$ over the first season.
The rise is not monotonic but has apparent peaks and valleys separated by about 50 days.  
The trajectory in color-color space is in reasonable agreement with the reddening vector, but not particularly straight.   
The $J$, $J-H$ trajectory (not shown) is clearly one of blueing (by 0.6 in $J-H \over{J}$).  
The $K$, $H-K$ color-magnitude is generally on a blueing trajectory as well, except that there are two distinct tracks, one which appeared for the first hundred days or so and a second, brighter by 0.2 $K$, which held for the next $\sim$ 50 days.

\begin{figure}
\plotone{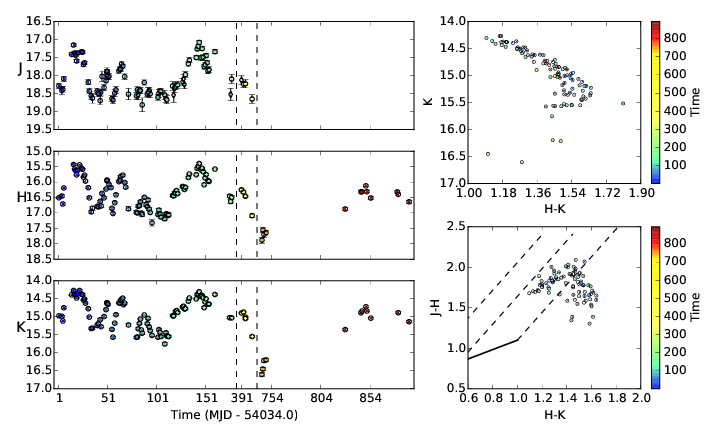}
\caption{ Light curve of the variable star ONCvar 218, which lacks a
  mid-infrared disk and shows ``compound variability'' in
  color-magnitude space.
This star's variability moves it back and forth along a
dust-reddening track in $H-K, K$ space when brighter than $K = 15.3$,
but moves along an accretion/disk-activity track when fainter than $K
= 15.3$, especially in the third observing season.
Its trajectory in the $J-H$ versus $H-K$ color-color diagram is dominated by noise at $J$ and should be disregarded.
\label{lc:218} }
\end{figure}

ONCvar 218 (Figure \ref{lc:218}), ``No disk'' class, displays behavior nearly opposite of ONCvar 494.  
It shows no long-term trend but has significant brightness peaks separated by $\sim$ 50 days. 
The $K$, $H-K$ color-magnitude diagram is well fit by ISM dust, albeit with a bit of scatter. 
The color-color data is dominated by the very large errors in $J$ band, and does not provide information.

\begin{figure}
\plotone{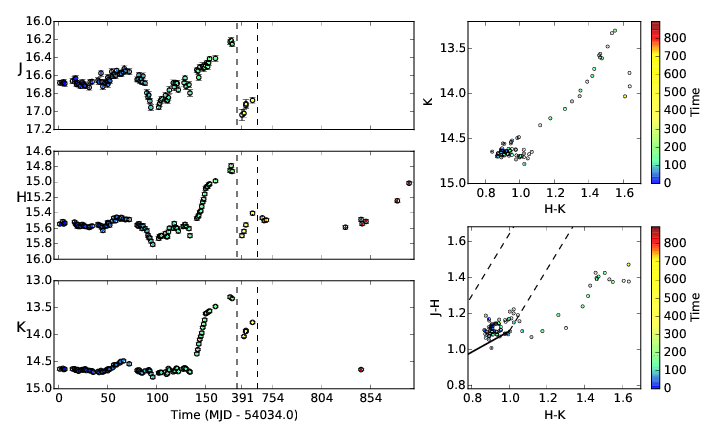}
\caption{ Light curves and color-magnitude and color-color diagrams for the protostar
  ONCvar 212, which shows ``compound variability'' in color-magnitude
  space.
This star occupies a ``quiescent'' region in $H-K, K$ space (which is
still significantly variable: $\Delta (H-K) = 0.2$) until $t=130$,
when it abruptly migrates by $\Delta (H-K) = 0.6$ along an
accretion/disk-activity track in $H-K, K$ space, brightening by 1.3
magnitudes at $K$.
Between the first and second observing seasons, the star dims by
$\Delta K = 0.6$,
moving along a dust reddening track.
In addition to its exceptional color behavior, this star shows
striking differences between the shape of its light curves at each band.
\label{lc:212} }
\end{figure}

ONCvar 212 (Figure \ref{lc:212}), Protostar class, was relatively calm for the first $\sim$ 50 days. 
This is followed by an event noticed mostly at J and H as a brightening followed by a significant dimming ($40\%$ at $J$), both of which appear reddening-related.
The system then brightens gradually at J and suddenly at K. 
The K band event seems to have started between days 130 and 135.  
The color-color track of the large-scale brightening event clearly traces along the CTTS locus.  
Interestingly, when the star is briefly re-observed in season 2 the color-color data are still far to the right of the extended CTTS locus, even though the brightness of the star is far fainter in all three bands. 

\begin{figure}
\plotone{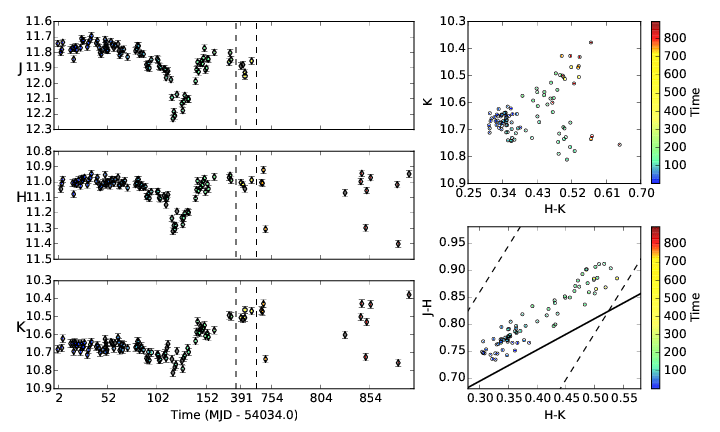}
\caption{ Light curve of the variable disked star ONCvar 466, which
  shows ``compound variability'' in color-magnitude space.
Between $t=0$ and $t=125$, the star moves in $H-K, K$ space along a
shallow dust-reddening track whose slope may be interpreted as
simultaneous reddening and disk/accretion activity. 
Then, at $t=130$,
the star brightens, moving up in $H-K, K$ space along a steeper
reddening track.
In the third observing season, this alternating behavior is seen to
repeat, suggesting that this star would show both mechanisms when
monitored over even longer timescales.
\label{lc:466} }
\end{figure}

ONCvar 466 (Figure \ref{lc:466}), Disked class, is an effective counterpoint to the previous figure. 
Again the star's light curve was relatively flat for the first $\sim$ 70 days. 
This is followed by an event noticed mostly at $J$ and then $H$ but this time only as dimming.  
As the star reaches its faintest at $J$ and $H$ it then brightens in all three bands.  
The star moves strongly along the CTTS locus starting on day $\sim 125$.
At $J$ and $H$, the star's behavior is consistent with only a reddening event; the addition of $K$ data help flesh out the full picture, where part of the star's behavior appears to be related to disk activity / accretion.

In addition to well-defined cases of compound color behavior such as those shown above, we also find an even larger subsample of stars with significant color variability that shows no clear linear trend but could be explained by a complicated mixture of variability mechanisms.

%\clearpage

%!TEX root = varm42.tex

\section{Variability Timescales}
\label{sec:timescales}

In this section we analyze how various parameters change with the length and number of observations in a near-infrared monitoring study.

\begin{figure}
	\plottwo{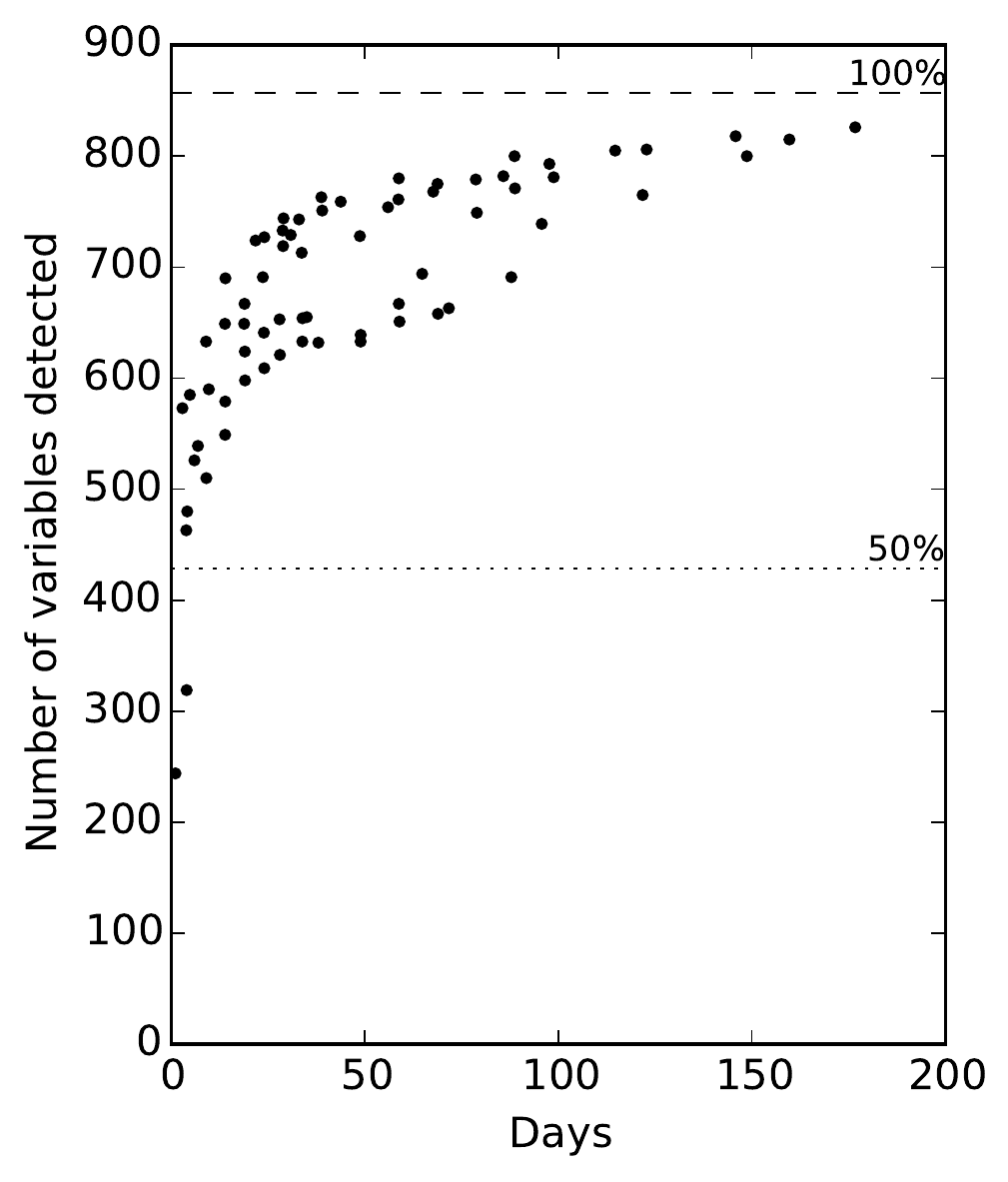}{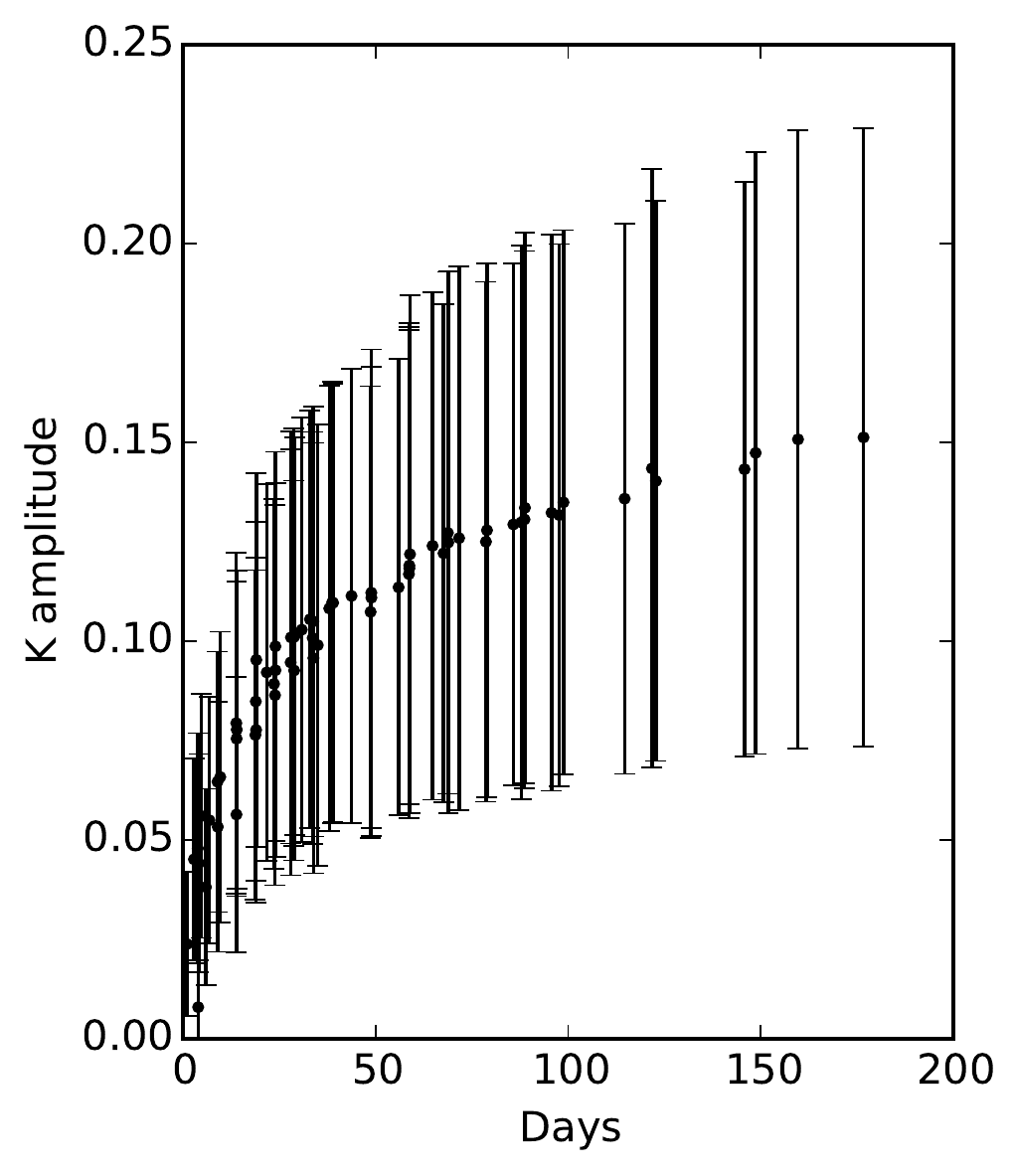}
	\caption{(left) Number of automatic variables detected as a function of sampled baseline. (right) Median $K$ amplitude as a function of sampled baseline.
	Error bars depict the median absolute deviation (MAD) of variable star amplitudes around the median.
	\label{fig:timescale_nvars_kamp}}
\end{figure}

We studied how the number of automatically-detected variables (i.e., only variables in quality classes $Q=1$ and $Q=2$; see Appendix \ref{sub:autosubj} for details)
changed as a function of time baseline (Figure \ref{fig:timescale_nvars_kamp}a). In the final dataset, there are 857 such ``automatic'' variables in our catalog of 1203 variables.
At the shortest baselines ($\sim 5$ days) hundreds of variables are detected; this increases to over 800 after $\sim 200$ days.
When the longest baselines (up to 900 days; not shown in this plot) are included, the number reaches the expected total of 857.
More than half of the final variables are detected within just 5 days of monitoring.

We studied how the median $K$ amplitude of variables changes as more datapoints are added (Figure \ref{fig:timescale_nvars_kamp}b). 
At the shortest baselines ($<$ 5 days), the amplitude is below $0.05$ mag, 
but climbs steadily to $\sim 0.15$ mag when the sampling baseline is $\sim 200$ days. 
By the end of the $\sim 900$ day study, the median amplitude climbs to $\sim 0.16$ (not shown in figure).
Half of the final variability amplitude is reached after about 20 days of monitoring.

\begin{figure}
	\plottwo{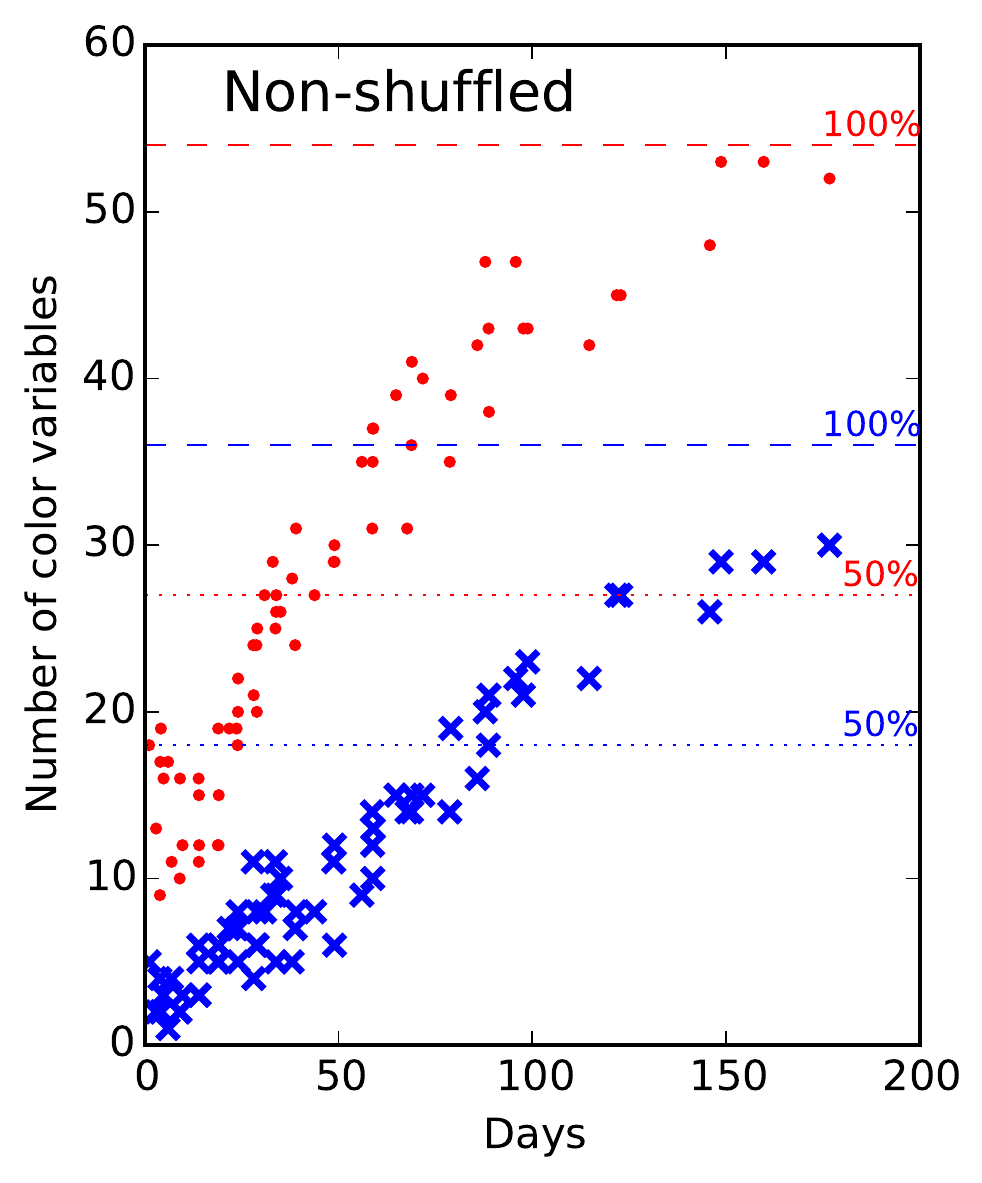}{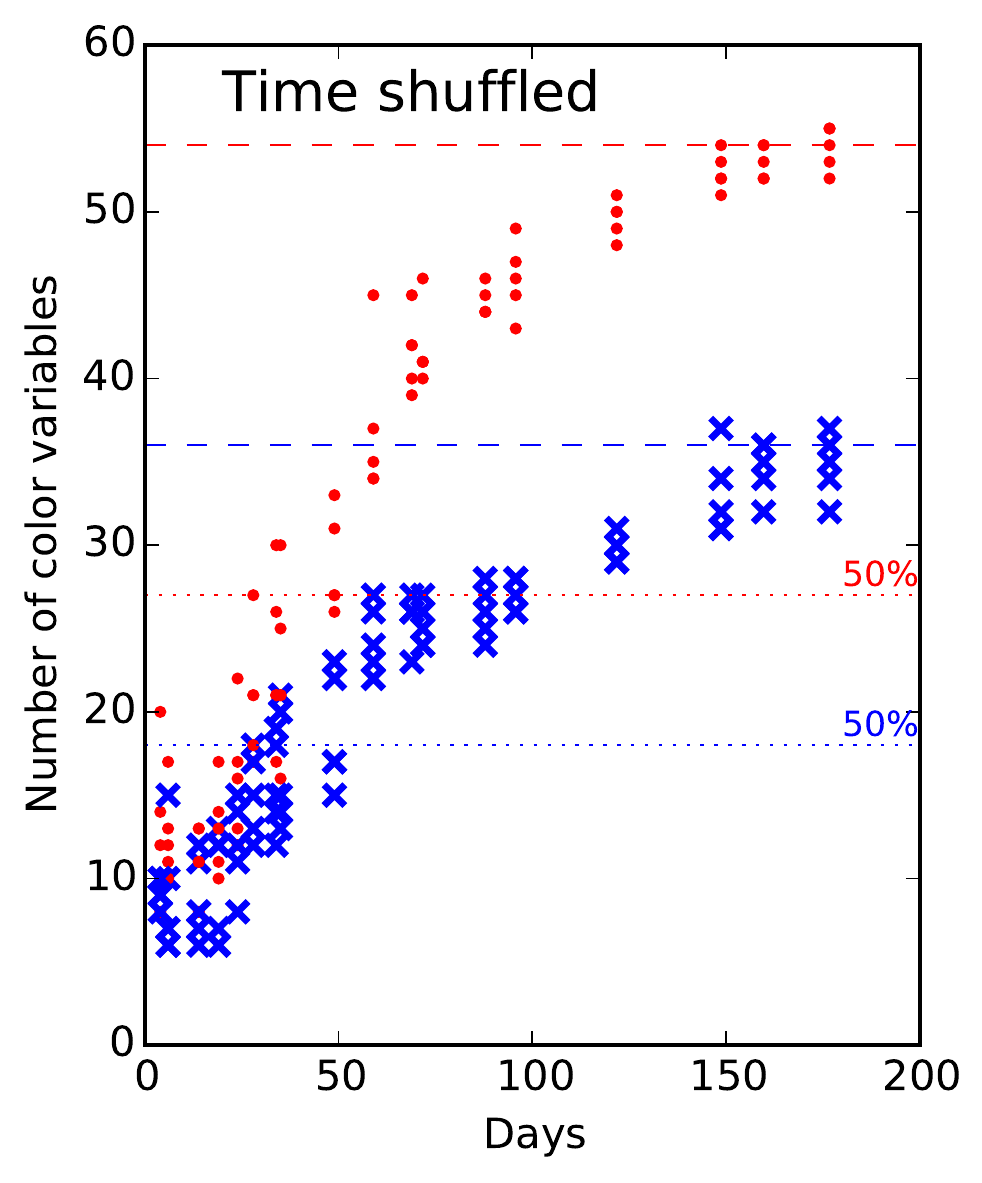}
	\caption{Number of reddening (red dots) and blueing (blue X's) objects as a function of time, for the shuffled and non-shuffled samples.
	There is a very significant difference between the shuffled and non-shuffled samples, indicating that the different shapes of the non-shuffled red and blue curves are due to differences in their intrinsic timescales.
	Dotted and dashed lines in each color show 50\% and 100\% of the final count of color variables of each type, respectively.
	Note that the non-shuffled blue curve takes much longer to reach its 50\% mark than the time-shuffled blue curve.
	In the shuffled data, a few points appear above the ``100\%'' line due to chance selection of non-consecutive color points during the shuffling process; this does not affect the interpretation.
	\label{fig:timescale_redblue}}
\end{figure}

Finally, we studied how the number of ``blueing'' versus ``reddening'' stars changed over time (Fig. \ref{fig:timescale_redblue}).
As a class, we refer to these stars as ``color variables''.
The following plots were made by comparing two samples: 
(a) the original data, un-altered, but sampled over time windows of varying length;
(b) the data with randomized timestamps, also sampled over time windows of varying length.
The inclusion of time-randomized data is to distinguish whether the signal simply emerges because additional datapoints bring down the noise, or because the addition of data sampled at specific times reveals a truly time-varying signal.
Multiple non-shuffled samples were created by sliding the time window by the size of window considered, allowing us to probe the same temporal space with independent subsets of the data.

\begin{figure}
\plotone{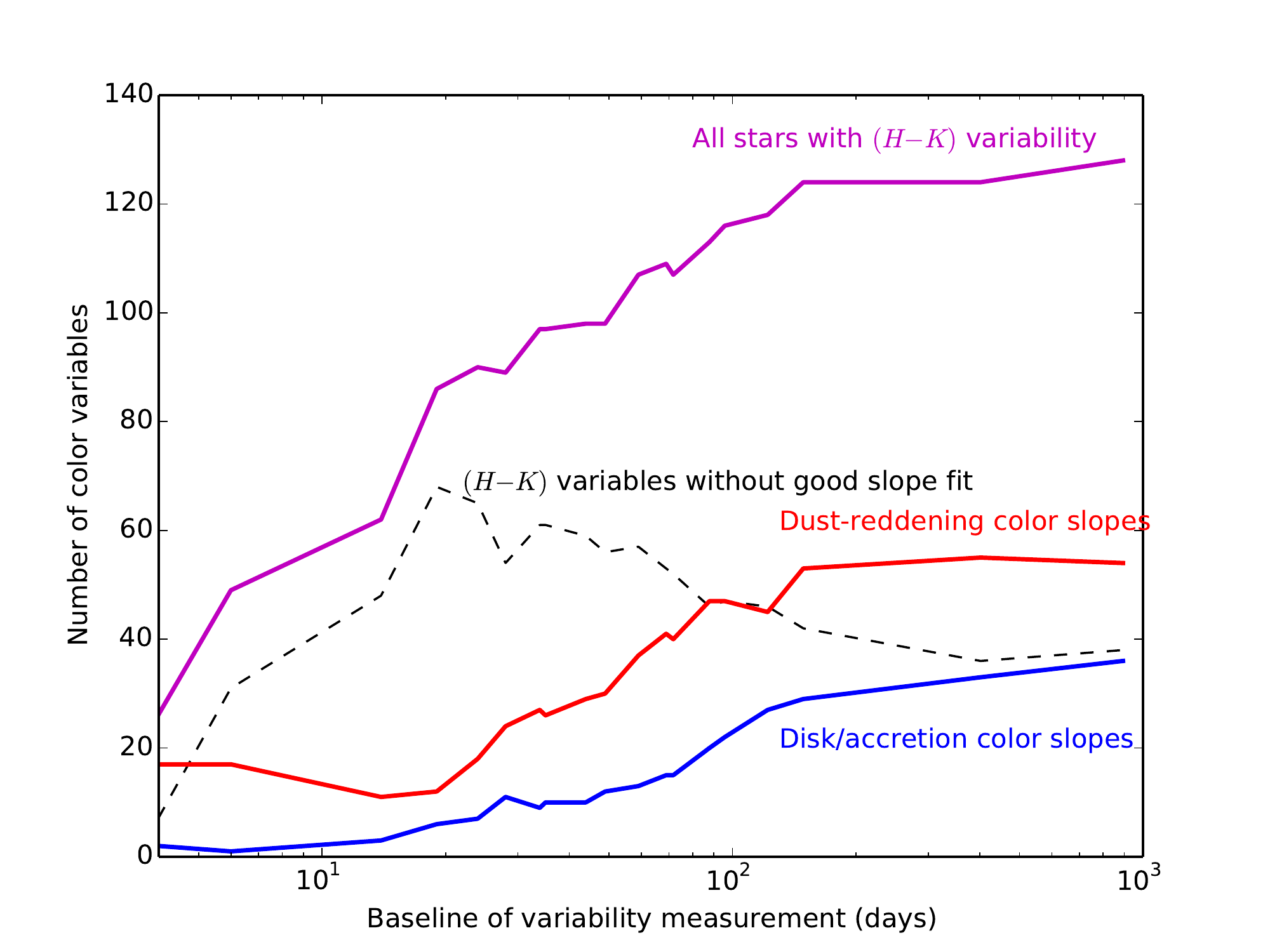}
\caption{ The number of variables showing significant $H-K$
  variability (and a `good' fit to a color slope, in the red and blue curves), as a function of observing baseline.
By analyzing progressively longer subsets of our data, we 
find that the number of ``color variables'' increases steadily with the
length of monitoring.
Notably, the disk/accretion variables show the strongest dependence
on observing baseline: for simulated campaigns spanning less than $\sim 80$ days,
less than half of the accretion/disk variables that we find in our
final analysis are recovered.
The length of our nearly continuous first observing season is $\sim
180$ days; in this plot, the slope of each line becomes shallower
after this interval, suggesting that if we had nightly observations
year-round for the same 894 day span, the $t > 200$ slope of each line
would be steeper.
Note that the red curve is not monotonically increasing because some stars initially show a reddening color trend that is washed out by later data (in some cases due to the compound variable behavior outlined in \S \ref{sec:compound}).
\label{fig:colorslope_temporal} }
\end{figure}

As noted in \S \ref{sec:aggregate}, we find a larger proportion of ``blueing'' to ``reddening'' color variables in this study compared to \citet{carpenter01}.
By analyzing color trends over progressively longer subsets of our
time-series data, we find that this difference is due primarily to our longer and more
complete time coverage (Figure \ref{fig:colorslope_temporal}).
We found that while 1/3 of the reddening-dominated stars could be identified with just a few days of data, at least a month was required to identify a similar fraction of disk/accretion dominated objects.
A little over one-fourth of the color variables never presented a linear slope which could be fitted with precision of better than 20\%. 
Within 5 days, more than 10 reddening objects were typically found, but less than 5 blueing objects are found in these short baselines.
By the end of the study, there were 54 ``reddening'' objects and 36 ``blueing'' objects identified;
half of the reddening objects were identified within 40 days, but it took 90 days to identify half of the blueing objects.
There is a substantial difference between the shuffled and non-shuffled samples: 
in the shuffled sample, the reddening and blueing curves have a similar shape, and hence a similar time-dependence;
but in the non-shuffled sample, the blueing curve lags substantially behind the reddening curve and has a lower slope until around 80 days.
Since the characteristic timescale of blueing objects is more than twice that of reddening objects, this is a strong indication that long, contiguous studies are necessary to better understand the blueing objects, even as studies around 40 days in length were able to sufficiently characterize the reddening objects.

%\clearpage

\section{Exceptional Variables}

%!TEX root = varm42.tex

\label{sec:exceptional}

The tremendous amount of data in a survey such as this has allowed us
to identify a number of lightcurves of special interest.
These systems include ``Dippers'' and AA Tau variables (\S \ref{sec:dippers}), stars which undergo widely-spaced periodic brightenings (\S \ref{sec:flareboys}), stars with particularly long periods or long-timescale non-periodic variability
(\S \ref{sec:longperiods}), and eclipsing binaries (\S \ref{sec:eb}).
Each is discussed below.

\subsection{Dippers and AA Tau type variables}
\label{sec:dippers}

\citet{morales11} note a number of stars that get
redder as they get fainter during abrupt ``dimming'' events that are
sometimes periodic, sometimes not; they use the term ``dippers'' for
these stars.
The interpretation of this behavior presented in \citet{morales11} is that these dips are either caused by high-latitude warps
in the inner disk (especially in periodic cases showing AA~Tau-type
variability) or that the stars are being extincted by overdense
clouds in the disk atmosphere.
In our dataset we also find a number of dippers.

\begin{figure}
\plotone{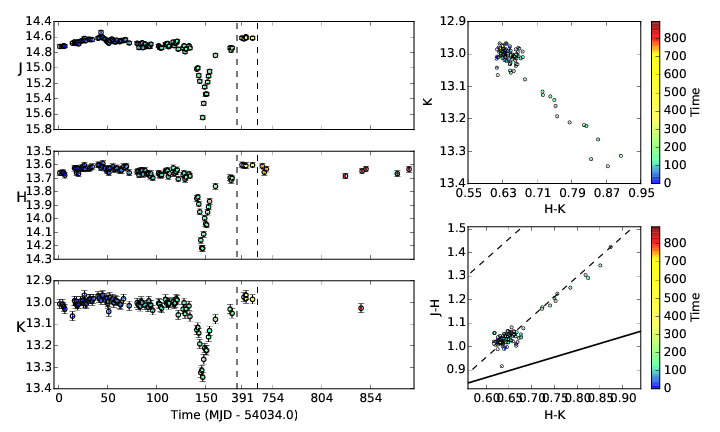}
\caption{Light curve for ``dipper'' star ONCvar 310.
It shows an abrupt dust eclipse from $t=135$ to $t=160$, dimming by
0.9 mag at $J$ and 0.3 mag at $K$.
\label{lc:310} }
\end{figure}

\begin{figure}
	\plotone{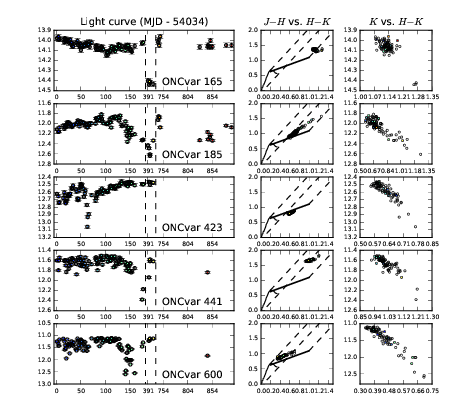}
	\caption{Five more ``dipper'' light curves, each somewhat irregular.
	Columns from left to right: $K$ band light curve; $J-H$ vs. $H-K$ color-color diagram; $K$ vs. $H-K$ color-mag diagram.	
	\label{lc:irregular_dippers} }
\end{figure}

To clarify terminology, we make a distinction between non-periodic dippers, which may or may not show repeated ``dipping'' events, and the periodic AA~Tau-type variables; we use only the term ``dipper'' for the former, and ``AA~Tau'' or ``AA~Tau-type variable'' for the periodic latter class.
Example dippers are shown in Figures \ref{lc:310} and \ref{lc:irregular_dippers}.
Example AA~Tau-type variables are shown in \ref{fig:aatau_tenpanel}.

Among individual, non-periodic light curves, a few stand out as deserving
particular attention as ``dippers''.
ONCvar~310 (Figure \ref{lc:310}) shows a dramatic, well-defined
``dip'' on top of mostly quiescent variability.
The quiescent state lasts from $t=0$ until $t=135$;
from $t=135$ to $t=160$, it decreases and then steadily
increases in brightness, with $\Delta J = 0.92$, $\Delta H = 0.53$, and
$\Delta K = 0.33$.
The observed wavelength dependence during eclipse closely matches
typical ISM exctinction laws. 

Five other prominent examples of one-off or irregular ``dippers'' are 
shown in Figure \ref{lc:irregular_dippers}.

\subsubsection{Periodic ``dippers'': AA Tau analogs}
\label{aatau}

\begin{figure}
	\plotone{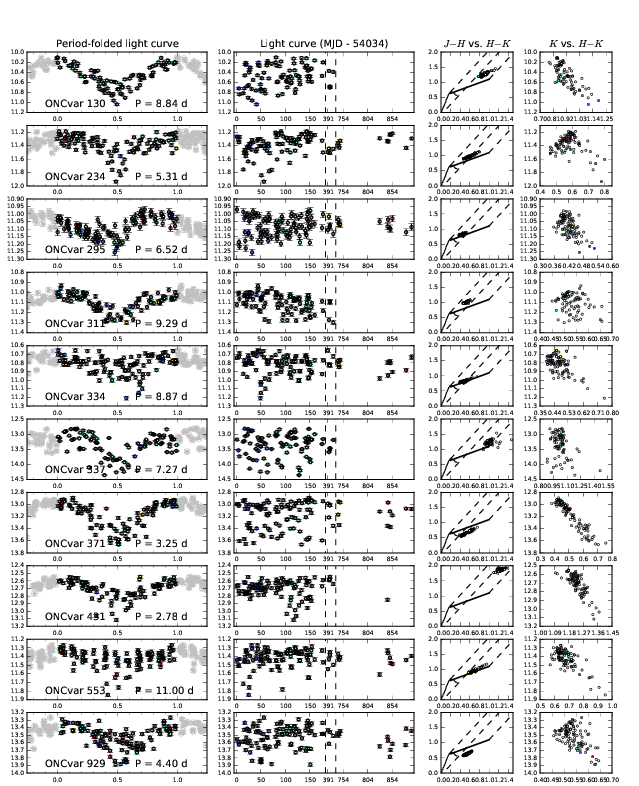}
	\caption{ Ten example AA Tau light curves.
	\label{fig:aatau_tenpanel} } 
\end{figure}

The variable star ``AA~Tau'' exhibits periodic dips in flux consistent with dust-based extinction, leading authors to suggest the presence of a magnetically-warped inner dust disk responsible for the dips in flux (Bouvier et al. 2003, 2007; Donati et al. 2010).
This class of variability appears frequently in our data.
We identify such variables in our data using three criteria: 
\textit{a)} They are periodic, as determined by the periodicity criteria presented in \S \ref{subsubsection:autoperiod} 
\textit{b)} Their color variability in $H-K, K$ space is significant and consistent with dust-based eclipses
\textit{c)} Their lightcurve morphology is not well-behaved (i.e., ``cleanly'' periodic) enough to be explained by rotational modulation of starspots; often, this means large variability amplitudes, an amplitude that changes over time, etc.
Several example AA Tau type light curves are shown in Figure \ref{fig:aatau_tenpanel}, with a range of periods and variability amplitudes.
Notably, each AA Tau type variable seems to sweep out a similar narrow line (parallel to the CTTS locus of \citealt{meyer97}) in $J-H, H-K$ space, even though $J-H, H-K$ behavior is not among the selection criteria.

By applying these criteria to all 539 periodic light curves and investigating them by eye, we identify 73 AA Tau variables.
This is 6.1\% of all variables and 13.5\% of periodic variables.
The class distribution of AA Taus is 2 protostars, 63 disked stars, 3 non-disked stars, and 5 of unknown class; clearly this group is by far dominated by disked stars.
There are 641 disked variables, and 229 periodic disked variables in this study.
Thus 9.8\% of all disked stars are of AA Tau type,
and 27.5\% of periodic disked stars are of AA Tau type.

\begin{figure}
	\plottwo{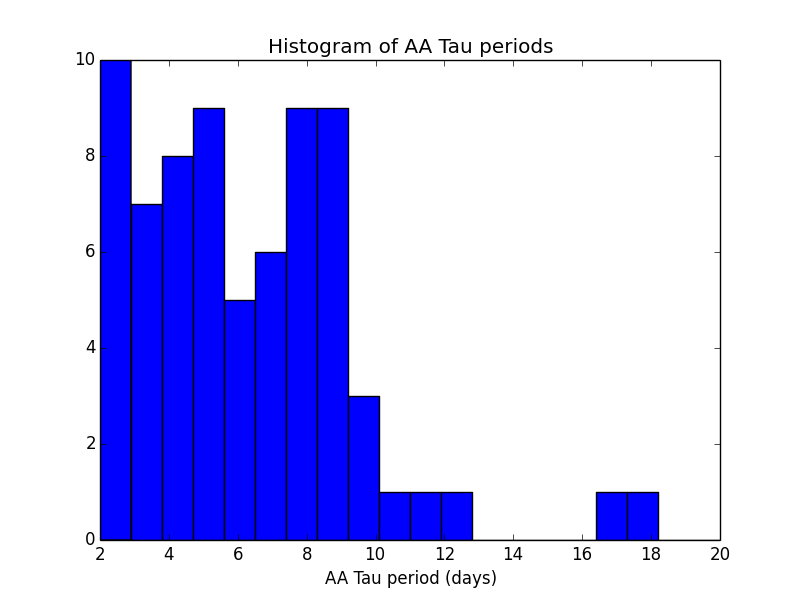}{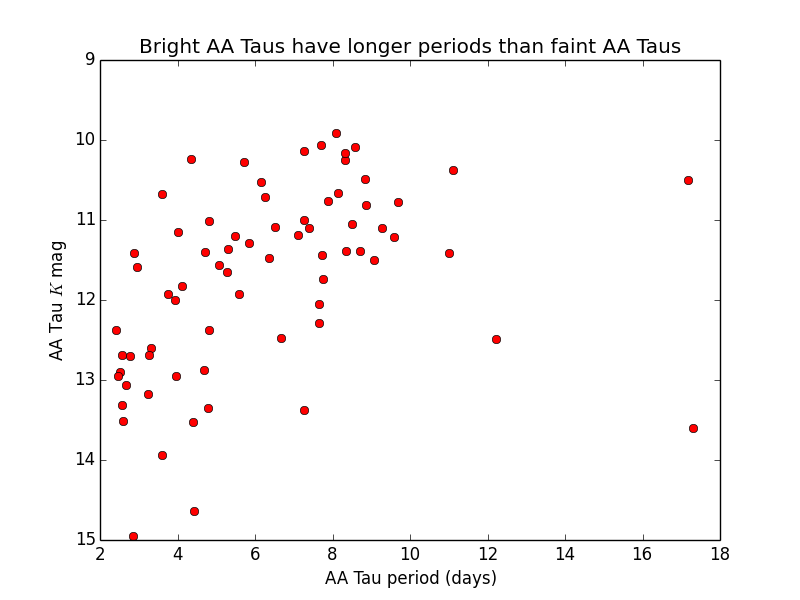}
	\caption{(left) Distribution of periods among AA Tau type variables. (right) Median $K$ magnitude of AA Tau type variables versus period. 
	The apparent trend in the data is likely an observational bias, as long periods are intriniscally easier to detect for bright stars.
	\label{fig:aatau_period_hist_mag} }
\end{figure}

The period distribution of the AA Tau type stars is shown in Figure \ref{fig:aatau_period_hist_mag}a. 
Most AA Tau type variables have periods less than 15 days, with a concentration below 6 days and another concentration between 7 and 10 days.
This slight bimodality is reminiscent of the overall disked period distribution from the middle panel of Figure \ref{fig:period_distributions}.
When plotted against median $K$ magnitude of each star (Figure \ref{fig:aatau_period_hist_mag}b), we see a strong trend for fainter stars to have shorter periods. 
The scarcity of AA Taus in the bottom-right quadrant of this plot is primarily a selection effect, explained by our low sensitivity to variables at these periods and brightnesses:
A noisy signal is easier to confirm observationally if it repeats many times. Since bright stars have less observational noise than faint stars, their periods are more easily detectable even when the periods are long. The worse signal-to-noise for fainter stars means more repetitions are needed to confirm a signal.
The lack of AA Taus in the top-left quadrant, however, is not explained by a selection effect, and may indicate that AA Taus of a given period have a maximum luminosity.
This is perhaps explained by noting that the brightest stars will sublimate the dust that would orbit at small distances, thus putting a minimum period for bright stars.
As discussed in Section \ref{sec:analysis}, 
the physical properties of such young stars and their disks have been measured by \citet{eisner09}: 
the inner disk walls are in the range $0.14-0.18$ AU, with disk wall temperatures of $1260-1850$ K.

\subsection{Long-Period Brighteners}
\label{sec:flareboys}

Most periodic variables either \textit{a)} spend equal amounts of time brighter and fainter than
the mean (e.g., starspots or any nearly sinusoidal variability) or \textit{b)} spend most of the time brighter than
the mean (e.g., eclipsing binaries or stars that undergo occultation).
In addition to these types of variables, we observe three periodic variables with cleanly ``inverted''
light curves, i.e., they spend most of their time \textit{fainter} than
their mean brightness; four more stars show similar behavior, but not as cleanly, and are discussed in Section \ref{sec:longperiods}. 
Among YSOs, this general class of near-infrared behavior has been previously seen in the
long-period variables
CHS 7797 in the Orion Nebula Cluster 
\citep[17.8 day period;][]{rodriguez12, rodriguez13},
WL 4 in $\rho$ Oph 
\citep[130.87 day period;][]{plavchan08b}, 
and YLW 16A also in $\rho$ Oph 
\citep[92.6 day period; ][]{plavchan13}, 
with KH 15D in NGC 2264 
(48 day period; \citealt{kearns98}, \citealt{windemuth14} and references therein) as an optical counterpart.
Importantly, all such variables show periods greater than 15 days, suggesting dynamics longer than either typical stellar rotation or the orbital period of the inner disk.
In this dataset, we find three unambiguous examples of long-period brighteners:
the previously noted CHS 7797, known here as ONCvar 497 (Figure \ref{lc:479}), 
ONCvar 149 (36.4 day period; Figure \ref{lc:149}),
and ONCvar 1226 (88.5 day period; Figure \ref{lc:1226}). 
This type of periodic behavior, sometimes called ``inverse eclipses'' (e.g., by \citealt{parks14}), has been suggested to be due to occultations of binary stars by their circumbinary disks 
(\citealt{parks14}, \citealt{plavchan08b, plavchan13}).
Such systems may provide key insights into circumbinary disks, where circumbinary planets such as 
Kepler-16b \citep{doyle11} would be expected to form.

KH 15D in NGC 2264 has a period of 48.37 days and is interpreted as a pair of T Tauri stars surrounded by an inclined, warped, disk-like ring of non-ISM dust \citep{windemuth14}.
Its periodic behavior was discovered in 1998 \citep{kearns98} and it has since been well-characterized at many wavelengths, with polarization and spectra.
\citet{windemuth14} note the possibility of a young planet shepherding the well-defined ring structure and possibly contributing to $J$ and $H$ excess flux seen toward the system.

These long-period brightening sources are reminiscent of peculiar stars seen in the embedded star forming region Lynds 1688 in $\rho$ Oph.
WL 4 and YLW 16A in $\rho$ Oph, the 131-day and 93-day periodic systems found, respectively, by \citet{plavchan08b} and \citet{plavchan13}, and discussed further by \citet{parks14}, have near-infrared
behavior remarkably similar to the examples displayed below (especially ONCvar 149 and 1226), so the analysis 
presented in these papers may serve as a useful guide for interpreting the two ONCvar objects.
In summary, the lightcurves and SEDs of these stars are well-modelled by a triple system including an inner binary pair eclipsed by their circumbinary disk; both systems have resolved the outer tertiary companion.
The color behavior during the ``bright'' part of the lightcurve gets redder at $J-H$ and $H-K_s$ in each system.

\subsubsection{ONCvar 479 (CHS 7797)}
ONCvar 479 (CHS 7797; \citealt{carpenter01}, \citealt{rodriguez12, rodriguez13})
shows an unusual periodic brightening cycle with period $\sim 18$
days and a large amplitude of $\sim 1.7$ mag.
This star has been characterized extensively for photometric
variability in optical and infrared ($[3.6]$ and $[4.5]$
\textit{Spitzer} bands) as well as spectroscopically in two phase
points in its periodic light curve.
This star's period was accurately derived to be 17.786 days by
\citet{rodriguez12}
using six years of $I$-band monitoring;
its variability was also characterized at $R$, $z^\prime$, $[3.6]$ and
$[4.5]$ bands, showing variability amplitudes
$\sim 1.7$ in $R$, $I$ and $z^\prime$ and decreasing slightly at longer
wavelengths. 
The star's SED indicates that it is a Class I object, and its spectral features show that it is a low-mass object near the stellar/substellar boundary \citep{rodriguez13}.
The star was noted to show colorless variability at wavelengths $\lambda \le 2 \mu m$,
but at $\lambda \ge 2 \mu m$ it became redder when fainter; this was
interpreted as evidence for dust grains that had grown to $1-2 ~\mu m$.
In \citet{rodriguez13}, the authors identify spectral differences between two light curve phases that suggest the system has a similar geometry to KH 15D.

\begin{figure}
\plotone{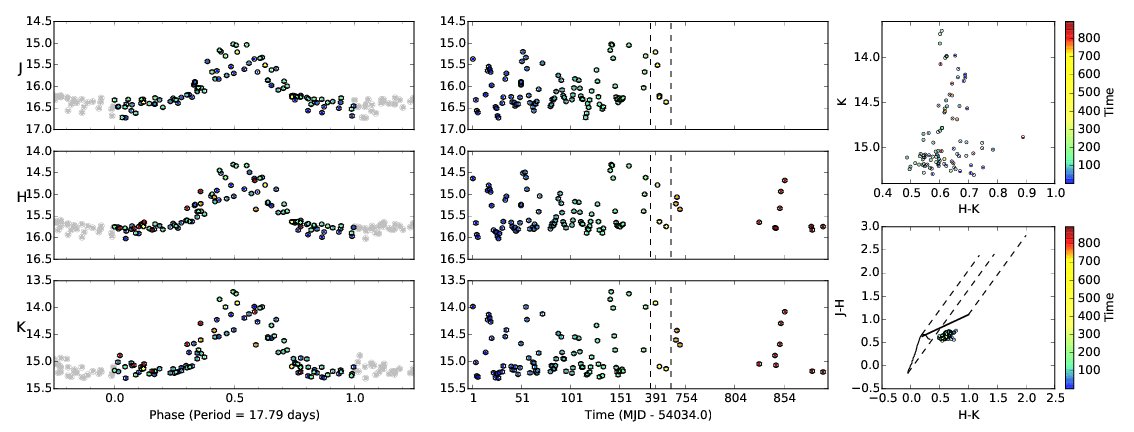}
\caption{Light curve for ONCvar 479 (also known as CHS 7797), an example of a ``Long-Period Brightener''.
\label{lc:479} }
\end{figure}

In our observations (Figure \ref{lc:479}), the star varies with amplitude $\Delta K \sim
1.5$, $\Delta J \sim \Delta H \sim 1.4$, within the fluctuations seen
in the optical monitoring, and its flux changes are on
average colorless, which is consistent with the \citet{rodriguez12} study.
The period we derive from these observations is 17.795 days, consistent within 0.01 day of the six-year monitoring study.
Many of our $JHK$
observations are contemporaneous with the six-year monitoring
of this star at $I$ \citep{rodriguez12}.

\subsubsection{ONCvar 149}
\begin{figure}
\plotone{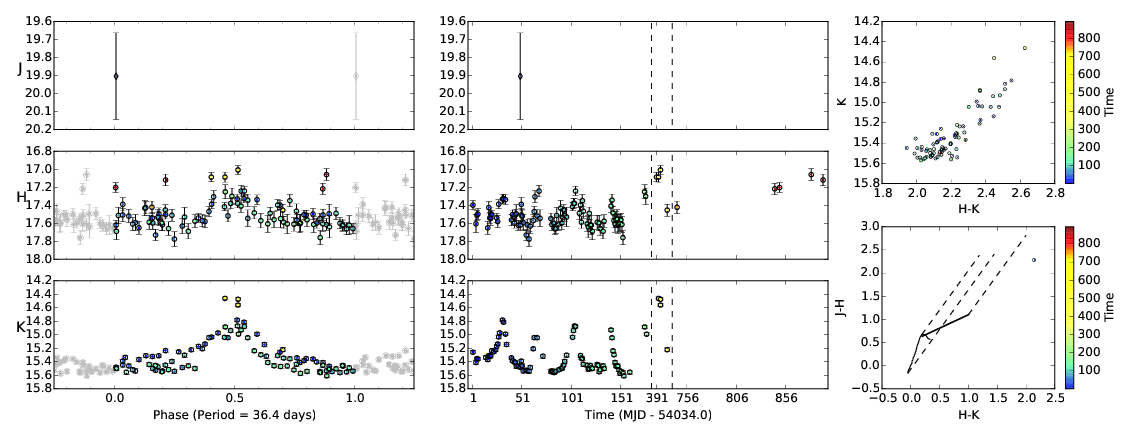}
\caption{ Light curve for ONCvar 149, an example of a ``Long-Period Brightener''.
\label{lc:149} }
\end{figure}

Another long-period brightener is ONCvar 149 (Figure \ref{lc:149}),
which is redder and has a much longer period of 36.40 days.
Its \citet{megeath12} classification is ``protostar'' and it is highly reddened, with $H-K$ varying around 2.3.
Due to this star's faintness at shorter wavelengths, our data contain only one measurement at $J$ ($J=19.90 \pm 0.24$).
Even though this star's faintness at $J$ and $H$ means that its
measurements at those bands have high uncertainties, it is still clear
that the variability amplitude is much higher at $K$ (one cycle: $\Delta K
= 0.64$) than at $H$ (one cycle: $\Delta H \le 0.3$), in stark contrast to
the colorless or weaker-with-long-wavelength variability seen in ONCvar 479 (CHS
7797).
This $K$, $H-K$ slope indicates that disk/accretion activity dominates the variability.
The differing wavelength dependence of variability may indicate that
the superficially similar light curve morphology seen in ONCvar 479 (CHS 7797) and
ONCvar 149 does not ultimately originate from analogous physical conditions.

\subsubsection{ONCvar 1226}
\begin{figure}
\plotone{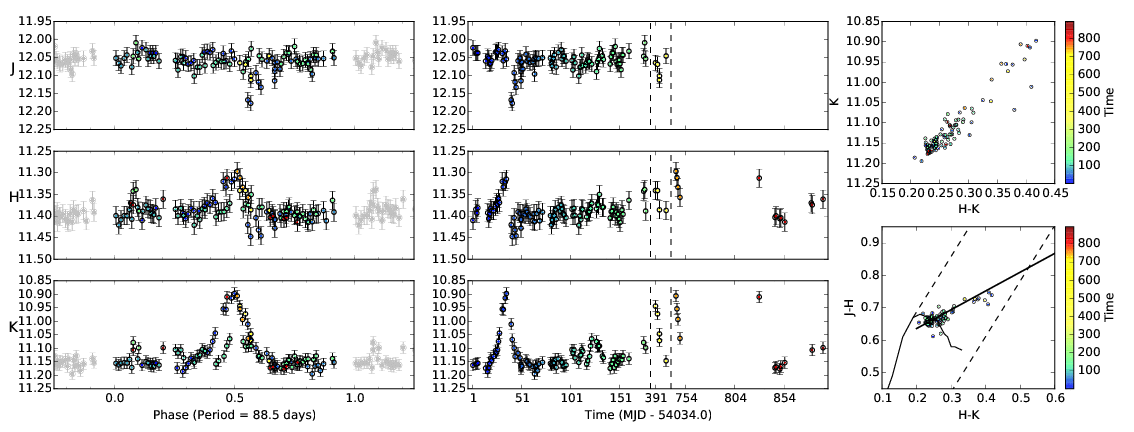}
\caption{ Light curve for ONCvar 1226, the longest-period example of a
   ``Long-Period Brightener'' found in this dataset.
\label{lc:1226} }
\end{figure}

ONCvar 1226 is another variable showing ``long-period brightener'' light curve
morphology (see Figure \ref{lc:1226}).
It shows the characteristic periodic bursts that fold on a steady
period with modulations in amplitude.
This light curve is particularly notable for two reasons: 
{\it a)}, its very long 88.5 day period is longer than previous NIR
monitoring campaigns of the ONC;
and {\it b)}, its bursts get stronger with increasing wavelength -- they
are clearly seen at $K$, and show an even larger amplitude at $[3.6]$
and $[4.5]$ bands, 
but the $J$ band light curve has weak anticorrelated (i.e., \textit{fading}) behavior instead;
this anticorrelation gives ONCvar 1226 a Stetson index of 0.97, just below the nominal 1.0 cutoff.
This also sets ONCvar 1226 clearly apart from ONCvar 479 (CHS 7797), whose
light curve in bands shorter than $2
\mu m$, have similar shape and amplitude including $R$, $I$ and $z^\prime$.
Unlike ONCvar 149, which is faint and highly reddened ($K \approx 15, J
\approx 20$), making optical followup impossible, ONCvar 1226 is bright
and relatively unreddened ($K \approx 11, J \approx 12$), so it would be an
excellent target for optical spectroscopic followup, as was done with
CHS 7797 in Rodriguez-Ledesma et al. (2013).

\begin{figure}
\plotone{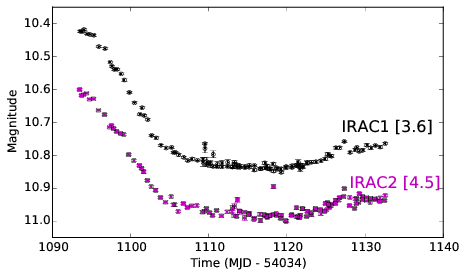}
\caption{ 40-day YSOVAR light curve for ONCvar 1226, showing a brightening event and possible secondary brightening roughly 7 months after the end of our UKIRT monitoring program. 
The $[3.6]$ and $[4.5]$ data show a continuation of the 88.5 day period seen at $H$ and $K$.
\label{ysovar:1226} }
\end{figure}

Based on the 88.5 day period, ONCvar 1226 is expected to have undergone ten brightening events during our 894-day observing window.
Five of these overlap with our actual observing epochs, and all five are confirmed in the light curve.
(This is fortuitous, as our 120 nights have only $\sim13\%$ coverage of the 894-day window, and yet we catch 50\% of the brightenings.)
The $K$ amplitude of each brightening event varies considerably, from $\Delta K \sim 0.11$ to $0.3$, but most are near $0.3$.
This star has not been previously noted to be periodic or exceptional, but fortunately,
the public \textit{Spitzer} YSOVAR light curve data in the Orion Nebula Cluster taken in late 2009 (\citealt{morales11}) coincide with the third brightening event after the last one we observed (Figure \ref{ysovar:1226}).
We accessed the YSOVAR data after finding the star to be of interest from our data, and were able to further confirm the $\sim 88.5$ day period using these IRAC data, which show the latter half of the star's brightening event down to its quiescent level in the $[3.6]$ and $[4.5]$ IRAC bands, with amplitude 0.38 and 0.45 respectively.

At $K$ band, hints of possible ``secondary'' brightenings at half-phase are seen around $t=167$ and $t=876$ with amplitude $\Delta K \sim 0.08$, but the data are not conclusive. 
A corresponding half-phase brightening is not seen at $t=79$, where one might be expected.
Notably, the IRAC $[3.6]$ and $[4.5]$ data from YSOVAR each indicate a clear, plateau-shaped rise in brightness of order 0.06 mag (well above the photometric noise limit) at half-phase (Figure \ref{ysovar:1226}).

Our $JHK$ data could also be consistent with a period twice as long, or 177.0 days; the three cleanest brightenings (at $t=35$, $t=389$, and $t=743$ all fall on each other when folded in this way.

\subsection{Long Periods and Long-Term Variability}
\label{sec:longperiods}

\subsubsection{Periodic}

\begin{figure}
	\plotone{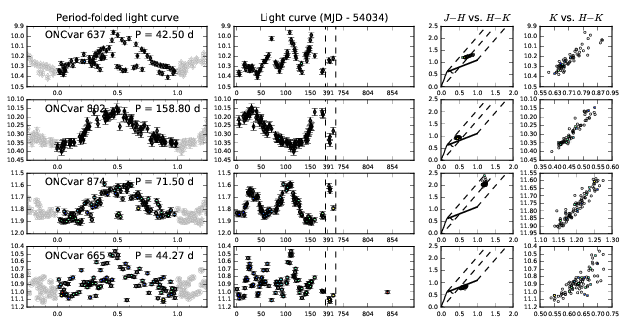}
	\caption{Composite light curve of four long-period variables
	that we identify as candidate long-period brighteners.
	Their color behavior is quite consistent, and similar to the color behavior of the definite long-period brighteners shown in Figures \ref{lc:1226} and \ref{lc:149}.
	\label{lc:longperiods}}
\end{figure}

In addition to the long periods seen in \S \ref{sec:flareboys} (ONCvar
149 and 1226, Figures \ref{lc:149} and \ref{lc:1226}), 12 other
stars with near-IR trends indicating cycles longer than $\sim 30$ days are apparent in our data
(see Figure \ref{lc:longperiods} for several examples).
The aforementioned stars ONCvar 149 and 1226 have periods likely to originate
in the orbital period of a binary companion;
in other cases, we speculate that slow-rotating spotted stars induce a clean long-term
period (such as in Figure \ref{lc:rotators}), while in other cases a more messy period is induced by dynamic behavior
associated with the inner edge of a disk interacting with its star (this is suggested because such systems often show ``blueing'' behavior in $K$, $H-K$ color-mag space). 
The observations presented in \citet{plavchan08b} and \citet{plavchan13} represent some of the first identifications of long-period
systems.

% rudiments of a table
\begin{deluxetable}{lcc}
\tablewidth{0pt}
  \tablecaption{ Long period periodic stars
\label{tab:longperiods} }
	\tablehead{
\colhead{ONCvar ID } & \colhead{ Period (d) }& \colhead{LPB}} 
  \startdata 
  231 & 70.48 & \\
  479 & 36.40 & Y \\
  529 & 92.51 & C \\
  637 & 42.5 & C \\
  665 & 44.27 & C \\
  801 & 96.21 & \\
  802 & 158.8 & C \\
  816 & 92.35 & \\
  819 & 37.0 & C \\
  826 & 30.22 & \\
  843 & 75.16 & C \\
  874 & 72.88 & C \\
  1027 & 37.82 & \\
  1226 & 88.5 & Y \\
  \enddata
  \tablecomments{A ``C'' in the last column denotes stars that are identified as candidate long-period brighteners, while a ``Y'' denotes a definite identification as such.}

\end{deluxetable}

We restricted our automated period search in \S
\ref{subsection:period} to periods below 50 days in order to guarantee
that the observed periods were reliably repeatable, as many irregular
variabilities in young stars may appear sinusoidal over some time frame
only to change its variability pattern later. 
In this section we focus on targets whose apparent periodicity is
nonetheless worthy of comment, although we cannot constrain the
absolute frequency of such systems.

Some stars show what might be classified as purely irregular
variability that, observed over long term, shows an overall periodicity.
ONCvar 637 (Figure \ref{lc:longperiods}, row 1) is a system showing quasi-periodic behavior
over a 42.5 day period, with varying amplitude each cycle, in which
the variability is dominated by changes in $K$ excess. 
Such a system is inconsistent with AA~Tau-type dust-occultation
behavior, and the 42.5 day period is also longer than the expected
rotation period of the inner disk that emits at $K$ band.
The period may be caused by the presence of a binary companion,
perhaps modulating the accretion rate or the disk structure via
gravitational interaction with an orbiting point mass.
Note that ONCvar 637 avoids a definitive ``long-period brightener'' classification as its ``bright'' state is not narrow compared to its ``faint'' state, and also its light curve is somewhat irregular, but its period and color behavior are otherwise comparable to the long-period brightners presented in the previous section (see Figure \ref{lc:1226} and \ref{lc:149}).
ONCvar 874 (Figure \ref{lc:longperiods}, row 3) shows similar behavior during a
roughly 50 day period, in which the amplitude modulates over the
course of our observations, but the color signature and qualitative
morphology is the same as in ONCvar 637.
ONCvar 665 (Figure \ref{lc:longperiods}, row 4) is a third example of a star showing
this approximate behavior, in which irregular variability dominated by
an accretion/disk track in $H-K, K$ space is modulated by a period
around 40-50 days. 
This one has a 44.27 day period.

\begin{figure}
\plotone{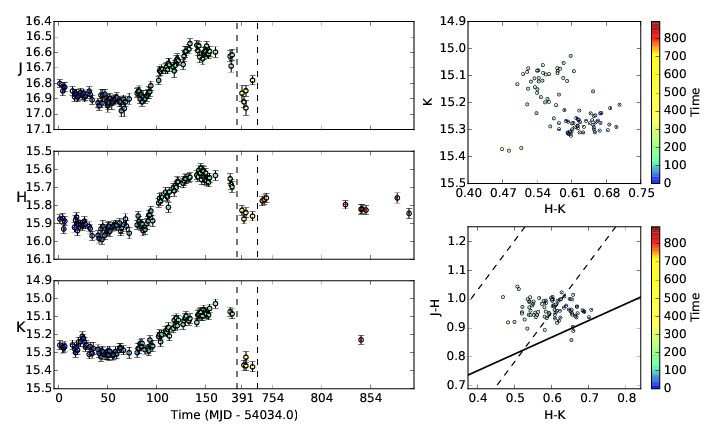}
\caption{Light curve for slowly-varying star ONCvar 663.
It seems to show continuous cyclic variations, but our observations are
unable to constrain whether its lightcurve repeats regularly.
\label{lc:663} }
\end{figure}

Some stars vary smoothly over such long terms that not even our long
time baseline can confirm that they are repeatably periodic.
ONCvar 663 (Fig. \ref{lc:663}) does not show definitively periodic behavior, but shows
nearly sinusoidal long-term changes continuously throughout our first
observing season. 
The light curve of ONCvar 802 (Fig. \ref{lc:longperiods}, row 2), in contrast, has a
believable 158.8 day period, where the modulation seen over the course
of the first season begins to repeat and the second-season
observations fall on top of season one when folded on 158.8 days.

\subsubsection{Non-periodic}

25 stars show dramatic long-term variability that is not periodic.

\begin{figure}
\plotone{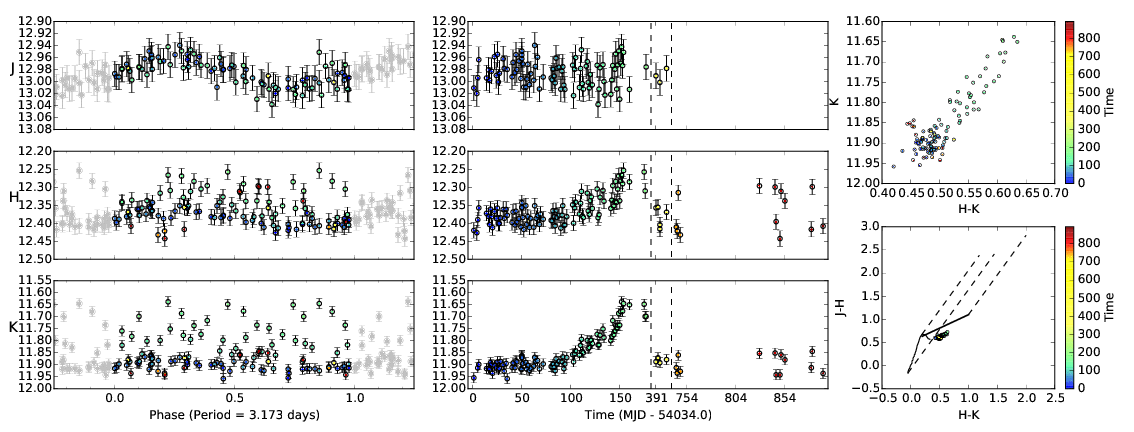}
\caption{ Light curve of ONCvar 957, which shows different qualitative
  behavior at different wavelengths: at $J$, it is neatly periodic,
  but at $K$, it has a long-term rise in brightness perhaps associated
  with a steady increase in accretion rate.
\label{lc:957} }
\end{figure}

ONCvar 957 (Figure \ref{lc:957}) 
shows a clean period at $J$ but a long-term trend at $K$.
The most natural way to explain such strange behavior is to assume
that the $J$-band short-period behavior originates on the stellar photosphere in
this system, analogous to what would be observed at optical
wavelengths, 
while the $K$-band long term variability originates in the disk.
Indeed, the $K$ variability seen in ONCvar 957 
shows an unambiguous disk/accretion track in $H-K, K$ space.

\begin{figure}
	\epsscale{0.75}
	\plotone{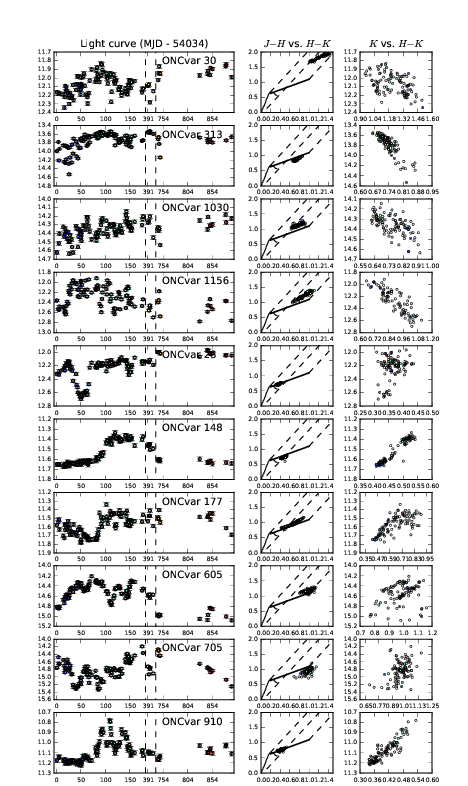}
	\caption{ Ten light curves of long-term non-periodic stars (labelled in panels). 
	Columns from left to right: $K$ band light curve; $J-H$ vs. $H-K$ color-color diagram; $K$ vs. $H-K$ color-mag diagram.	
	The stars are roughly organized by $H-K$, $K$ color-magnitude slope: stars that redden when faint are sorted to the top, followed by stars that become blue when faint.
	\label{lc:longnonpers} 	}
\end{figure}

Further shown in Figure \ref{lc:longnonpers} are ten examples of prominent, high-amplitude 
long-term variables which do not show strong signatures of periodicity.

Finally, it is worth noting that many or most of the non-variable objects monitored in this survey
could show non-periodic variability on timescales longer than 900 days, if only they could be monitored for long enough.

%\clearpage

%!TEX root = varm42.tex

\section{Eclipsing Binaries}
\label{sec:eb}

\begin{figure}
\plotone{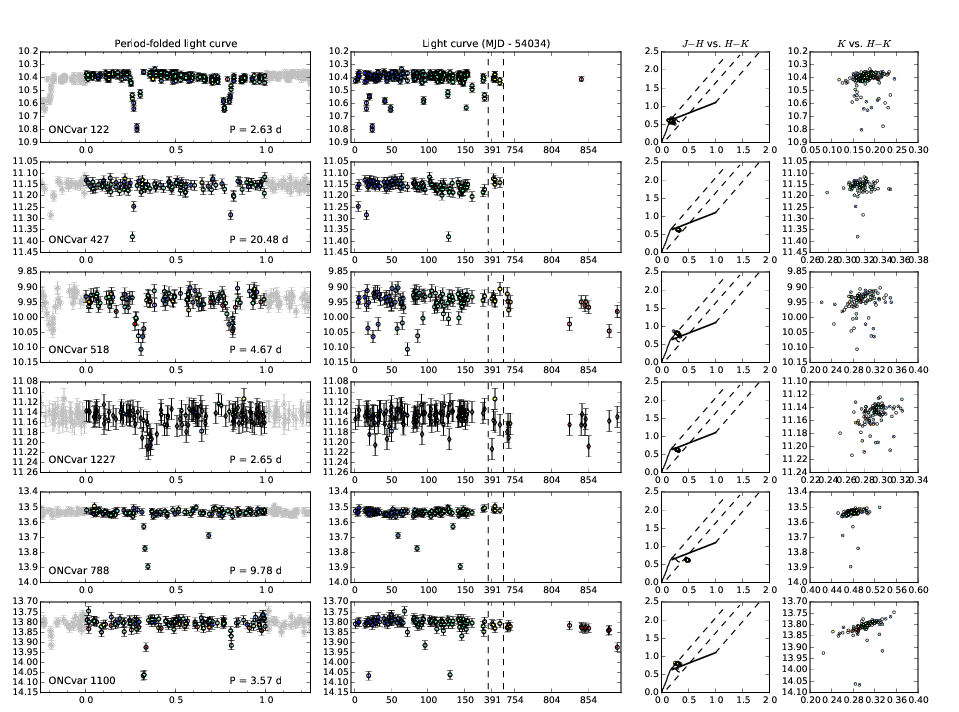}
\caption{
  Six previously-discovered eclipsing binaries that have light curves in our data. 
  See Table \ref{table:EB} for period references.
  \label{3oldEBs}
}
\end{figure}

Eclipsing binaries are astrophysically important objects because they
can yield fundamental stellar parameters such as masses, radii, and
temperature ratios.  Our survey is suited to find eclipsing
binaries in the direction towards the ONC. In the field, less than one
percent of stars are eclipsing binaries, but young stars have
considerably larger radii than on the main sequence, and hence the
chance that PMS binaries show eclipses should be correspondingly
higher. Our survey has monitored several thousand stars, so we would
expect to find a number of eclipsing binaries. Indeed, we uncovered 10
eclipsing binaries in the area studied, six of which were previously
known (shown in Figure~\ref{3oldEBs}). 
This is in fact fewer than we had expected given the larger radii
of PMS stars. In principle, the paucity of eclipsing binaries
might result if the in-spiral phase required to form a close
binary would on average take longer than the mean age of the
current population of the ONC. 
Alternatively, this could, at least in part, be a detection problem, since the irregular
variability of many of the stars in our study could impact our
ability to recognize all but the most pronounced eclipses. An even
longer monitoring campaign may be required to identify additional
eclipsing binaries in the ONC.

In the following we discuss the four newly discovered systems.
Unfortunately, we have been unable to determine conclusively whether the
four new eclipsing binaries are young. Two (V1916 Ori and V2716 Ori)
were reported to show low-level irregular variability by \citet{carpenter01}; 
marginal levels of variability were seen out-of-eclipse
in our data, not exceeding about 0.1 mag in each band. None of the
four systems show evidence of strong infrared excess based on
near-infrared and WISE photometry. However, it is perhaps not
surprising that a young eclipsing binary would not have much infrared
excess, depending on the prior orbital evolution after the binary was
formed, and many of the known PMS eclipsing binaries similarly do not
have infrared excess, e.g., JW~276, JW~380, V1174~Ori, and Par~1802.
Spectroscopy may be required to settle whether these four systems are
young or only line-of-sight associations with the ONC.

In Table~\ref{table:EB} we list the four newly discovered eclipsing binaries and
their basic data, together with nine other eclipsing binaries that
fall within our WFCAM field. Of these nine, three are saturated in all
three filters in our data, but for the other six our data confirm them
as eclipsing binaries. Eight of these eclipsing binaries
are proven young objects 
\citep[e.g., ][]{stassun04, stassun06, irwin07, cargile08, morales12}.

The period fits for these eclipsing binaries were obtained using the Fast Chi-Squared method (F$\chi^2$)
discussed in Section \ref{subsection:period}. 
Unlike the Lomb-Scargle periodogram, which searches for sinusoidal signals
and is not sensitive to periods below twice the sampling rate (here, 2 days), the F$\chi^2$ method
is able to recover periods for non-sinusoidal light curves even below two days.

\subsection{V1916 Ori }

This new eclipsing binary has a short period for which our best
estimate from the limited observations is 1.112403 days, seen in Figure~\ref{3newEBs}, top row. The secondary
minimum is at phase 0.5, indicating a circular orbit. The minima
appear to be slightly unequal, suggesting a surface temperature
difference of a few hundred degrees. A variable star name was given to
this star, since \citet{carpenter01} found the star to be
consistently variable throughout their observations, with amplitudes
in the range 0.11-0.12 and a Stetson index of 1.34. The variability we
see is much less than that, and only marginally larger than expected
from noise.  If real, perhaps the variability seen by Carpenter et al.
was due to transient spots.  The 2MASS colors suggest little or no
infrared excess. The star was also observed by WISE, and when placed
in a W1$-$W2 vs W2$-$W3 color-color diagram it falls in an area where very
few young stars are found \citep{koenig12}. Further study is
required to establish whether the star is young or not.

\subsection{ONCvar  1147}

This newly discovered object has a characteristic W~UMa-type light
curve \citep[see e.g.,][]{lucy68} with no constant parts, indicating that the stars are highly
deformed. The period is extremely short, only 9.492720 hours, which
accounts for the deformation of the stars. The light curve has been
modelled and Figure~\ref{ONCvar1147} shows a model fit to the $J$-band light curve.
The solution indicates a circular orbit with an inclination of
74.3$^\circ$ and a mass ratio of unity. The sum of the radii of the
stars is 0.75 times the semimajor axis, indicating that the system is
very near contact. The secondary minima get gradually deeper relative
to the primary minima from $J$ over $H$ to $K$.  The near-infrared colors
show no evidence for an infrared excess and little or no extinction.
The star was also observed by WISE, and when placed in a W1$-$W2 vs
W2$-$W3 color-color diagram it falls in a region where few young stars
are found \citep{koenig12}. The star's lightcurve is shown with color information in Figure~\ref{3newEBs}, middle row. 

\subsection{ONCvar 1190}

\begin{figure}
\plotone{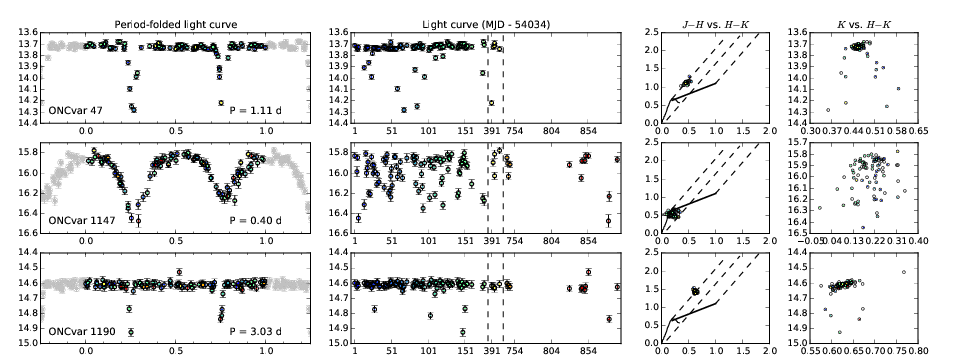}
\caption{
  Three of the four newly discovered eclipsing binaries.
  \label{3newEBs}
}
\end{figure}

\begin{figure*}
\epsscale{0.4}
\plotone{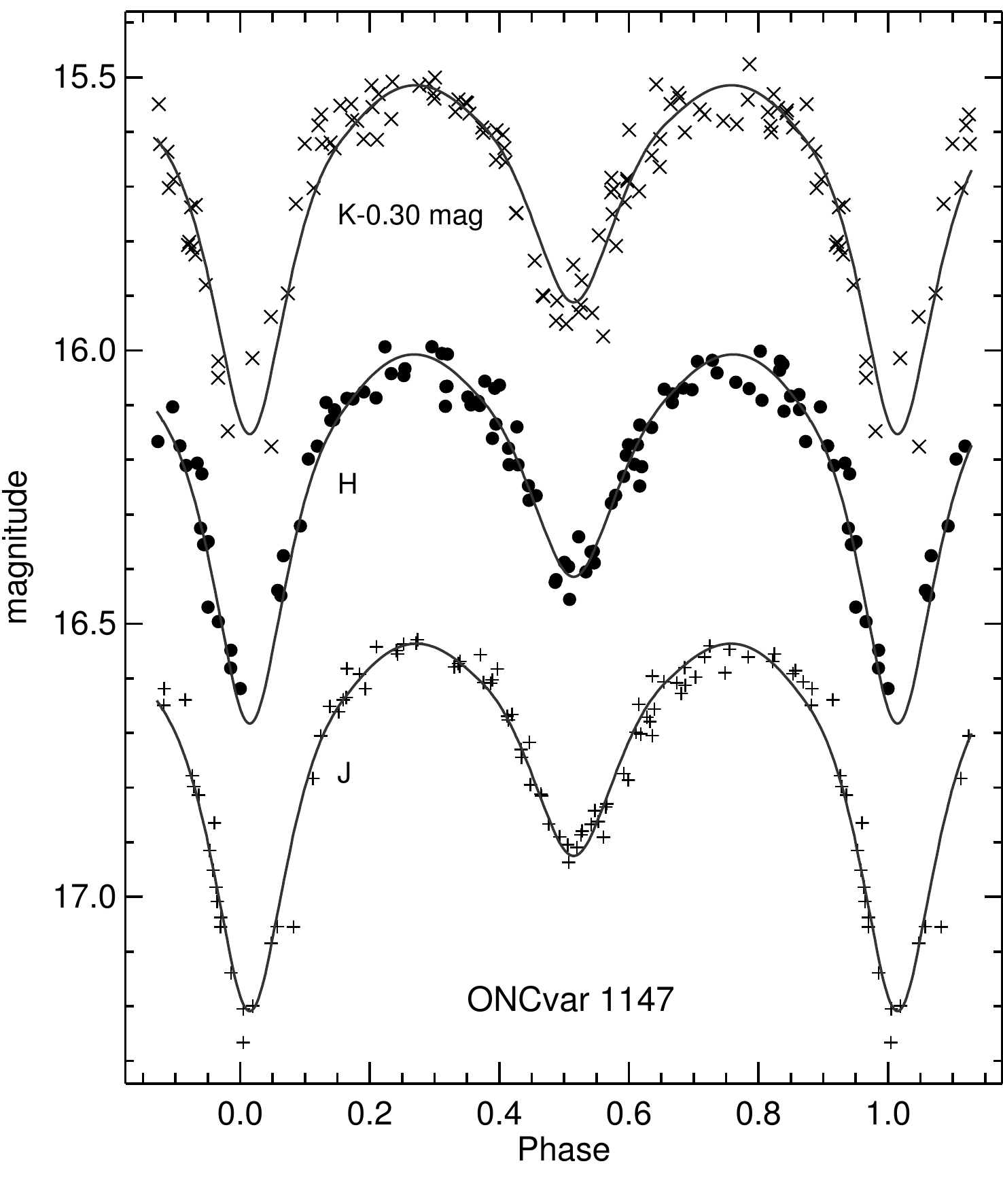}
\caption{The newly discovered eclipsing variable ONCvar 1147 shows a W UMa-type light
curve, indicating highly deformed components near contact. A WD model
(\citealt{wilson71}; \citealt{wilson79, wilson93}; \citealt{vaz07}
and references given therein) of the system is here laid over the $JHK$-light curves.
The $K$ magnitude values shown were displaced by $-0.3$ mag, in order to prevent
overlap with the $H$-light curve.
  \label{ONCvar1147}}
\end{figure*}

This newly discovered eclipsing binary has poor phase coverage, with
only half a dozen observations within the eclipses, as seen in Figure~\ref{3newEBs}, bottom row. Nonetheless the
coverage of the constant parts of the light curve is good, and the
data show well-separated narrow minima indicating non-interacting stars.  
Our best estimate for the period
is 3.031 days. The star is not included in the 2MASS point source
catalog, although it is visible on individual frames.  The position of
the star in the $J-H$ vs. $H-K$ color-color diagram indicates no evidence
for circumstellar material, only a considerable extinction, in fact
the star is not optically visible on a SERC $I$-band sky survey image. We
can thus not exclude that this is a background field star unrelated to
the ONC, although it is located close to young stars, the nearest
young star (V1775 Ori) is only 23 arcsec away.

\subsection{V2716 Ori}

\begin{figure}
\epsscale{1}
\plotone{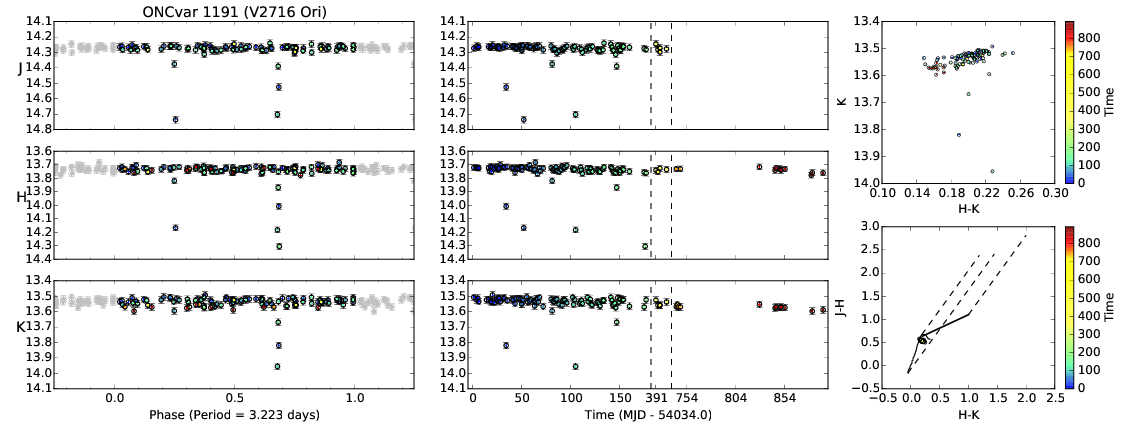}
\caption{
  One of the four newly discovered eclipsing binaries, with a tentative period of 3.223 days.
  \label{onc1191}
}
\end{figure}

While the eclipses of this object are pronounced with narrow deep
minima up to half a magnitude, the eclipses are poorly covered; see Figure \ref{onc1191}. The
minimum spacing between the observed eclipses is about 20 days, and
the period of 3.22 days is highly uncertain. The star has a GCVS
variable star name because \citet{carpenter01} found long-term
variability during their survey, with a Stetson index of 1.18. We see
little evidence for variability outside eclipses. The infrared colors
give no evidence for infrared excess, and the system suffers little or
no extinction.

%%%%%%%%%%%%%%%%%%%%%%%%%%%%%%%%%%%%%%%%%%%%%%%%%%%%%%%%%%%%%%%%%%
%%%%%%%%%%%%%%%%%%%%%%%%%%%%%%%%%%%%%%%%%%%%%%%%%%%%%%%%%%%%%%%%%%
\begin{deluxetable}{rlcccrrrrrcl}
%\tabletypesize{\normalsize}
\tabletypesize{\scriptsize}
\tablecaption{Eclipsing Binaries towards M42 within WFCAM field \label{table:EB}}
\tablewidth{0pt}
\tablehead{
\colhead{ONCvar} &
\colhead{Other ID} &  
\colhead{2MASS} & 
\colhead{$\alpha_{2000}$} &
\colhead{$\delta_{2000}$} &   
\colhead{I$^a$} &
\colhead{J$^a$} & 
\colhead{H$^a$} &
\colhead{K$^a$} &
\colhead{P[d]$^b$} &
\colhead{Ref.$^c$} &
}
\startdata
 47 &V1916 Ori&J05340205-0517256&05:34:02.06&-05:17:25.7&17.30&15.29&14.18&13.72&1.11&this paper\\
1147&         &J05362022-0514214&05:36:20.21&-05:14:21.7&18.18&16.74&16.19&15.97&0.40&this paper\\
1190&         &J05365393-0525550&05:36:53.94&-05:25:55.1&    &16.67&15.24&14.61&3.03&this paper\\
1191&V2716 Ori&J05365496-0458198&05:36:54.97&-04:58:19.9&16.18&14.27&13.73&13.54&3.22:&this paper\\
\cline{1-12}
 122&V1174 Ori&J05342785-0541378&05:34:27.85&-05:41:37.8&12.36&11.16&10.57&10.39&2.63 &Sta04 \\
 427&JW 276   &J05350571-0523540&05:35:05.71&-05:23:54.1&13.66&12.10&11.47&11.15&20.48&Mor12 \\
 518&Par 1802 &J05351113-0536511&05:35:11.14&-05:36:51.2&12.44&11.04&10.24& 9.94&4.67&Car08 \\
1227&JW 380   &J05351214-0531388&05:35:12.15&-05:31:38.9&13.55&12.06&11.44&11.15&2.65&Irw07 \\
    &V1501 Ori&J05351554-0525140&05:35:15.55&-05:25:14.1&11.55&10.02& 9.05& 8.73&?&Mor12 \\
    &$\theta^1$ Ori E&J05351577-0523100&05:35:15.76&-05:23:09.9& 6.63& 6.64& 6.24& 6.06&9.89&Mor12 \\
    & [H97b]9209&J05351802-0522052&05:35:18.03&-05:22:05.4&14.69&11.69&10.43& 9.54&5.62&Mor12 \\
 788&V2384 Ori&J05352184-0546085&05:35:21.84&-05:46:08.5&17.28&14.64&14.02&13.53&9.78&Sta06 \\
1100&         &J05360595-0500413&05:36:05.95&-05:00:41.2&17.60&14.90&14.11&13.80&3.57&Mor12 \\
\enddata

%% Text for table notes should follow after the \enddata but before
%% the \end{deluxetable}. Make sure there is at least one \tablenotemark
%% in the table for each \tablenotetext.

\tablenotetext{a} {$I$-magnitudes are from the literature, $J,H,K$-magnitudes are mean values from our photometry, or, in the case of the three stars V1501 Ori, $\theta^1$ Ori E, and [H97b]9209, which are all saturated in our data, from the 2MASS point source catalog or \citet{carpenter01}}
\tablenotetext{b}{Periods rounded to two decimals. Periods for the
  four new eclipsing binaries are best estimates from the current
  data; a trailing colon on (3.22:) indicates a tentative period fit.} 
\tablenotetext{c}{Car08: Cargile et al. (2008), Irw07: Irwin et al.
  (2007), Mor12: Morales-Calder\'on et al. (2012), Sta04: Stassun et
  al. (2004), Sta06: Stassun et al. (2006), }

\end{deluxetable}

%\clearpage

\section{Concluding Commentary}
\label{sec:discussion}

%!TEX root = varm42.tex

In addition to the previous studies of infrared variability in the Orion Nebula Cluster 
carried out by \citet{carpenter01} and \citet{morales11}, a number of other recent 
variability studies of young stars have been published.
In this section we relate our results to other such studies,
which include near-infrared monitoring in 
$\rho$ Oph (\citealt{parks14}),
and Cyg OB7 
\citep{rice12, wolk13b},
as well as \textit{Spitzer} mid-infrared monitoring in
NGC 2264 
\citep{cody14, stauffer14} and
IC 348 \citep{flaherty13},
Additionally, disk-related long-term variability in young stars was studied by \citet{findeisen13} in the North America Nebula at $R$ band.

Carpenter \e (2001; $JHK_s$) and Morales-Calder\'on \e (2011; $[3.6]$ and $[4.5]$) made the first attempts to get a detailed handle on variability in their respective wavelength regimes.
\citet{carpenter01} monitored $\sim 18,000$ stars at $JHK_s$ in a wide region overlapping the current study, and found 1235 variables, noting that much of the variability seen could be explained by cool or hot starspots, with some contribution from variable dust obscuration and accretion/disk activity.
\citet{morales11} monitored 2069 objects (1249 known Orion YSOs plus 820 other likely Orion members) and found 1259 variables; 
they took special note of a class of stars which showed periodic or aperiodic flux dips consistent with short dust eclipses, called ``dippers''.
\citet{flaherty13} found 125 variables among 269 known cluster members in IC 348, and used models of the inner disk structure to conclude that at mid-infrared wavelengths, structural perturbations (kinks, warps, and dents) in the inner disk could drive much variability over periods of days to weeks.
In NGC 2264, \citet{cody14} and \citet{stauffer14} monitored $\sim 1500$ stars including 490 cluster members and $> 1000$ field stars;
\citet{stauffer14} found 23 ``burst-dominated'' light curves among them, while \citet{cody14} studied the variability of 162 T Tauri stars,
presenting ``stochasticity'' (or departures from perfect periodicity) as an important metric when classifying YSO light curve morphology.
In the North America Nebula, \citet{findeisen13} monitored $\sim 59,000$ stars and found 3000 variables, including 41 distinct ``bursters'' or ``faders''.
\citet{rice12} studied $\sim 9200$ stars towards Cyg OB7 and identified $\sim 200$ variables, of which 30 showed a near-infrared excess;
using the Cyg OB7 data, Wolk \e (2013) built on the work of \citet{carpenter01} to identify physical mechanisms that could be associated with the observed color trends.

An advantage to our focus on a region such as the Orion Nebula Cluster, with its rich history of detailed studies, is that we have independent determination of the evolutionary classes of nearly all of our variables (primarily via \citealt{megeath12}).
Thus we can comment directly on the near-infrared variability characteristics of young stars in different classes.
\citet{morales11} monitored a mostly overlapping region and also used the same data for evolutionary class assignment, so we can directly compare our near-infrared variability results to their mid-IR results, with the caveats that our study has a much longer time baseline.
Of their 126 Class I sources, 106 were variable, 71 with amplitude $> 0.2$ mag (67\%), 15 with amplitude $> 0.5$ mag (14\%), 3 with amplitude $> 1$ mag (3\%), and 3 with $[4.5]-[3.6]$ color amplitude $> 0.2$ (3\%).
Among $Q=2$ stars in this study, we identify 11 Class I stars, all of which are variable: 9 with amplitude $> 0.2$ mag ($82 \pm 27\%$), 6 with amplitude $> 0.5$ mag ($55 \pm 22\%$), 3 with amplitude $> 1$ mag ($27 \pm 16\%$), and 8 with $H-K$ color amplitude $> 0.2$ ($73 \pm 26\%$).
These numbers suggest two different interpretations: 
perhaps the most extremely variable Class I stars are also brighter in the near-infrared than less-variable ones (reflecting an observational bias in our data),
or that their measured variability increases dramatically over longer timescales.

\citet{carpenter01} measured the fraction of significant color variables that showed either a reddening behavior (associated with variable dust extinction) or blueing behavior (associated with accretion/disk activity); they found $\sim 6$ times more reddening-type stars than blueing-type stars.
In this study we identify a ratio of 1.5 times as many reddening to blueing stars (\S \ref{sec:aggregate}), thus finding substantially more blueing-type relative to the previous study. 
Part of this difference is likely to be from the fact that the region we study is smaller, more focused on the youngest stars in Orion, and deeper at $H$ and $K$ by several magnitudes, giving us more sensitivity to the types of stars likely to show large color changes.
But a large part of it comes from the fact that our observed baseline is longer:
we analyze in \S \ref{sec:timescales} the number of color variables detected as a function of time, and find that while the number of reddening sources rise quickly to their final value, the number of blueing sources takes longer to rise, indicating a longer intrinsic timescale for the underlying variability mechanism.
This shows that monitoring programs longer than 1-2 months are necessary to accurately study variability, and that the conclusions reached by \citet{carpenter01} on the nature of infrared variability in young stars are only true on shorter timescales.
This may reveal a population of disks whose structures are quasi-unstable and that these systems take 1-2 months to reveal themselves.
A possible origin of such instability may lie near, but not at, the inner disk boundary.

While infrared monitoring studies such as \citet{carpenter01}, \citet{morales11}, \citet{flaherty13}, and \citet{cody14} are limited by their relatively short contiguous time coverage ($\sim 1-2$ months), some longer continuous or nearly-contiguous infrared surveys of young stars have been carried out, especially Parks \e (2014; 2.5 years in $\rho$ Oph), 
Rice \e (2012; 1.5 years in Cyg OB7), and the current study (2.5 years).
These longer studies have enabled a better understanding of changes in young stars on timescales significantly longer than the stellar rotation period or the orbital period of the inner disk; since infrared studies are best suited to studying disk activity, this longer timescale is well warranted.
\citet{rice12} showed that over 1.5 years, 36\% of stars that show a $K$-band excess will drift in color-color space between ``having an excess'' and ``not having an excess''.
\citet{parks14} show that some stars have aperiodic variability with timescales between 64 and 790 days.
They also discuss seven stars in $\rho$ Oph which clearly demonstrate multiple variability mechanisms over their $\sim 2.5$ year data.
They focus on stars which show different, overlapping periodic signatures with different amplitudes and/or morphology (such as rotation versus orbital disk phenomena), or periodic versus aperiodic variability in the same light curve.
Similar behavior is seen in this study, e.g., ONCvar 957 (\S \ref{sec:longperiods}, Figure \ref{lc:957});
in addition, we note individual stars which show different variability mechanisms that impart distinct, orthogonal color behavior in $H-K, K$ space at different times (\S \ref{sec:compound}).
In this study we also identify long-period systems, and demonstrate that the timescales involved for disk/accretion variability is typically longer than many previous YSO infrared monitoring studies.
In addition to these infrared studies, the optical study done by \citet{findeisen13} lasted 3.3 years and focused on disk-related variability.
They study a number of ``bursters'' and ``faders'', finding that fading behavior is twice as common as bursting behavior, though individual objects change the duty cycle of their bursts and fades from year to year.
We now know that $JHK$ variability, unlike most optical variability, happens on many different timescales: while ``early'' studies focused on variability tied to the rotation of the star and the orbit of the inner disk edge, 
these more recent results, especially from the present study, indicate that significant changes in accretion rate and disk structure occur on one- to many-month timescales.
If much of this month(s)-timescale ``blueing'' variability is primarily accretion-based, it may imply that a significant fraction of disks have density enhancements in the inner $0.1-0.2$ AU capable of influencing accretion.
Such density enhancements have typically not been considered in
previous models of young stellar variability, which usually consider stable accretion.
Overall, these examples of ``non-stationary'' variability 
\citep{feigelson12} indicate
that short-duration monitoring is not sufficient to characterize the
near-infrared variability of individual pre-main sequence stars, as many stars
change the typical amplitude, morphology, and mechanism of their
variability over timescales of months to years.

The conclusions of this analysis of near-infrared variability in the
Orion Nebula Cluster over a $\sim 900$ day span:

\begin{enumerate}

  \item We have identified 1203 variable stars in a nearly square degree in
    the Orion Nebula Cluster; among an unbiased subset of these, about
    24\% are periodic variables.
  \item Guided by infrared and X-ray evolutionary class determinations from Megeath \e 
    (2012) and Pillitteri \e (2013), we find that NIR variability amplitude in all bands and
    colors is a strong function of class, with most variability detected
    in the least evolved stages.
  \item Color behavior in $H-K, K$ color-magnitude space contains
    information about physical processes behind variability, with the dominant
    color-variable processes involving dust extinction and
    accretion/disk activity:
    \begin{itemize} 
    \item We observe $1.5 \times$ more color variability among YSOs due
      to dust extinction than accretion/disk activity, but the
      proportion of color variables showing accretion/disk activity increases
      steadily with observed time baseline, as does the total number of
      color variables.
    \item A substantial number of stars show ``compound'' color
      variability in which multiple modes of color behavior are seen
      separated in time for an individual star. This suggests that
      many more stars may undergo such behavior over longer
      timescales, and hints that perhaps ``primary mode of
      variability'' is appropriate only for shorter time-intervals in
      some cases.
    \end{itemize}
  \item We identify 73 ``AA Tau'' type periodic variables, which constitute 6.1\% of variables and 13.5\% of all periodic variables.
    They are predominantly disked stars; 9.8\% of all disked stars exhibit periodic, AA Tau type variability.
  \item We identify a number of long-term periodic variables,
    including three ``long-period brighteners'' in the class similar to the
    130-day and 93-day
    periodic objects identified in Plavchan \e (2008, 2013) (termed ``Inverted Eclipse'' therein) which are
    conjectured to arise from eclipses of a circumbinary disk. 
    Other special variables of interest are also discussed.
  \item We have found four previously unidentified eclipsing binaries, but it is 
    not established whether they are members of the ONC or just line-of-sight associations.

\end{enumerate}

\acknowledgments

We are grateful to Andy Adamson and the staff at UKIRT for 
their dedication in obtaining this major set of observations. 
We thank the referee for detailed comments, which improved this paper.
We thank Marcelo Guimar\~aes for his interest in the early phases of
this project.
BR acknowledges support from the NASA Origins program under grant number NNX11AK55G. 
This material is based upon work supported by the National Aeronautics
 and Space Administration through the NASA Astrobiology Institute under
 Cooperative Agreement No. NNA09DA77A issued through the Office of Space
 Science.
This research has made use of the SIMBAD database,
 operated at CDS, Strasbourg, France, and of NASA's Astrophysics Data
 System Bibliographic Services.
This research made use of Astropy, a community-developed core Python
package for Astronomy \citep{astropy13}.
This research made use of Astroquery, a suite of tools for querying online astronomical data services. 
This research has made use of the AstroBetter blog and wiki.
This research made use of APLpy, an open-source plotting package for Python hosted at \url{http://aplpy.github.com}.
The authors wish to recognize and acknowledge the very significant cultural role and reverence that the summit of Mauna Kea has always had within the indigenous Hawaiian community. We are most fortunate to have the opportunity to conduct observations from this mountain.

% REFERENCES
%\begin{thebibliography}{}

%\item[]{} ..............

\bibliography{tom}
  
%\clearpage 

%% Use the figure environment and \plotone or \plottwo to include 
%% figures and captions in your electronic submission.

% \begin{figure*}
% \epsscale{0.7}
% %\plotone{X.ps}
% %\plotone{Y.ps}
% %\plotone{Z.ps}
% \epsscale{1.0}
% \caption{...text...
% \label{fig0}}
% \end{figure*}

\appendix

%!TEX root = varm42.tex

\section{Automatic and Subjective Selection Criteria}
\label{sub:autosubj}
 
Our original data had a varying range of photometric quality issues for a number of reasons; these quality issues were typically flagged in the reduced photometry catalog we retrieved from the WSA (see \S \ref{subsec:data}).
These reasons include the bright, complex nebulosity in the ONC, the presence of many bright stars, the crowding towards the central Trapezium cluster, the fact that some stars are near chip edges, the variable weather from night-to-night, the fact that we were observing a huge number of stars, and the fact that some stars were near our saturation (brightness) or detection (faintness) limits for the exposure times we used. 
The process we used to select variables reflects these varying quality issues. 
Our criteria to bin stars into quality bins $Q=2$, $Q=1$, and $Q=0$ is detailed below, and is primarily a function of how many bands (out of the three $J$, $H$, $K$ bands) were deemed ``pristine'', i.e., free of photometric error flags and within brightness and faintness limits that we considered safe.

$Q=2$ is reserved for stars with ``pristine'' data in all three bands.
Stars were placed in the $Q=2$ bin if they met \textit{all} of these criteria:
\textit{a)} their PSF appeared starlike (i.e., not extended),
\textit{b)} they were observed between 50 and 125 times in each band \footnote{Notably, some stars lay on the overlap regions between detectors and were observed twice per night; sometimes these stars had erratic light curves due to a mismatch between the absolute flux calibrations of the two detectors (likely because they lay on an edge of one of these detectors).},
\textit{c)} their mean $J$ magnitude was between 11 and 17, 
$H$ between 11 and 16, and
$K$ between 11 and 16,
\textit{d)} each band had no datapoints flagged for photometric quality issues, and
\textit{e)} they did not lie in the easternmost 3\% of the observed field, which had anomalous data.

$Q=1$ is for stars with ``pristine'' data in one or two bands.
Stars were placed in $Q=1$ if they did not meet the $Q=2$ criteria but
\textit{a)} their PSF appeared starlike (i.e., not extended), they did not lie in the easternmost 3\% of the observed field, 
and one or two of their $J$, $H$, $K$ bands met the following criteria:
\textit{b)} they were observed between 50 and 125 times in this band,
\textit{c)} their mean magnitude in this band was between 11 and 16 (for $H$ and $K$) or between 11 and 17 (for $J$), 
\textit{d)} no datapoints in this band had been flagged for photometric quality issues.

$Q=0$ is for stars with some suspicious or flagged data in all three bands.
Stars were placed in $Q=0$ if they did not meet $Q=2$ or $Q=1$ but they were detected at least 50 times in at least one single band (i.e., at least 50 $J$ or 50 $H$ or 50 $K$ detections). 

Stars in $Q=2$ and $Q=1$ were added to our final variable list if they met Stetson cuts, but no stars from the $Q=0$ bin were included in our final variable list without individual, by-eye scrutiny.
Once stars are placed into the above quality bins, we select them as variable or nonvariable according to the following criteria. 
Note that $Q=2$ and $Q=1$ stars are selected automatically, while every $Q=0$ star is classified by eye.
$Q=2$ stars are classified as variable if, in addition to meeting all of the $Q=2$ criteria, their Stetson index exceeds 1.0.
$Q=1$ stars are classified as variable if their $S \ge 1.0$ and their one or two ``pristine'' data bands are variable independently of the rest.
$Q=0$ stars are classified as variable if they have $S \ge 1.0$ and pass individual scrutiny where they are inspected for:
{\it a)} bands co-vary in a consistent way through clear multiple-day trends, OR
{\it b)} bands co-vary in a clearly periodic way. 
These two criteria are intended to exclude any variability that could feasibly be induced by instrumental or processing artifacts; as a side-effect, no short-term irregular variables were included via this criterion.

To find periodic signals in $Q=0$ stars, all $Q=0$ stars with $S \ge 1.0$ were run through the period-finding search described in \S \ref{subsubsection:autoperiod}. 
$Q=0$ stars that returned a strong period through this search were considered ``subjective periodic candidates'' and those that did not return a strong period were considered ``subjective nonperiodic candidates''. 

The total pool of $Q=0$ sources numbered 11,136. 
Of these, 2,273 had $S \ge 1.0$, though this Stetson value was influenced by corrupted data in a large number of cases. 
Our algorithm described in \S \ref{subsubsection:autoperiod} detected periods in 231 of the $Q=0, S \ge 1$ sources.
After going through all of these 2273 light curves by eye, we confirmed 153 non-periodic and 192 periodic $Q=0$ stars for our final variable list.

%!TEX root = varm42.tex

\section{LOW-AMPLITUDE VARIABLE CANDIDATES}
\label{sec:lowvars}

While the focus of our present study is strong variables with $S \ge 1.0$, we examined possible variability in stars whose Stetson values lay between 0.55 and 1.0. 
The exact value of $S$ which defines variability is difficult to assign, and depends on the coverage and noise properties of the data. 
For example,  \citet{carpenter01} use $S > 0.55$ as the cutoff based on visual inspection of light curves along with concurrence from the $\chi^2$ value, which independently indicates that many of these stars are variable.
\citet{flaherty13} used a similar method but selected $S \ge 0.45$ for their dataset.
For the bulk of this study we focus on strongly variable stars with $S > 1$. 
However, some stars with Stetson index between 0.55 and 1.0 are certainly variable as well.  
We discuss this subsample as low-amplitude variable candidates, with the term ``candidate'' intended to convey that we may be seeing weak correlated instrumental or observational effects as well (such as were suspected by \citealt{carpenter01}).

Figure \ref{fig:stetson_vs_hmag} shows the Stetson index $S$ for all 2348 $Q=2$ sources included in the initial study. 
Among these $Q=2$ stars, 506 had a Stetson value within $0.55 < S < 1.0$.
We decided that, with only one exception, we could safely ignore all of these stars' variability with no impact on our study's objective.
The single exception was a star whose very unusual anticorrelated variability between its $J$ and $K$ bands pushed its Stetson value down to $S=0.97$; we found its light curve at $K$ band to have a period of 88.5 days after further analysis. 
We included this star as ONCvar 1226 in our final variable list, and it is called out for discussion in \S \ref{sec:flareboys} and \S \ref{sec:longperiods}.

The 506 low-amplitude variable candidate ($1 < S < 0.55$) light curves were period-searched in the same manner as described in \S \ref{subsubsection:autoperiod}; 55 of them (10.8\%) showed mathematically significant periods.
This is roughly one-fourth the rate (42.1\%) seen among $Q=2,~S\ge 1$ variables;
these disparate periodicity fractions may be an indication that many of the $1 < S < 0.55$ stars are not true variables.
This is especially true since we expect most stars with such small variability amplitudes to be diskless stars, and diskless stars are noted in \S \ref{sec:crossmatching} to have a \textit{higher} periodicity fraction than less-evolved, more stochastic higher-amplitude variables.

\end{document}